\documentclass[letterpaper]{article} 
\usepackage[utf8]{inputenc}
\usepackage[usenames,dvipsnames]{color}
\usepackage{bm}
\usepackage[title]{appendix}
\usepackage{graphicx,fancyhdr,lscape,dsfont,amsmath,amssymb} 
\usepackage{color, amsthm}
\usepackage{longtable}
\usepackage{psfrag}
\usepackage{subfigure}
\usepackage{tabularx}
\usepackage{boldline,multirow}
\usepackage{stmaryrd}
\usepackage[sl]{caption}
\usepackage{appendix}
\usepackage{soul}
\usepackage{cancel}
\usepackage[noend]{algpseudocode}
\usepackage{etoolbox}
\usepackage{algorithm2e}
\usepackage{mathtools}
\usepackage{sidecap}
\usepackage{enumitem}
\sidecaptionvpos{figure}{c}

\allowdisplaybreaks

\renewcommand{\em}[1]{\it{#1}}
\newcommand{\iu}{\mathrm{i}\mkern1mu}

\begin{document}

\title{An FFT framework for simulating non-local ductile failure in heterogeneous materials\footnote{Accepted in Computer Methods in Applied Mechanics and Engineering}}
\author{M. Magri$^{1}$, S. Lucarini$^{1}$, G. Lemoine$^{2}$, L. Adam$^{2}$ and J. Segurado$^{1,3\footnote{Corresponding author}}$
\\
\\
\begin{small}
$^{1}$ Fundaci\'on IMDEA Materiales, C/ Eric Kandel 2, 
28906, Getafe, Madrid, Spain
\end{small}
 \\
 \\
\begin{small}
$^{2}$ e-Xstream Engineering, Axis Park-Building H,
Rue Emile Francqui 9, B-1435 Mont-Saint-Guibert, Belgium
 \end{small}
 \\
 \\
\begin{small}
$^{3}$ Department of Materials Science, Technical University of Madrid, E.T.S. de Ingenieros de Caminos,
28040, Madrid, Spain
\end{small}
}

\maketitle

\begin{abstract}

The simulation of fracture using continuum ductile damage models attains a pathological discretization dependence caused by strain localization, after loss of ellipticity of the problem,  in regions whose size is connected to the spatial discretization. Implicit gradient techniques suppress this problem introducing some inelastic non-local fields and solving an enriched formulation where the classical balance of linear momentum is fully coupled with a Helmholtz-type equation for each of the non-local variable. Such Helmholtz-type equations determine the distribution of the non-local fields in bands whose width is controlled by a characteristic length, independently on the spatial discretization. The numerical resolution of this coupled problem using the Finite Element method is computationally very expensive and its use to simulate the damage process in 3D multi-phase microstructures becomes prohibitive. 

In this work, we propose a novel FFT-based iterative algorithm for simulating gradient ductile damage in computational homogenization problems. In particular, the Helmholtz-type equation of the implicit gradient approach is properly generalized to model the regularization of damage in multi-phase media, where multiple damage variables and different characteristic lengths may come into play. In the proposed iterative algorithm, two distinct problems are solved in a staggered fashion: (i) a conventional mechanical problem via a FFT-Galerkin solver with mixed macroscopic loading control and (ii) the generalized Helmholtz-type equation using a Krylov-based algorithm combined with an efficient pre-conditioner. The numerical implementation is firstly validated on simple two-dimensional microstructures, showing identical responses for different spatial discretizations and reproducing a ductility change dependent on the characteristic length. Finally, the robustness and efficiency of the algorithm is demonstrated in the simulation of failure of complex 3D particle reinforced composites characterized by millions of degrees of freedom. 

\end{abstract}

\section{Introduction}

The fracture process of ductile metals has been profusely studied in the last decades and several well established models are available for the prediction, in a more or less phenomenological manner, of nucleation and evolution of damage \cite{Besson2010}. Some of the most popular examples include the Gurson model \cite{GURSON1977} (or its adaptation by Tvergaard and Needleman \cite{TVERGAARD1984157}), the Rousselier model \cite{ROUSSELIER198797}, and other models based on damage mechanics \cite{Lemaitre1985}.
Nevertheless, it is well known that the numerical solution of boundary value problems with this class of constitutive laws --- for example using Finite Elements (FE) ---  results in a pathological discretization dependence due to loss of ellipticity of the problem after strain softening  \cite{JIRASEKNOTES}. Different regularization techniques have been proposed to overcome this limitation and formulate well-posed failure models \cite{PEERLINGS1996}. Among them, the approaches based on a non-local continuum \cite{ERINGEN1966179}, where additional averaging equations provide extra unknown fields to the original mechanical problem, have been studied extensively. The interest of these models is in their ability to control the size of the localization region through an uniform intrinsic length scale associated with the considered averaging equations, $\ell$. Non-local continuum mechanics was first applied to prevent damage localization in the 80s for quasi-brittle materials \cite{bazant87,bazant88}, and since then it has been widely used for the regularization of  damage for different kinds of material response, including ductile damage \cite{JBLeblond94}. Non-local approaches can be formulated as integral or gradient models  \cite{JIRASEKNOTES}. The latter is the most used formulation since it exploits the differential form of the non-local approach and, therefore, is more prone to the numerical implementation through techniques as FE. In their classical implementation \cite{PEERLINGS1996}, higher-order or implicit gradient approaches enhance the constitutive equations through a non-local field, typically the equivalent plastic strain $\bar{\varepsilon}_{eq}$, which is obtained from the corresponding local field ${\varepsilon}_{eq}$, by solving the Helmholtz-type partial differential equation 

\begin{equation} \label{eq:non-localPeerlings}
\bar{\varepsilon}_{eq} - \ell^2 \nabla^2  \bar{\varepsilon}_{eq} = \varepsilon_{eq} \, . 
\end{equation}

\noindent
In this way, the resulting formulation consists of a system of two coupled partial differential equations: the equation for the mechanical equilibrium -- i.e. $\text{div} [{\boldsymbol{\sigma}}] = \vec{0}$ with $\boldsymbol{\sigma}$ denoting the Cauchy stress tensor --  and the additional equation for computing the non-local field, i.e. Eq. \eqref{eq:non-localPeerlings}. It is worth nothing that, from a mathematical view point, the implicit gradient regularization resembles phase-field fracture models as a particular case of averaging equation \cite{DEBORST201678,Steinke2017}.  Non-local damage mechanics via the implicit gradient approach has been mainly implemented in FE by means of a monolithic scheme of the overall problem as pursued in many relevant works that studied the application of such a regularization technique on different types of damage models \cite{LANGENFELD2020112717, NGUYEN2020103891}.
 However, the elevated computational cost of the FE implementation limits the geometrical complexity and the discretization level of the numerical simulations that, in most cases, are restricted to two-dimensional problems \cite{THAMBURAJA2019871,SEUPEL2020106817,doi:10.1002/nme.6074}.
 
Simulation of ductile damage has a lot of interest in micromechanics in order to capture the effect of the microstructure -- for example in metal matrix composites \cite{SHAKOOR2018110,LLORCA2004267} and metallic porous materials or foams \cite{weck07,ZYBELL20148,Amani2018,AMANI2018395} -- on the nucleation and development of failure. In micromechanical simulations, the boundary value problem is solved for a multi-phase representative volume element of the microstructure (RVE). This RVE usually contains a complex geometrical arrangement of the phases to statistically represent the microstructure, and this complexity typically requires a fine discretization in the numerical scheme. In such a scenario, the computational cost of the non-local regularization of damage, in the realm of FE solvers, becomes particularly expensive so that many studies just exploit standard local damage models that, therefore, limit the validity of the results to the adopted level of the spatial discretization \cite{SHAKOOR2018110,LLORCA2004267,Amani2018,AMANI2018395}. Moreover, the use of periodic boundary conditions in micromechanics (the most accurate approach for computational homogenization \cite{Geers2010}) introduces additional conditions for meshing and further increases the computational cost of FE models \cite{Bohm2004}.  Only a few examples that implement non-local approaches in the context of FE homogenization can be found in the literature \cite{DRABEK2005329,DRABEK200629,reusch08}. An integral non-local regularization of several ductile damage models is developed in \cite{DRABEK2005329,DRABEK200629}  to account for the effect of particle distribution in the fracture of metal matrix composites. However, the resulting formulation only partially alleviates the mesh dependence of the numerical predictions as it relies on an explicit integration of the non-local relation. On the other hand, Reusch et al. \cite{reusch08} implemented a rate-dependent implicit gradient extension of the Gurson model for the simulation of metal matrix composites of two dimensional RVEs in an embedded-cell approach. In this case, the integration is implicit but limited to two dimensional problems containing only a few particles. 


Efficient alternatives to FE in the field of computational homogenization are the methods based on the Fast-Fourier-Transforms (FFT) pioneered by H. Moulinec and P. Suquet in the 90s \cite{MOULINEC1994,MOULINEC1998}.  FFT based methods -- also referred as spectral solvers -- offer many advantages with respect to traditional FE and currently represent well established and mature approaches for micromechanics. For instance, periodic boundary conditions arise naturally in FFT solvers and no meshing is necessary so that digital images of the microstructure can be directly used. 
Spectral solvers were first applied to computational homogenization of linear elastic problems at small strains, where the resulting Lippmann-Schwinger equation was resolved iteratively by means of the so-called \textit{basic scheme} \cite{MOULINEC1994}. To accelerate the rate of convergence, particularly poor in case of high phase stiffness contrast, the basic scheme has been object of modifications resulting in the form of accelerated and augmented Lagrangian schemes \cite{Michel2001}. In addition, this class of FFT iterative solvers was also extended to the solution of mechanical problems with non-linear constitutive behavior \cite{Michel2001} and in finite strains analysis \cite{KABEL2014}, thus allowing the implementation of a broad class of mechanical problems in solid mechanics. More recently, an alternative FFT approach was developed from the Galerkin method by Vond\v{r}ejc et al. \cite{VONDREJC2014} and later extended to non-linear problems \cite{ZEMAN2017,DEGEUS2017}. Such an approach is extremely advantageous in the context of computational mechanics since it is derived starting from the same variational scheme that FE solvers are based on. Indeed, upon algorithmic linearization, the FFT-Galerkin scheme results in the same material residuals and tangent operators that characterize the counterpart FE implementation. This simple adaptation of complex material models together with the extension of the method by Lucarini and Segurado \cite{LUCARINI2019} to generic macroscopic loading histories combining stress or strain control, lead to a formally identical computational homogenization framework to FE in terms of material models, microstructure, and loading history.


In spite of the great potential of FFT-based solvers for fracture problems in a micromechanical context, only a few works can be found in this regard \cite{Li2012,Diehl2017,ERNESTI2020112793,MA2020,BOEFF2015373}. Li et al. \cite{Li2012} proposes a model based on non-local damage mechanics employing an integral approach to define non-local stresses. However, the model is not properly a non-local approach since damage does not evolve naturally but is applied locally to somehow reproduce a crack of a voxel width. On the other hand, Diehl et al. \cite{Diehl2017}, Ernesti et al. \cite{ERNESTI2020112793}, and Ma et al. \cite{MA2020} focus on phase-field brittle fracture. In \cite{Diehl2017} it is proposed, in the context of polycrystals, a hybrid FE--FFT method due to the numerical artifacts caused by oscillations in a pure spectral approach. Nevertheless, a hybrid method does not fully exploit the numerical performance of a pure spectral solver and implies the use of regular cubic meshes, thus losing the benefit of mesh adaptivity of the finite element method. In \cite{ERNESTI2020112793,MA2020} pure FFT based solvers are proposed for phase-field fracture. In both cases the formulation is implemented through a staggered scheme. In \cite{MA2020} the focus was made on simulating the fracture propagation in three-dimensional polycrystals while in \cite{ERNESTI2020112793} the model was applied to matrix brittle damage on composites. 
To our knowledge, the only spectral implementation of a non-local ductile damage approach is presented in the paper of Boeff et al. \cite{BOEFF2015373}. In this work an iterative algorithm is proposed for the solution of an implicit gradient regularization of a simple damage model. Despite the undeniable innovative character of this work, the proposed algorithm presents significant limitations. Firstly, the basic scheme of Moulinec and Suquet is exploited in the FFT implementation. This algorithmic choice strongly limits the applicability of the proposed approach since, even for small phase property contrast, the development of damage introduces regions with very low stiffness making the convergence really poor. Due to this limitation, the proposed approach allows only for the simulation of relatively simple geometries in a two-dimensional setting.  Secondly, the framework does not account for the material heterogeneity at microstructural level as the non-local regularization is applied to the full simulation domain, including the regions occupied by phases which do not consider damage, and assuming a uniform characteristic length of the regularization. This approximation simplifies the implementation of the regularized model but results in a wrong spatial distribution of the non-local damage showing a non physical diffusion of damage through the interface between damaged and undamaged phases.

Motivated by the aforementioned limitations, in this work we present a general, robust and efficient algorithmic implementation suitable for the numerical solution of non-local ductile fracture in heterogeneous media by means of FFT. The proposed scheme is based on implicit gradient regularization, as discussed in Section \ref{sec:non-local-theories}, and is  applied to two different ductile damage models, i.e. the micro mechanical model by Gurson-??Tvergaard-??Needleman (GTN) \cite{TVERGAARD1984157} and the Lemaitre \cite{Lemaitre1985} model. The problem consists of an enriched continuum formulation where the classical balance of linear momentum is coupled with auxiliary equations of Helmholtz-type. To model effectively the non-local extension of the considered damage laws for multi-phase RVEs, the implicit gradient averaging equation is properly generalized to the case of heterogeneous materials by prescribing a non-uniform characteristic length. The resulting model is solved by means of the iterative staggered algorithm presented in Section \ref{sec:FFTimplementation} which exploits a sequential usage of FFT-Galerkin and conjugate gradient schemes. In particular, the introduction of a non-uniform characteristic length in the proposed non-local regularization implies that the Helmholtz-type equation becomes implicit. This particular fact renders the solution of the associated problem far from being trivial and not yet attempted in a similar framework to our knowledge. To get an efficient stable solver, the proposed conjugate gradient scheme combines discrete Fourier derivatives, a Krylov solver, and an \textit{ad-hoc} preconditioner. Finally, in Section \ref{sec:examples}, numerical examples of representative two dimensional and three dimensional RVEs are carried out to study the impact of the non-local regularization on heterogeneous materials.

\section{Non-local regularization of ductile damage} \label{sec:non-local-theories}

\subsection{Review of some models for ductile damage}

In this section, the original models for ductile fracture to which the non-local regularization will be applied are recalled in their essentials. For further details the reader is invited to refer to the relevant literature.  For the scope of the present study the infinitesimal strain theory is employed. Therefore, we assume a standard additive decomposition of the rate of strain tensor $\dot{\boldsymbol{\varepsilon } } $ of type

\begin{equation*}
\dot{\boldsymbol{\varepsilon}} = \dot{\boldsymbol{\varepsilon}}^e + \dot{\boldsymbol{\varepsilon}}^p 
\end{equation*}

\noindent
where $\boldsymbol{\varepsilon}^e$ and $\boldsymbol{\varepsilon}^p$ are the elastic and inelastic part of the strain respectively while a superposed dot indicates a partial derivative with respect to time (or, more precisely, \textit{pseudo-time}). Mechanical equilibrium is imposed by the local form of the balance of linear momentum in the absence of inertial and body forces

\begin{equation} \label{eq:balance_of_forces}
\text{div} \left[ \boldsymbol{\sigma} \right] = \vec{0} \, ,
\end{equation}

\noindent
where $\boldsymbol{\sigma}$ is the nominal Cauchy stress tensor.

\subsubsection{Gurson type model} \label{sec:gurson-type}

We consider the phenomenological extension proposed by Tvergaard and Needleman \cite{TVERGAARD1984157} of the physically-based model for void growth of Gurson \cite{GURSON1977}. The model is based on the definition of the so-called effective porosity $f_*$, which quantifies the level of damage induced by the presence of voids in the material. The mechanical degradation related to the presence of porosity is taken into account by the definition of the following yield surface

\begin{equation} \label{eq:yield_gurson}
\phi \left( \boldsymbol{\sigma}, \varepsilon_0^p, f_* \right) = \left( \frac{s}{\sigma_0}\right)^2 + 2 \, f_* \, q_1 \, \text{cosh} \left( - \frac{3}{2} \frac{q_2 \, p }{\sigma_0}\right) - \left( 1 + q_3 \, f_*^2  \right) \, , 
\end{equation}

\noindent
where $\varepsilon_0^p$ is the matrix equivalent plastic strain and $\sigma_0$ the matrix flow stress. Symbols $q_1$, $q_2$, $q_3$ are phenomenological coefficients, while $s$ and $p$ refer to the Mises equivalent stress and hydrostatic pressure, respectively, i.e.

\begin{equation*}
s = \sqrt{\frac{3}{2} \, \boldsymbol{s} : \boldsymbol{s} } \quad \text{and} \quad p = - \frac{1}{3} \text{tr}\left[ \boldsymbol{\sigma}\right] \, ,
\end{equation*}

\noindent 
being $\boldsymbol{s}= \text{dev}\left[ \boldsymbol{\sigma} \right]$ the deviatoric stress tensor. Equation \eqref{eq:yield_gurson} implies that the plastic behavior is pressure dependent for non-zero values of the effective porosity. The latter affects the yield surface by decreasing the set of admissible stress states as $f_*$ increases, leading eventually to a complete loss of bearing capacity of the material matrix. This condition is reached for a limit value of the effective porosity $f_V^*$ that can be estimated by imposing $\phi=0$ at zero stress, namely

\begin{equation*}
f_V* = \frac{q_1+\sqrt{q_1^2 - q_3}}{q_3} \, .  
\end{equation*}

\noindent
The stress tensor $\boldsymbol{\sigma}$ is defined according to an isotropic linear elastic law in rate form

\begin{equation} \label{eq:elasticity_gurson}
\dot{\boldsymbol{\sigma}} = K \, \text{tr} \left[ \dot{\boldsymbol{ \varepsilon }} - \dot{\boldsymbol{ \varepsilon }}^p \right] \boldsymbol{I} \, + 2 \, \mu \, \text{dev} \left[\dot{\boldsymbol{\varepsilon}}-  \dot{\boldsymbol{ \varepsilon }}^p \right] \, ,
\end{equation}

\noindent where $K$ and $\mu$ are the bulk and shear moduli. The evolution of the plastic strain tensor $\boldsymbol{\varepsilon}^p$ is derived from application of the normality rule

\begin{equation} \label{eq:flow_rule_gurson}
\dot{\boldsymbol{\varepsilon}}^p = \lambda \, \frac{\partial \phi}{\partial \boldsymbol{\sigma} } = \lambda \, \left[ \frac{\partial \phi}{\partial p}  \, \frac{\partial p }{\partial \boldsymbol{\sigma}} + \frac{\partial \phi}{\partial s}  \, \frac{\partial s}{\partial \boldsymbol{\sigma}} \right] \, ,
\end{equation}

\noindent 
with $\lambda$ being the plastic multiplier such that the \textit{persistency condition} holds, i.e. $\dot{\phi} \left( \boldsymbol{\sigma}, \varepsilon_0^p, f_* \right)=0$ if $\lambda>0$ \cite{SIMOHUGHES}. Conversely, the evolution of $\varepsilon_0^p$ does not follow a normality rule, but it is instead derived from the identity

\begin{equation} \label{eq:micro-macro}
\boldsymbol{\sigma} : \dot{\boldsymbol{\varepsilon}}^p = \left( 1- f \right) \, \sigma_0 \, \dot{\varepsilon}_0^p \, , 
\end{equation}

\noindent
where $f$ is the actual void volume fraction defined in the spirit of Gurson \cite{GURSON1977}. Assuming a hardening law of type $\sigma_0 = \sigma_0 \left( \varepsilon_0^p \right)$ for the undamaged material, equation \eqref{eq:micro-macro} can be easily solved for $\dot{\varepsilon}_0^p$  as

\begin{equation} \label{eq:constitutive_eqpstr_gurson}
\dot{\varepsilon}_0^p = \frac{\boldsymbol{\sigma} : \dot{\boldsymbol{\varepsilon}}^p}{ \left( 1- f \right) \, \sigma_0 \left( \varepsilon_0^p \right) } \, .
\end{equation}

\noindent The kinetics of evolution of the effective porosity $f_*$ is established to model the physical processes taking place in ductile fracture of metals, namely (i) void nucleation, (ii) void growth, and ultimately (iii) formation of macro cracks due to void coalescence. The latter process is prescribed using the following  phenomenological law

\begin{equation} \label{eq:effective_porosity}
f_* \, ( \, f  \, )= \left\{
\begin{aligned}
& \, f  \qquad && \text{if} \quad f < f_C  \, , \\
& \, f_C + \frac{f_V^* - f_C}{f_F - f_C} \, (f -f_C) \qquad \, && \text{if} \quad f_C \le f  < f_F \, ,\\
& \, f_V^* \qquad \, && \text{if} \quad f  \ge f_F \, ,
\end{aligned}
\right.
\end{equation}

\noindent
where $f_C$ is a critical void volume fraction above which void coalescence activates, while $f_F$ refers to the void volume fraction at fracture. Finally, the evolution of $f$ reflects the combination of the nucleation and growth processes 

\begin{equation} \label{eq: local_porosity_gurson}
\dot{f} = \mathcal{A}_N \left(\varepsilon_0^p \right) \dot{\varepsilon}_0^p + \left( 1- f  \right) \, \text{tr} \left[  \dot{\boldsymbol{\varepsilon}}^p \right] \, ,
\end{equation}

\noindent  where $\mathcal{A}_N $ is a strain rate controlled nucleation rate that, according to Chu and Needleman \cite{CHU-NEEDLEMAN1980}, can be specified as
 
\begin{equation} \label{eq:nucleation}
\mathcal{A}_N = \frac{f_N}{s_N \, \sqrt{2 \, \pi }} \, \text{exp}\left[ - \frac{1}{2} \left( \frac{\varepsilon_0^p - \varepsilon_N}{ s_N }\right)^2 \right] \,.
\end{equation}

\noindent
In equation \eqref{eq:nucleation}, $f_N$, $\varepsilon_N$ and $s_N$ are material parameters. $f_N$  represents the volume fraction of void nucleating particles while $\varepsilon_N$ and $s_N$ are material parameters.

\subsubsection{Lemaitre type model} \label{sec:lemaitre-type}

According to the pioneering work of Lemaitre \cite{Lemaitre1985}, the presence of microscopic defects, such as microcracks and cavities, is described at the macroscale by defining an intrinsic damage variable $D$. For an undamaged material $D=0$, $D=1$ implies a state of rupture, while $0<D<1$ reflects a damaged material. The influence of $D$ on the mechanical response is specified through the definition of the so-called effective stress $\tilde{\boldsymbol{\sigma}}$

\begin{equation} \label{eq:effective_stress}
\tilde{\boldsymbol{\sigma}} = \frac{\boldsymbol{\sigma}}{\left( 1- D \right)} \, ,
\end{equation}

\noindent
which identifies the actual stress experienced by the pristine material that effectively carries  a mechanical load. The effective stress is then specified as function of the elastic deformation by means of a constitutive equation. In the simple case of standard $J_2$ plasticity with isotropic hardening it results 

\begin{equation*}
\dot{\tilde{\boldsymbol{\sigma}}} = K \, \text{tr} \left[ \dot{\boldsymbol{ \varepsilon }} - \dot{\boldsymbol{ \varepsilon }}^p \right] \boldsymbol{I} \, + 2 \, \mu \, \text{dev} \left[\dot{\boldsymbol{\varepsilon}}-  \dot{\boldsymbol{ \varepsilon }}^p \right] \, ,
\end{equation*}

\noindent where the evolution of the plastic strain is derived from the yield function

\begin{equation} \label{eq:elasticity_lemaitre}
\phi( \tilde{\boldsymbol{\sigma}} , \, \epsilon_p ) = \| \tilde{\boldsymbol{s}} \| - \sqrt{\frac{2}{3}} \, \sigma_0 \left( \epsilon_p \right) \, ,
\end{equation}

\noindent
being $\| \tilde{\boldsymbol{s}} \| = \sqrt{\text{dev}[\tilde{\boldsymbol{\sigma}}] : \text{dev}[\tilde{\boldsymbol{\sigma}}]}$, $\sigma_0\left( \epsilon_p \right)$ the flow stress, and $\epsilon_p$ the equivalent plastic strain. It thus follows that

\begin{subequations} \label{eq:flow_rule_lemaitre}
\begin{align}
&\dot{\boldsymbol{\varepsilon}}^p = \lambda \, \frac{\tilde{\boldsymbol{s}}}{ \| \tilde{\boldsymbol{s}} \|} \, ,  \\
&\dot{\epsilon}_p = \lambda \,  \sqrt{\frac{2}{3}} \, .
\end{align} 
\end{subequations}

\noindent
A simple linear constitutive law is employed to describe damage evolution as function of the equivalent plastic strain
\begin{equation*}
D = \left\{
\begin{aligned}
& 0 \qquad && \text{if} \quad \epsilon_p < \epsilon_C  \, ,\\
&\frac{\epsilon_p - \epsilon_C}{\epsilon_R - \epsilon_C} \qquad && \text{if} \quad \epsilon_C \le \epsilon_p < \epsilon_R  \, , \\
& 1  \qquad && \text{if} \quad \epsilon_p \ge \epsilon_R  \, ,
\end{aligned}
\right.
\end{equation*}

\noindent where $\epsilon_C$ is a damage strain threshold and $\epsilon_R$ a damage strain at failure.

\subsection{Application of implicit gradient type regularization to ductile fracture} \label{sec:nonlocal_regularization}

The aforementioned constitutive laws for ductile damage formulated in a conventional local framework suffer from ill-posedness and lack of objectivity, which results in non-converging numerical solutions upon grid refinement \cite{JIRASEKNOTES}. Generally speaking, the simulated strain field distributes in highly localized regions whose size is connected to the spatial discretization of the problem, thus leading to a mesh dependence of the obtained mechanical response. From a mathematical point of view, this aspect is related to the loss of ellipticity of the associated boundary value problem that occurs when the projection of the tangent stiffness tensor in some direction (the so-called \textit{acoustic tensor}) becomes singular. It is well known that this condition can be reached in the two ductile damage models reviewed. 

In order to formulate well-posed failure models, several regularization techniques have been applied to the original local models, including viscoplastic regularization, micro-polar and micro-morphic theories and the use of a non-local continuum approach. Focusing the attention in the non-local continuum regularization,  three categories can be recognized in the context of damage mechanics \cite{PEERLINGS1996}, namely (i) non-local integral type, (ii) explicit gradient type, and (iii) implicit gradient type. A common feature among them is that one, or more, local internal variables are made non-local through the solution of additional equations of integral or differential type. This artefact has been proven to inhibit spurious strain localization in the numerical solution of damage models if the non-localization is applied to the appropriate variables.

In the class of non-local integral models \cite{BAZANT2002}, a generic local variable $\alpha$, defined in a material point $\vec{x}$, is replaced by its non-local counterpart $\overline{\alpha}$ obtained by means of weighted averaging over a surrounding volume $V$ of $\vec{x}$, namely

\begin{equation} \label{eq:non-local-integral}
\overline{\alpha} \left(\vec{x} \right) = \int_V g ( \, \vec{x}, \vec{z} \,  )  \, \alpha \left(  \vec{z}  \right) \,  \text{d}\vec{z} \, , 
\end{equation}

\noindent
where $g$ is a suitable non-local weight function such that

\begin{equation} \label{eq:constraint_weight}
\int_V g ( \, \vec{x}, \vec{z} \,  )  \,  \text{d}\vec{z} = 1  \qquad \forall \vec{x} \in V \, , 
\end{equation}

\noindent in order to ensure that the non-local field corresponding to a uniform local field preserves its local value in the vicinity of the boundary of $V$. If isotropy is assumed in the non-local averaging, the argument of the weight function is the distance to the center

\begin{equation}
 \label{eq:isotropy} g ( \, \vec{x}, \vec{z} \,  )  = g ( \| \vec{z}- \vec{x} \|  ) \, . 
\end{equation} 

\noindent
The weight function typically contains at least one parameter with the dimension of length that characterizes the length scale of the resulting non-local model.

On the other hand, non-local gradient models are based on averaging equations of differential type that can be simply obtained as approximation of the integral equation \eqref{eq:non-local-integral}, as recalled next. Consider a second-order Taylor expansion of the local field $\alpha$ 

\begin{equation} \label{eq:taylor-exp}
\alpha \left( \, \vec{z} \, \right) \simeq \alpha \left( \, \vec{x} \, \right) + \nabla \left[ \alpha \left( \, \vec{x} \, \right) \right] \cdot \left( \vec{z} - \vec{x} \right) + \frac{1}{2} \, \nabla \left[ \nabla \left[ \alpha \left( \, \vec{x} \, \right) \right] \right] : \left( \vec{z} - \vec{x} \right) \otimes \left( \vec{z} - \vec{x} \right) \, .
\end{equation}

\noindent The non-local variable $\overline{\alpha}$, obtained through weighted volume averaging, can be approximated by substitution of \eqref{eq:taylor-exp} into \eqref{eq:non-local-integral} as

\begin{equation} \label{eq:approx_integral}
\begin{gathered}
\overline{\alpha}\left( \, \vec{x} \, \right) \simeq \int_V g ( \, \vec{x}, \vec{z} \,  )  \,   \text{d}\vec{z} \,  \alpha \left( \, \vec{x} \, \right) +  \int_V g ( \, \vec{x}, \vec{z} \,  )  \left( \vec{z} - \vec{x} \right) \,   \text{d}\vec{z} \, \cdot \nabla \left[ \alpha \left( \, \vec{x} \, \right) \right] + \,  \\
 + \frac{1}{2} \,  \int_V g ( \, \vec{x}, \vec{z} \,  )  \left( \vec{z} - \vec{x} \right) \otimes \left( \vec{z} - \vec{x} \right)  \,   \text{d}\vec{z} : \nabla \left[ \nabla \left[ \alpha \left( \, \vec{x} \, \right) \right] \right] \, .
 \end{gathered}
\end{equation}

\noindent
Exploiting equations \eqref{eq:constraint_weight} and \eqref{eq:isotropy}  and assuming that approximation \eqref{eq:taylor-exp} is sufficient, leads from \eqref{eq:approx_integral} to the characteristic averaging equation of explicit gradient models 

\begin{equation} \label{eq:exp_averaging}
\overline{\alpha} \left( \, \vec{x} \, \right) = \alpha \left( \, \vec{x} \, \right) + \ell^2 \, \nabla^2 \left[ \alpha \left( \, \vec{x} \, \right) \right] \, ,  
\end{equation}

\noindent where $\ell$ is the characteristic length, a constant parameter with units of length that is related with the size of the averaging region $V$ in Eq. (\ref{eq:constraint_weight}). Such a parameter provides an internal length scale of the model which is related to the ability of a local field to diffuse towards neighbouring grid points\footnote{Here, the length parameter $\ell$ is treated exclusively as a parameter related to the non-local regularization of damage models. However, $\ell$ might be specified in view of the physical mechanisms underlying ductile failure thus involving aspects that are out of the scope of the present contribution.}.

In the context of explicit non-local damage models, the numerical solution of equation \eqref{eq:exp_averaging} requires the usage of $\mathcal{C}^1$ continuous finite elements. Moreover, specific boundary conditions should be specified at the elasto-plastic interface (i.e. the boundary between a plastified region and an elastic one) in order to formulate physically admissible solutions, thus rendering explicit gradient models not well suited for damage mechanics. These drawbacks can be avoided simply by considering the equality

\begin{equation*}
\overline{\alpha} \left( \, \vec{x} \, \right) - \ell^2 \, \nabla^2 \left[ \overline{\alpha} \left( \, \vec{x} \, \right) \right] = \alpha  \left( \, \vec{x} \, \right)+ \ell^2 \, \nabla^2 \left[ \alpha \left( \, \vec{x} \, \right) \right] - \ell^2 \, \nabla^2 \left[  \alpha \left( \, \vec{x} \, \right) + \ell^2 \, \nabla^2 \left[ \alpha \left( \, \vec{x} \, \right) \right] \right] \, ,
\end{equation*}

\noindent
which simplifies to give the implicit gradient equation of  Helmholtz-type

\begin{equation} \label{eq:Helmholtz_hom}
\overline{\alpha}  \left( \, \vec{x} \, \right)  - \ell^2 \, \nabla^2 \left[ \overline{\alpha} \left( \, \vec{x} \, \right) \right] = \alpha \left( \, \vec{x} \, \right) \, ,
\end{equation}

\noindent
if the fourth-order term is considered negligible. Differently from the explicit gradient model \eqref{eq:exp_averaging}, the boundary condition of equation \eqref{eq:Helmholtz_hom} is specified simply as a natural boundary condition, i.e.

\begin{equation} \label{eq:bc_helmholtz}
\nabla \overline{\alpha} \cdot \vec{n} = 0 \, ,
\end{equation}

\noindent
where $\vec{n}$ denotes the external normal unit vector. This boundary condition is usually imposed in non-local damage models in order to avoid artificial damage generation at the boundary of the body.  The alternative use of Dirichlet-type boundary conditions, typically exploited in regularized brittle damage models using phase-field fracture, does not seem adequate in this class of problems as it implies spurious damage growth not correlated with plastic deformation. In addition, it can be easily proved that the solution of the differential problem \eqref{eq:exp_averaging} - \eqref{eq:bc_helmholtz}  is equivalent to the non-local integral averaging with the  Green function associated with the Helmholtz-type equation playing the role of the weight function $g$.

\subsubsection{Implicit gradient regularization in heterogeneous materials} \label{sec:Implicitgradientregularizationforheterogeneous materials}

In the derivation of the averaging partial derivative equations for gradient models, it has been made the tacit assumption that the characteristic length of the non-local regularization, $\ell$, is uniform throughout the domain. This assumption is valid only for macroscopically homogeneous media. However, while dealing with micromechanics the behavior of each phase constituting the microstructure is modeled individually and so does the damage regularization of each damageable material. Given a RVE of volume $\Omega$ containing $Q$ different phases, we denote with $\Omega_q$ the the volume occupied by the $q$-th phase, so that 

\begin{equation*}
\Omega = \bigcup_{q=1}^Q \, \Omega_q \, . 
\end{equation*}

\noindent
Under the assumption that each phase $q$ can develop damage, let $\alpha_q$ and  $\overline{\alpha_q}$ be the local and non-local variables characterizing damage in that phase. Note that each phase might be represented by a different damage model and that the choice of the physical variable  $\alpha_q$ to be regularized can also be different for each phase. The implicit gradient regularization -- Eqs. \eqref{eq:Helmholtz_hom} and \eqref{eq:bc_helmholtz} -- applies separately to all the constituting phases as follows

\begin{equation} \label{eq:damage_multi}
\left\{
\begin{aligned}
&\overline{\alpha_q} - \ell_q^2 \, \nabla^2 \left[ \, \overline{\alpha_q} \, \right] = \alpha_q &&\qquad \text{in} \quad \Omega_q \, , \\ 
\\
&\nabla \overline{\alpha_q} \cdot \vec{n} = 0 &&\qquad \text{on} \quad \partial \Omega_q \, , 
\end{aligned}
\right.
\end{equation}

\noindent
for $q=1,\, 2,\, ...\, Q$. In Eq. \eqref{eq:damage_multi}, $\ell_q$ is the characteristic length involved in the regularization of the variable $\alpha_q$, while\ $\partial\Omega_q$ refers to the boundary of $\Omega_q$. It is assumed here that the local damage variables are independent and only defined in their corresponding regions, i.e. $\alpha_q = 0 $ in $\Omega_p$ with $p \neq q$. Accordingly, the boundary condition in Eq. \eqref{eq:damage_multi} prescribes the normal derivative of the non-local variable $\overline{\alpha_q}$ to be zero at the material interfaces thus preventing propagation of damage from $\Omega_q$ to other phases. This is the typical case in which either one of the two materials in contact is not damageable (e.g. in particle reinforced metal matrix composites) or different damaging mechanisms rule the behavior of the materials at the interface (e.g. a ductile metal in contact with a brittle material). A different scenario arises if two different phases share the same damage mechanism, i.e. both materials follow the same constitutive laws but with different material parameters. In such a case, damage can diffuse across their mutual interface and no interface condition is needed for the Helmholtz-type equation of the non-local regularization.

\begin{figure}[htbp]
\centering
\includegraphics[width=12cm]{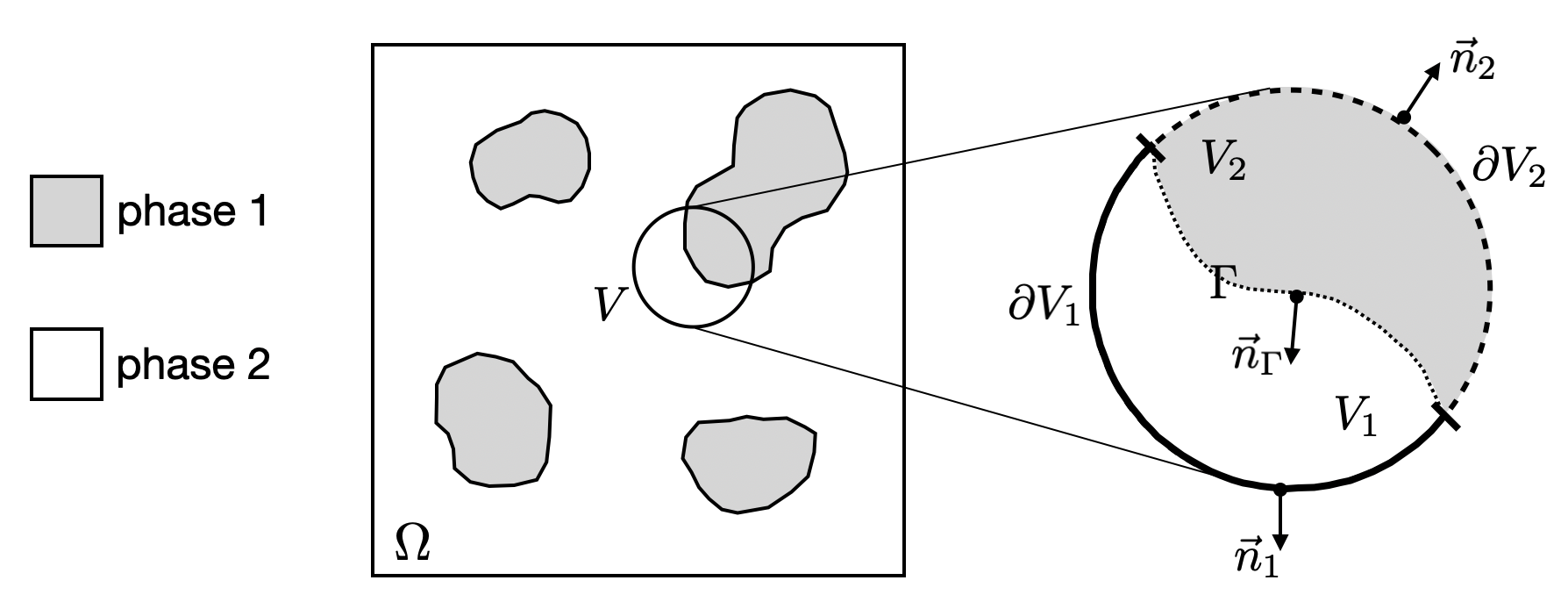}
\caption{\em{Schematic of the microscopic arrangement of an idealized biphasic composite material. In the right side of the figure it is reported an enlargement of the control volume $V$ where there is specified the presence of the subvolumes $V_1$ and $V_2$ along with and their mutual interface $\Gamma$. }}
\label{fig:interface_cond}
\end{figure}

The implementation of implicit damage regularization in FFT-based solvers requires an adaptation of the differential problem Eq. \eqref{eq:damage_multi} since the whole domain $\Omega$ have to be considered and regularly discretized. Therefore, each non-local variable $\overline{\alpha_q}$ has to be defined in the whole domain of the RVE and Eq. $\eqref{eq:damage_multi}$ is rewritten as

\begin{subequations} \label{eq:FFThetreg}

\begin{equation} \label{eq:Helmholtz}
\overline{\alpha_q} - \text{div} \left[ \widetilde{\ell_q}^2 \left( \vec{x} \right)  \, \nabla \overline{\alpha_q} \right] = \alpha_q  \qquad \text{in} \quad \Omega \, ,
\end{equation}

\noindent
where

\begin{equation} \label{eq:piecewiseell}
\widetilde{\ell_q} \left( \vec{x} \right) = \left\{ 
\begin{aligned}
&\ell_q \qquad &&  \text{if} \quad \vec{x} \in \Omega_q \, , \\
 &0 && \text{otherwise} \, , 
\end{aligned}
\right.
\end{equation}

\end{subequations}
\noindent
for $q=1,\, 2,\, ...\, Q$. Proving that Eq. \eqref{eq:FFThetreg} is equivalent to Eq. \eqref{eq:damage_multi} is straightforward and it is addressed here for the particular case of a two-phase medium as idealized in Fig. \ref{fig:interface_cond}. To this end, it is convenient to rewrite Eq. \eqref{eq:Helmholtz}, for $q=1$, in the corresponding integral form 

\begin{equation} \label{eq:int_helmholtz}
\int_V \overline{\alpha_1} - \text{div} \left[ \widetilde{\ell_1}^2 \left( \vec{x} \right)  \, \nabla \overline{\alpha_1} \right] - \alpha_1 \, \text{d}V = 0  \, ,  
\end{equation}

\noindent
where $V \in \Omega$ is a generic control volume. In a general scenario, $V$ occupies both phase 1 and phase 2 so that integral \eqref{eq:int_helmholtz} can be split in the two different regions. By doing so, and after application of the divergence theorem, Eq. \eqref{eq:int_helmholtz} reduces to

\begin{equation}\label{eq:int_helmholtz2}
\begin{gathered} 
\int_{V_1} \overline{\alpha_1}    - \alpha_1  \, \text{d}V +  \int_{V_2} \overline{\alpha_1}   \, \text{d}V   - \int_{\partial V_1}  \ell^2_1 \, \nabla  \overline{\alpha_1} \cdot \vec{n}_1 \, \text{d}S  \,
+ \int_{\Gamma} \ell^2_1 \, \nabla  \overline{\alpha_1} \cdot \vec{n}_\Gamma  \, \text{d}S  \, = 0 \, ,
\end{gathered}
\end{equation}

\noindent 
where Eq. $\eqref{eq:piecewiseell}$ has been exploited along with the condition that $\alpha_1=0$ in $\Omega_2$.
Since  Eq. \eqref{eq:int_helmholtz2} must hold for every $V$, and therefore for any $\Gamma$ in the boundary between 1 and 2, the following local equations can be derived

\begin{equation*}
\left\{
\begin{aligned}
&\overline{\alpha_1} - \ell_1^2 \, \nabla^2 \left[ \overline{\alpha_1} \left( \, \vec{x} \, \right) \right] = \alpha_1  &&\qquad \text{in} \quad \Omega_1 \, , \\
\\
&\nabla \overline{\alpha_1} \cdot \vec{n} = 0 &&\qquad \text{on} \quad \partial \Omega_1 \, , \\
\\
&\overline{\alpha_1} = 0 &&\qquad \text{in} \quad \Omega_2 \, , \\
\end{aligned}
\right.
\end{equation*}

\noindent 
which corresponds to Eq. \eqref{eq:damage_multi}, for $q=1$, in the particular case of the considered biphasic medium along with the additional condition for $\overline{\alpha_1}$ in $\Omega_2$.

\medskip 

\noindent

{\bf{Remark - }}
The zero value of the parameter $\widetilde{\ell_q}$ out of region $\Omega_q$ given by Eq. \eqref{eq:piecewiseell} satisfies the condition of no diffusion of $\overline{\alpha_q}$ in phases $p\neq q$ but leads to convergence issues in the numerical solution of Eq. \eqref{eq:Helmholtz} via spectral solvers \cite{TO2020113160}. A practical solution consists in selecting non-zero values for $\widetilde{\ell_q}$ in $\Omega_p$ ($p \neq q$) such that the prescribed contrast between different phases controls the diffusion of non-local variables at the interfaces. Focusing on the biphasic medium of Fig. \ref{fig:interface_cond} for simplicity, we would write
\begin{equation*}
\widetilde{\ell_1} \left( \vec{x} \right) = \left\{ 
\begin{aligned}
&\ell_{11} \qquad &&  \text{if} \quad \vec{x} \in \Omega_1 \, , \\
 &\ell_{12} && \text{if} \quad \vec{x}  \in \Omega_2 \, ,
\end{aligned}
\right.
\end{equation*}

\noindent 
 where the choice $\ell_{12}=0$ recovers Eq. \eqref{eq:piecewiseell}. With this new definition of $\widetilde{\ell_1}$ the integral form of Eq. \eqref{eq:FFThetreg}, for $q=1$, becomes

\begin{equation}\label{eq:int_helmholtz22}
\begin{gathered} 
\int_{V_1} \overline{\alpha_1}  - \alpha_1  \, \text{d}V - \int_{\partial V_1}  \ell^2_{11} \, \nabla  \overline{\alpha_1} \cdot \vec{n}_1 \, \text{d}S  \, + 
\int_{V_2} \overline{\alpha_1}  \, \text{d}V - \int_{\partial V_2}  \ell^2_{12} \, \nabla  \overline{\alpha_1} \cdot \vec{n}_2 \, \text{d}S  \, + \\
+ \int_{\Gamma} \ell^2_{11} \, \nabla  \overline{\alpha_{11}} \cdot \vec{n}_\Gamma -  \ell^{2}_{12} \, \nabla  \overline{\alpha_{12}} \cdot \vec{n}_\Gamma \, \text{d}S  \, = 0 \, ,
\end{gathered}
\end{equation}

\noindent 
where 

\begin{equation*}
\nabla \overline{\alpha_{11}} = \lim_{\vec{x} \in V_1 \to \Gamma}  \nabla \overline{\alpha_1} \qquad \text{and} \qquad \nabla \overline{\alpha_{12}} = \lim_{\vec{x} \in V_2 \to \Gamma}  \nabla \overline{\alpha_1} \, ,
\end{equation*}

\noindent
Since  Eq. \eqref{eq:int_helmholtz22} must hold for every $V \in \Omega$, and therefore for any $\Gamma$ in the boundary between 1 and 2, the internal boundary integral must vanish (second line of Eq. \eqref{eq:int_helmholtz22}), leading to a relation between the gradients of the non-local variable at the interface

\begin{equation*}
\frac{\nabla  \overline{\alpha_{11}} \cdot \vec{n}_\Gamma}{\nabla  \overline{\alpha_{12}} \cdot \vec{n}_\Gamma} = \left(\frac{\ell_{12}}{\ell_{11}}\right)^2 \, ,
\end{equation*}

\noindent which depends on the ratio $\ell_{12}/\ell_{11}$. The multi-phase problem introduced in Eq. (\ref{eq:damage_multi}), which assumes independent non-local fields for each phase and free Neumann boundary conditions on every interface, is recovered when $\ell_{12}/\ell_{11} \to 0$. In this case the relevant interface condition reads

\begin{equation*} 
\frac{\nabla  \overline{\alpha_{11}} \cdot \vec{n}_\Gamma}{\nabla  \overline{\alpha_{12}} \cdot \vec{n}_\Gamma} \to 0 \, , 
\end{equation*}

\noindent
that corresponds to the free Neumann boundary for original Helmholtz-type problem at the interface $\Gamma$. Therefore, to emulate this condition at a phase-interface is sufficient to chose $\ell_{12}<<\ell_{11}$ as it will force the value of the smoothed non-local variable $\overline{\alpha_1}$ to zero when approaching a point in the domain $\Omega_2$. Other values of the ratio $\ell_{12}/\ell_{11}$ can be considered from a mathematical view point but do not have a physical meaning since they allow for the diffusion of a variable out of the domain in which is defined. 
%
%
%


\subsubsection{Non-local extension of Gurson type model} \label{sec:non-local-gurson}

In the context of ductile fracture, the implicit gradient regularization can be applied either to the damage variable or to the scalar strain measures from which the evolution of damage is estimated. However, as discussed by Jir\'asek et al. \cite{JIRASEK20031553}, the first type of regularization would lead to a locking effect that ultimately renders the numerical solution unreliable. Additionally, a non-local averaging of plastic deformation would provide a more efficient regularization scheme \cite{NGUYEN2020103891}, and will be applied to the models considered in this paper.

Therefore, the Helmholtz-type equation  \eqref{eq:Helmholtz} is exploited for the regularization of both the equivalent plastic strain $\varepsilon_0^p$ and the trace of plastic strain $\text{tr} \left[ \boldsymbol{\varepsilon}^p \right]$. In this way, the law defining the porosity rate of non-local Gurson rewrites from \eqref{eq: local_porosity_gurson} as

\begin{equation} \label{eq: non-local_porosity_gurson}
\dot{f} = \mathcal{A}_N \left(\overline{\varepsilon_0^p} \right) \,  \dot{\overline{\varepsilon_0^p}} + \left( 1- f  \right) \,  \dot{\overline{\text{tr}  \, \left[ \boldsymbol{\varepsilon}^p \right]}} \,  , 
\end{equation}

\noindent where the non-local strain plastic strains $\overline{\varepsilon_0^p}$ and $\overline{\text{tr} \left[ \boldsymbol{\varepsilon}^p \right]}$ are solutions of

\begin{subequations} \label{eq:helmholta_gurson}
\begin{gather}
\overline{\varepsilon_0^p} - \text{div} \left[ \ell^2 \left( \vec{x} \right) \, \nabla \overline{\varepsilon_0^p} \right] = \varepsilon_0^p \, , \\
\overline{\text{tr} \left[ \boldsymbol{\varepsilon}^p \right]} - \text{div} \left[ \ell^2 \left( \vec{x} \right) \, \nabla \overline{\text{tr} \left[ \boldsymbol{\varepsilon}^p \right]} \right] = \text{tr} \left[ \boldsymbol{\varepsilon}^p \right] \, .
\end{gather}
\end{subequations}



\noindent
The non-local extension of Gurson model discussed in Section \ref{sec:gurson-type} thus consists of balance equations \eqref{eq:balance_of_forces} and \eqref{eq:helmholta_gurson}, completed by yield condition Eq. \eqref{eq:yield_gurson} and constitutive laws for stress \eqref{eq:elasticity_gurson}, plastic strain tensor \eqref{eq:flow_rule_gurson}, equivalent plastic strain  \eqref{eq:constitutive_eqpstr_gurson}, void volume fraction \eqref{eq: non-local_porosity_gurson}, and effective porosity \eqref{eq:effective_porosity}. 

\subsubsection{Non-local extension of Lemaitre type model} \label{sec:non-local_lemaitre}

Following the same reasoning adopted for the regularization of the Gurson model, the non-local extension of the damage variable appearing in the simplified Lemaitre model discussed in Section \ref{sec:lemaitre-type} becomes

\begin{equation} \label{eq:non-local_damage_lemaitre}
D = \left\{
\begin{aligned}
& 0 \qquad && \text{if} \quad \overline{\epsilon_p} < \epsilon_C  \, ,   \\
&\frac{\overline{\epsilon_p} - \epsilon_C}{\epsilon_R - \epsilon_C} \qquad && \text{if} \quad \epsilon_C \le \overline{\epsilon_p} < \epsilon_R  \, , \\
& 1  \qquad && \text{if} \quad \overline{\epsilon_p} \ge \epsilon_R  \, ,
\end{aligned}
\right.
\end{equation}

\noindent where the non-local equivalent plastic strain $\overline{\epsilon_p}$ is calculated from

\begin{equation} \label{eq:helmholta_lemaitre}
\overline{\epsilon_p} - \text{div} \left[ \ell^2\left( \vec{x} \right) \, \nabla \overline{\epsilon_p} \right] = \epsilon_p \, . \\
\end{equation}


\noindent
 Note that the same type of regularization has been pursued by Boeff et al. \cite{BOEFF2015373} for spectral solvers, but considering a uniform characteristic length in the Helmholtz-type equation. To sum up, the non-local extension of the considered Lemaitre-type model consists of governing equations \eqref{eq:balance_of_forces} and \eqref{eq:helmholta_lemaitre} with constitutive laws \eqref{eq:effective_stress}, \eqref{eq:elasticity_lemaitre}, \eqref{eq:flow_rule_lemaitre}, and \eqref{eq:non-local_damage_lemaitre}.

\section{Numerical implementation in FFT-based solvers} \label{sec:FFTimplementation}

 \subsection{General form of the governing equations}
In this section, the algorithm for the numerical solution of the non-local damage models analyzed in Section \ref{sec:nonlocal_regularization} will be described.  For the sake of generality, the governing equations of non-local damage mechanics are first rephrased in abstract setting. A single damaging phase will be considered and,  following the reasoning of the previous section, the variables that will be regularized, $\alpha_j$, are set to zero in the non-damaging regions. In order to recover the free Neumann boundary conditions for the non-local variables $\overline{\alpha_j}$ in the internal interphases, a non-homogeneous characteristic length $\ell(\vec{x})$ --  as defined in  Eq. \eqref{eq:piecewiseell} --  is introduced. Given a RVE of domain $\Omega$ and a time interval $[t_0, \, t_f]$ in which the macroscopic loading history is defined,  the problem consists of the following system of coupled partial differential equations

\begin{subequations} \label{eq:governing_non-local}
\begin{align} 
&&&\text{div} \left[ \boldsymbol{\sigma} \right] = \vec{0} \, , \label{eq:Lmimentum_abstract}  \\
&&&\overline{\alpha_j} - \text{div} \left[ \ell_j^2 \left( \vec{x} \right) \, \nabla \overline{\alpha_j}  \right] = \alpha_j  \,   \qquad j=1, \,  2,\,  ... \, J \,, \label{eq:helmholtz_abstract}
\end{align}
\end{subequations}

\noindent where $\overline{\alpha_j}$ is a general $j$-th non-local internal variable. Equations \eqref{eq:governing_non-local} are solved for a prescribed mixed loading history given by a combination of components of the macroscopic strain $\mathbf{E}(t)$ and stress $\boldsymbol{\Sigma}(t)$ tensors, such that

\begin{equation*}
E_{hk} (t)=  \frac{1}{V_\Omega} \, \int_\Omega \, {\varepsilon}_{hk}  \left( \vec{x} , \, t\right)\, \text{d}V\, ,   \qquad \Sigma_{HK} (t) = \frac{1}{V_\Omega} \, \int_\Omega \, {\sigma}_{HK}  \left( \vec{x} , \, t\right)\, \text{d}V\, ,  \qquad h,k \cap H,K = \emptyset \, ,
\end{equation*}

\noindent along with periodic boundary conditions for all the fields involved in the solution

\begin{subequations} \label{eq:periodicity}
\begin{align} 
&&&\boldsymbol{\varepsilon}(\vec{x})=\boldsymbol{\varepsilon}(\vec{x}+\vec{n}L) \, , \\
&&&\overline{\alpha_j}(\vec{x})=\overline{\alpha_j}(\vec{x}+\vec{n}L)  \,   \qquad j=1, \,  2,\,  ... \,  J \,,
\end{align}
\end{subequations}

\noindent
with $\vec{n}L$ referring to any vector with components obtained as the product of an integer number by the periodicity of the cell $L$.

The stress tensor $\boldsymbol{\sigma}$, plastic strain $\boldsymbol{\varepsilon}^p$, and the generic $i$-th local internal variable $\alpha_i$ follow their relevant constitutive specifications of type

\begin{subequations}\label{eq:kinetics_non-local}
\begin{align}
&&&\dot{\boldsymbol{\sigma}} = \boldsymbol{f} \left( \boldsymbol{\varepsilon}, \, \boldsymbol{\varepsilon}^p , \vec{\alpha}, \vec{\overline{\alpha}} , \dot{\boldsymbol{\varepsilon}}, \, \dot{\boldsymbol{\varepsilon}}^p , \dot{\vec{\alpha}}, \dot{\vec{\overline{\alpha}} } \right) \, , \\
&&&\dot{\boldsymbol{\varepsilon}}^p = \boldsymbol{g}  \left( \boldsymbol{\varepsilon}, \, \boldsymbol{\varepsilon}^p , \vec{\alpha}, \vec{\overline{\alpha}} , \dot{\boldsymbol{\varepsilon}}, \, \dot{\boldsymbol{\varepsilon}}^p , \dot{\vec{\alpha}}, \dot{\vec{\overline{\alpha}} } \right)  \, ,  \\\, 
&&&\dot{\alpha}_i = h_i  \left( \boldsymbol{\varepsilon}, \, \boldsymbol{\varepsilon}^p , \vec{\alpha}, \vec{\overline{\alpha}} , \dot{\boldsymbol{\varepsilon}}, \, \dot{\boldsymbol{\varepsilon}}^p , \dot{\vec{\alpha}}, \dot{\vec{\overline{\alpha}} } \right)  \,   \qquad i=1, \,  2,\,  ... \, I \,
\end{align}
\end{subequations}

\noindent which are completed by the choice of the following initial conditions

\begin{subequations}
\begin{align*}
&&& \left. \boldsymbol{\sigma} \right|_{t=t_0}  =  \boldsymbol{\sigma}^0 \, , \\
&&& \left. \boldsymbol{\varepsilon}^p \right|_{t=t_0}  =  \boldsymbol{\varepsilon}^0_p \, , \\
&&& \left. \alpha_i \right|_{t=t_0} = \alpha_i^0 \qquad   i=1, \,  2,\,  ... \,  I \,. 
\end{align*}
\end{subequations}

The evolutive equations \eqref{eq:kinetics_non-local} are approximated according to the Backward Euler method. The (pseuo) time interval $[t_0, \, t_f]$ is then divided in $N_t$ increments, $\Delta t_n = (t_{n+1} - t_{n})$ for which the discretized fields read
\begin{equation}
\left. \left( \bullet \right) \right|_{n} = \left( \bullet \right) \left( t =  t_n   \right) \, , \qquad  \Delta\left( \bullet \right) = \left. \left( \bullet \right) \right|_{n+1} - \left. \left( \bullet \right) \right|_{n} \, \qquad n=0, \, 1, \, 2, \, ... \, N_t-1 .
\end{equation}

\noindent Therefore, for any time step $n=0, \, 1, \, ... \, N_t-1$, the discretized form of equations \eqref{eq:kinetics_non-local} in time reads

\begin{subequations}  \label{eq:disctetized_const_law}
\begin{align}
&&&\Delta {\boldsymbol{\sigma}} = \Delta t_n \, \boldsymbol{f} \left( \left. \boldsymbol{\varepsilon} \right|_{n+1}, \, \left. \boldsymbol{\varepsilon}^p \right|_{n+1}, \left. \vec{\alpha} \right|_{n+1}, \left. \vec{\overline{\alpha}} \right|_{n+1},\left. \boldsymbol{\varepsilon} \right|_{n}, \, \left. \boldsymbol{\varepsilon}^p \right|_{n}, \left. \vec{\alpha} \right|_{n}, \left. \vec{\overline{\alpha}} \right|_{n} \right) \, , \\
&&&\Delta {\boldsymbol{\varepsilon}}^p = \Delta t_n \, \boldsymbol{g} \left( \left. \boldsymbol{\varepsilon} \right|_{n+1}, \, \left. \boldsymbol{\varepsilon}^p \right|_{n+1}, \left. \vec{\alpha} \right|_{n+1}, \left. \vec{\overline{\alpha}} \right|_{n+1},\left. \boldsymbol{\varepsilon} \right|_{n}, \, \left. \boldsymbol{\varepsilon}^p \right|_{n}, \left. \vec{\alpha} \right|_{n}, \left. \vec{\overline{\alpha}} \right|_{n} \right)  \, ,  \\ 
&&&\Delta {\alpha}_i = \Delta t_n \, h_i \left( \left. \boldsymbol{\varepsilon} \right|_{n+1}, \, \left. \boldsymbol{\varepsilon}^p \right|_{n+1}, \left. \vec{\alpha} \right|_{n+1}, \left. \vec{\overline{\alpha}} \right|_{n+1},\left. \boldsymbol{\varepsilon} \right|_{n}, \, \left. \boldsymbol{\varepsilon}^p \right|_{n}, \left. \vec{\alpha} \right|_{n}, \left. \vec{\overline{\alpha}} \right|_{n} \right)  \,   \qquad i=1, \,  2,\,  ... \, I \,. 
\end{align}
\end{subequations}

\noindent Specification of the time discretization of the considered models is reported in Appendix \ref{app:time_discretization}.

 \subsection{Numerical Algorithm }

The system of differential equations \eqref{eq:governing_non-local} is, in general, fully coupled because of the implicit form of equations \eqref{eq:kinetics_non-local}. In this system, the scalar Helmholtz-type equation \eqref{eq:helmholtz_abstract} 
 has to be solved as many times as the number of internal variables $\alpha_j$ that are made non-local. In FE, this class of differential problems is usually solved in a monolithic scheme, being the degrees of freedom per node the components of the displacement vector and the non-local variables $\overline{\alpha_j}$. Under this scheme, the governing equations are linearized with respect to all the field variables, forming a unique tangent operator for each iteration of the Newton-Raphson algorithm. The benefit of the monolithic implementation is in its efficiency, since using a consistent tangent theoretically leads to a quadratic convergence of the numerical solver. Nevertheless, the algorithmic performance is affected (i) by the increase of the number of unknowns  $N$ in the linear system, with a corresponding increase of computational time of order $\mathcal{O}(N^2-N^3)$\cite{HughesBookFE}; (ii) because of the resulting unsymmetric tangent matrix \cite{PEERLINGS1996,Steinke2017} that implies an increase of memory and number of operations required for the solution of the linearized problem. Moreover,  monolithic algorithms may not converge in static cases during the damage propagation, as observed in \cite{delorenzis15} for phase-field fracture models. For these reasons, staggered approaches have also been occasionally used in FE implementations of implicit gradient models \cite{WU2018612}.

In the context of FFT solvers, staggered approaches are the common choice for any type of coupled problem, see for example \cite{Shanthraj2019,BERBENNI2020103808}. This type of solver is the common choice because spectral methods do not necessarily rely on solving linearized systems of equations but might use a non-linear iterative approach. Moreover, the use of a staggered solver allows combining different spectral methods for the solution of different field equations. For these reasons, an iterated staggered scheme is proposed here to solve equations \eqref{eq:governing_non-local} as schematically presented in Algorithm \ref{al:staggered_algorithm}.

\RestyleAlgo{boxed}
\begin{algorithm}
\DontPrintSemicolon
 {\it{(a) \underline{time incrementation}}} \\
\For { $n$ \text{in} $N_t$ } {
INPUT: $\left. \boldsymbol{\varepsilon} \right|_{n}, \, \left. \boldsymbol{\varepsilon}^p \right|_{n}, \left. \vec{\alpha}\right|_{n}, \left.\vec{\overline{\alpha}}\right|_{n} $ \\
set macroscopic goal strain and stress components $ \left. E_{hk} \right|_{n+1}, \,   \left. \Sigma_{HK} \right|_{n+1}$  \\

\smallskip
{\it{(b) \underline{staggered scheme mechanics/Helmholtz equations}}} \\
set initial guess variables $\left. \boldsymbol{\varepsilon} \right|^{(0)}_{n+1}, \, \left. \boldsymbol{\varepsilon}^p \right|^{(0)}_{n+1}, \left. \vec{\alpha} \right|^{(0)}_{n+1}, \left. \vec{\overline{\alpha}} \right|^{(0)}_{n+1} $ \\
\smallskip
\While {$ERR < TOL $} {
{\it{(i) \underline{Galerkin FFT solver for the mechanics}}} \\
compute  $\left. \boldsymbol{\varepsilon} \right|^{(k+1)}_{n+1}, \, \left. \boldsymbol{\varepsilon}^p \right|^{(k+1)}_{n+1}, \left. \vec{\alpha} \right|^{(k+1)}_{n+1}$ by solving
 \begin{equation*}
          \left\{
            \begin{aligned}
             &\text{div} \left[ \left. \boldsymbol{\sigma} \right|_{n+1}^{(k+1)} \right] = \vec{0} \,  \\
            &\Delta {\boldsymbol{\varepsilon}}^p = \Delta t \, \boldsymbol{g} \left( \left. \boldsymbol{\varepsilon} \right|_{n+1}^{(k+1)}, \, \left. \boldsymbol{\varepsilon}^p \right|_{n+1}^{(k+1)}, \left. \vec{\alpha} \right|_{n+1}^{(k+1)}, \left. \vec{\overline{\alpha}} \right|_{n+1}^{(k)},\left. \boldsymbol{\varepsilon} \right|_{n}, \, \left. \boldsymbol{\varepsilon}^p \right|_{n}, \left. \vec{\alpha} \right|_{n}, \left. \vec{\overline{\alpha}} \right|_{n} \right)  \,   \\ 
            &\Delta {\alpha}_i = \Delta t \, h_i \left( \left. \boldsymbol{\varepsilon} \right|_{n+1}^{(k+1)}, \, \left. \boldsymbol{\varepsilon}^p \right|_{n+1}^{(k+1)}, \left. \vec{\alpha} \right|_{n+1}^{(k+1)}, \left. \vec{\overline{\alpha}} \right|_{n+1}^{(k)},\left. \boldsymbol{\varepsilon} \right|_{n}, \, \left. \boldsymbol{\varepsilon}^p \right|_{n}, \left. \vec{\alpha} \right|_{n}, \left. \vec{\overline{\alpha}} \right|_{n} \right)  \,   \qquad i=1, \,  2,\,  ...  \, I \,
            \end{aligned}
            \right.
            \end{equation*}
            
{\it{(ii) \underline{Spectral solver for heterogeneous Helmholtz-type equations (see Algorithm \ref{al:Helmholtz_algorithm})}}}  \\
compute $\left. \overline{\alpha_j} \right|^{(k+1)}_{n+1} \quad j=1, \,  2,\,  ... \,  J \,$ by solving 
\begin{equation*}
            \begin{aligned}
            &&&\left. \overline{\alpha_j}\right|^{(k+1)}_{n+1} - \text{div} \left[ \ell_j^2 \left( \, \vec{x} \, \right) \, \nabla \left. \overline{\alpha_j} \right|^{(k+1)}_{n+1} \right] =  \left. {\alpha}_j\right|^{(k+1)}_{n+1} \, 
            \end{aligned}
            \end{equation*}
 {\it{(iii) \underline{compute error}}}
  \begin{equation*}
             ERR = \text{max} \left[  \frac{\text{max} \left( \left|  \,  \left.\boldsymbol{\varepsilon} \right|_{n+1}^{(k+1)} - \left. \boldsymbol{\varepsilon} \right|_{n+1}^{(k)}  \, \right| \right)} {< \left. \boldsymbol{\varepsilon} \right|_{n+1} ^{(k+1)}> } , \frac{\| \left. \overline{\alpha_j}\right|^{(k+1)}_{n+1} - \left. \overline{\alpha_j}\right|^{(k)}_{n+1}  \|}{\| \left. \overline{\alpha_j} \right|^{(k+1)}_{n+1} \|} \quad j=1, \,  2,\,  ...  \, J \right]
             \end{equation*}     
} 
}
\caption{Staggered algorithm for the solution of a generic non-local problem.}\label{al:staggered_algorithm}
\end{algorithm}

Assuming that the solution at time $t_{n}$ is known, the aim is to obtain the value of the strain and the non-local fields at time $t_{n+1}$, i.e. $\left. \boldsymbol{\varepsilon} \right|_{n+1}$ and  $\left. \vec{\overline{\alpha}} \right|_{n+1}$, which fulfil \eqref{eq:governing_non-local}, periodicity \eqref{eq:periodicity}, and constitutive equations \eqref{eq:disctetized_const_law}. For the sake of readability, the subindex ${n+1}$ will be omitted from now on. The solution fields in the considered time increment are calculated by exploiting an iterative staggered scheme where two different spectral solvers are employed. Denoting with $k$ the current iteration counter of the staggered procedure, $\left. \boldsymbol{\varepsilon} \right|^{(k)}$ and  $ \left. \vec{\overline{\alpha}} \right|^{(k)}$ refer to the solution fields obtained in this iteration. Then, a new iteration $k+1$ consists in solving first the balance of linear momentum \eqref{eq:Lmimentum_abstract} along with constitutive equations \eqref{eq:disctetized_const_law} in which the non-local variables are taken as known fields from the previous iteration, i.e $\vec{\overline{\alpha}}$=$ \left. \vec{\overline{\alpha}} \right|^{(k)}$ . 
Such a problem has the same mathematical structure of conventional continuum mechanics, so that its numerical solution can be pursued by implementation of conventional algorithms as well. In this case, the Galerkin based FFT algorithm for mixed control proposed by Lucarini et al. \cite{LUCARINI2019} has been employed. The outcomes of this purely mechanical operation are the updated strain tensor $\left. \boldsymbol{\varepsilon} \right|^{(k+1)}$, plastic strain  $\left. \boldsymbol{\varepsilon}^p \right|^{(k+1)}$, and local internal variables $\vec{\alpha}^{(k+1)}$. The latter will enter in the set of the Helmholtz-type equations as source terms to provide the updated non-local fields $\vec{\overline{\alpha}}^{(k+1)}$ by means of the second FFT solver appearing in the proposed staggered scheme. For this purpose, Eq. \eqref{eq:helmholtz_abstract} is transformed into a linear system that will be solved by means of a Krylov-based algorithm as schematically reported in Algorithm \ref{al:Helmholtz_algorithm} . Since equations \eqref{eq:helmholtz_abstract} are uncoupled, their numerical solutions are computed independently.  The updated non-local fields $\left. \vec{\overline{{\alpha}}}\right|^{(k+1)}$ will be then plugged-in the Galerkin based mechanical solver in the next iteration of the staggered scheme. Such a recursive algorithm continues until convergence in the considered time step, which requires fulfilling mechanical equilibrium in addition to having a correction of strain and non-local fields between two consecutive iterations sufficiently small. For completeness, the FFT-based algorithms employed in the staggered scheme, along with the adopted discretization of the Fourier space, will be described briefly below. 

It is worth nothing that the proposed staggered algorithm is fully implicit and its solution is identical to the one obtained with a monolithic approach integrated implicitly using the  Backward Euler. This type of implicit schemes are also known as \emph{iterative staggered} in contrast to other more relaxed approaches, such as \textit{simple staggered}   \cite{Steinke2017}. In this second case the coupled problem is solved sequentially only once per time step (first mechanical equilibrium  and then non-local fields) and the resulting fields are directly passed to the next time step. This numerical scheme is obviously faster but it provides just a rough approximation of the solution since it does not ensure mechanical equilibrium at the end of the time step.

\subsubsection{Discretization}
The simulation domain is a periodic RVE of the microstructure embedded in a cuboidal domain $\Omega$ with edges lengths ${L_1,L_2,L_3}$. The RVE is discretized with a regular array of $N_1 \times N_2  \times N_3$ voxels, where each voxel belongs to any of the phases represented. The unknown discrete fields correspond to the value at the center of each cell, with position $\vec{x}$ given by
$$x_i = (\frac{1}{2}+n_i) \frac{L_i}{N_i}, \qquad i=1,2,3 \ \text{and} \ n_i \ \in [0, N_i-1] \,  . $$

\noindent
The Fourier transform of a discrete field corresponds to the discrete Fourier transform that can be obtained with a very efficient FFT algorithm. The corresponding $N_1 \times N_2 \times N_3$ discrete frequencies in the Fourier space, $\vec{\xi}$, are given by
\begin{equation}\label{frequencies}
\xi_i=\frac{2\pi}{L_i}\left\{ 
\begin{aligned}
& n_i-\frac{(N_i-1)}{2}  && \text{if} \ N_i \, \text{odd}\\
& n_i-\frac{N_i}{2}  && \text{ if}\ N_i \, \text{even}
\end{aligned}
\right.  \qquad \text{ for } \quad  i=1,2,3 \ \text{and} \ n_i \in [0, N_i-1]  \, .
\end{equation}

\noindent
It must be noted that in order to improve the convergence and reduce the well-known Gibbs oscillation phenomena in the presence of large contrast between phases, discrete derivative operators in the Fourier space are used. In particular, the finite difference rule proposed by Willot \cite{WILLOT2015232} (\emph{rotated-scheme}) 
 is adopted in this paper. The alternative definition of the Fourier frequencies can be found in  \cite{WILLOT2015232}.

\subsubsection{Mechanical problem}
The solution of equation \eqref{eq:Lmimentum_abstract} consists in finding, for the current time $t_{n+1}$, the compatible microscopic periodic strain distribution in the RVE that is in equilibrium with the stress field. The Galerkin FFT method \cite{VONDREJC2014,ZEMAN2017} is used to solve this problem together with the procedure to incorporate stress and mixed control proposed by Lucarini and Segurado in \cite{LUCARINI2019}. This control technique sets a combination of  components of the macroscopic strain ($h,k$) and/or the macroscopic stress ($H,K$) history, i.e. $E_{hk}(t)$ and $\Sigma_{HK}(t)$, respectively.  The boundary value problem is solved expressing the weak form of the linear momentum balance for the current time increment

\begin{equation}\label{weakeq}
\left\{ \begin{array}{cl}
\int_{\Omega} \mathbb{G}^\ast \ast \boldsymbol{\zeta}(\vec{x}): \boldsymbol{\sigma}\left(\boldsymbol{\varepsilon}(\vec{x}),\vec{\alpha})\right)\mathrm{d}V & =0 \, , \\
 \left<\sigma \right>_{HK} &= \left. \Sigma_{HK} \right|_{n+1} \, ,  \\
 \left<\varepsilon \right>_{hk} & = \left. E_{hk} \right|_{n+1} \, ,
\end{array} \right. 
\end{equation}

\noindent
where $\left<\cdot\right>$ represents the volume average, $\boldsymbol{\zeta}(\vec{x})$ are second order tensor test functions, $\mathbb{G}^*$ stands for the projector operator that enforces the compatibility of the test functions,  $\ast$ is the convolution operation, and the indices of the macroscopic stress and strain obey $i,j \cap I,J = \emptyset$. In \eqref{weakeq} the original projection operator for small strain proposed in \cite{DEGEUS2017} is replaced by a modified $\mathbb{G}^\ast$  which includes modified zero frequencies to enforce the value of stress averages and has a closed-form expression in the Fourier space $\widehat{\mathbb{G}^*}$ \cite{LUCARINI2019}.

After the problem discretization in voxels, the weak form of the equilibrium equation \eqref{weakeq} can be expressed as an algebraic system of non-linear equations  \cite{VONDREJC2014}  

\begin{equation}\label{eqdiscrete2}
\mathcal{G}^*\left(\boldsymbol{\sigma}\right):=\mathcal{F}^{-1}\left\{ \widehat{\mathbb{G}^*}: \mathcal{F} \left( \boldsymbol{\sigma}\left(\boldsymbol{\epsilon}(\vec{x}),\vec{\alpha}\right) \right)  \right\} =\mathbf{0} \, , 
\end{equation}

\noindent
with symbol $\mathcal{F}$ indicating the Fourier transform. As customary, a Newton-Raphson algorithm is adopted for  the solution of the resulting non-linear problem so that the stress tensor is linearized with respect to the total strain as 

\begin{equation} \label{eq:lin_stress}
\boldsymbol{\sigma}^{(r+1)}=\boldsymbol{\sigma}^{(r)} +
 \frac{\partial \boldsymbol{\sigma} }{\partial \boldsymbol{\varepsilon} } :\delta  \boldsymbol{\varepsilon}^{(r+1)} \, . 
\end{equation}

\noindent where $\delta  \boldsymbol{\varepsilon}^{(r+1)}$ is the strain correction for Newton-Raphson iteration $r+1$. A linear problem is finally obtained by substitution of \eqref{eq:lin_stress} into \eqref{eqdiscrete2}

\begin{equation*}
\mathcal{G}^\ast\left(\frac{\partial \boldsymbol{\sigma} }{\partial \boldsymbol{\varepsilon} }:\delta  \boldsymbol{\varepsilon}^{(r+1)} \right)=-\mathcal{G}^\ast\left( \boldsymbol{\sigma}^{(r)}-\overline{\boldsymbol{\Sigma}}_{n+1}\right) \, , 
\end{equation*}

\noindent where $ \left.\overline{\boldsymbol{\Sigma}} \right|_{n+1}$ is a tensor containing the non-zero $HK$ components of the imposed stress at time $t_{n+1}$. This resulting linear system of equations is solved using the conjugate gradient method, whose convergence rate, efficiency, and memory allocation are optimal for this problem.

%
%
%
%

\subsubsection{Helmholtz  equation}
Consider the Helmholtz equation for heterogeneous materials

\begin{equation} \label{eq:FFThelmholtzeq}
\overline{\alpha} - \text{div} \left[ \ell^2(\vec{x}) \, \nabla \overline{\alpha} \right] = \alpha \, , 
\end{equation}

\noindent
where $\alpha$ is a known source term, while $\overline{\alpha}$ is the solution function. Note that in the case of non-uniform length parameter $\ell$, the Helmholtz-type equation of the considered non-local regularization becomes implicit. As a result of that, the relevant solution cannot be computed directly, as for example pursued in \cite{BOEFF2015373} with a uniform $\ell$, but an iterative solving scheme must be employed. It is convenient to rephrase Eq. \eqref{eq:FFThelmholtzeq} in a more convenient form as

\begin{equation} \label{eq:HelmholtzL}
\text{Find }   \, \overline{\alpha} \text{ such that } \mathcal{L} \left( \overline{\alpha} \right) = \alpha \,  , 
\end{equation}

\noindent where $\mathcal{L} = \left( \bullet \right) - \text{div} \left[ \ell^2(\vec{x}) \, \nabla \left( \bullet \right) \right] $ is a linear differential operator. In the context of FFT, problem \eqref{eq:HelmholtzL} can be easily solved in the frequency domain because it reduces to the following linear problem 

\begin{equation} \label{eq:linear_helmholtz}
\mathcal{F} \left( \mathcal{L} \left( \overline{\alpha} \right) \right) = \mathcal{F} \left( \alpha \right) \,
\end{equation}

\noindent with symbol $\mathcal{F}$ indicating the Fourier transform operator. The left hand side of equation \eqref{eq:linear_helmholtz} is simply derived by application of the differentiation rule in Fourier space, leading to 

\begin{equation*}
 \mathcal{F} \left( \mathcal{L}  \left( \overline{\alpha} \right)  \right) =  \mathcal{F} \left( \overline{\alpha} \right) -  \iu \, \vec{\xi} \cdot \mathcal{F} \left( \ell^2(\vec{x}) \, \mathcal{F}^{-1}   \left(  \,  \iu \, \vec{\xi}  \mathcal{F}  \left( \overline{\alpha} \right)\right) \, \right) =  \mathcal{F} \left( \overline{\alpha} \right) +  \vec{\xi} \cdot \mathcal{F} \left(  \ell^2(\vec{x}) \, \mathcal{F}^{-1}   \left(  \, \vec{\xi}  \mathcal{F}  \left( \overline{\alpha} \right)\right) \, \right) \, , 
\end{equation*}

\noindent where $\iu$ refers to the imaginary unit, while $\vec{\xi}$ is the frequency vector.  Therefore, the linear operator in the Fourier space $\widehat{\mathcal{L}}$ can be simply rewritten as 

\begin{equation}
\widehat{\mathcal{L}} = \left( \bullet \right)+\vec{\xi} \cdot \mathcal{F} \left(  \ell^2(\vec{x}) \, \mathcal{F}^{-1}   \left(  \, \vec{\xi}\  \left( \bullet \right)  \right) \, \right) \, . 
\label{eq:LhelmFourier}
\end{equation}

\noindent
Accordingly, the linear problem \eqref{eq:linear_helmholtz} can also be written in the Fourier space as

\begin{equation} \label{eq:linear_helmholtz_complez}
\widehat{\mathcal{L}} \left(\mathcal{F} \left({\overline{\alpha}}\right)\right)=\mathcal{F} \left({\alpha} \right) \, .
\end{equation}

\noindent
The resulting linear system is symmetric and has a unique solution that can be easily obtained if the characteristic length is uniform since the operator $\widehat{\mathcal{L}}$ can be inverted analytically.  On the contrary, if $\ell(\vec{x})$ is not uniform, the inverse operator cannot be obtained in a closed-form expression and  Eq. ($\ref{eq:linear_helmholtz_complez}$) has to be solved numerically. Due to the properties of $\widehat{\mathcal{L}}$, a conjugate gradient algorithm is exploited here to solve the problem. In addition, following the approach derived in \cite{LUCARINI2019103131} for a displacement based FFT homogenization algorithm, the conjugate gradient is sped up by using a preconditioner of type

\begin{equation}
\mathcal{M} \left( \mathcal{F} \left( \overline{\alpha} \right)  \right) = \left[ \frac{1  }{1 + < \ell^2 > \, \vec{\xi} \cdot  \vec{\xi} }\right]  \mathcal{F} \left( \overline{\alpha} \right) \,,
\label{eq:preconditioner}
\end{equation}

\noindent
with $< \ell^2 >$ indicating the average characteristic length of the non-local regularization. Note that, for a uniform $\ell$, the preconditioner Eq. (\ref{eq:preconditioner}) corresponds to the inverse of the linear operator. It is worth nothing that the usage of the proposed preconditioner is crucial since it significantly accelerates the solution of the modified Helmholtz-type equation, thus making it possible its application to problems with a high number of degrees of freedom (e.g. see Section \ref{sec:3Dexamples}).

%

\RestyleAlgo{boxed}

\begin{algorithm}

\DontPrintSemicolon
{\bf{Data:}} $ \left. {\alpha}_j \right|_{n+1}^{(k+1)}$, $\ell_j \left( \, \vec{x} \,  \right)$, $TOL$ \\
 \smallskip
{\bf{Result:}} non-local field $\left. \overline{\alpha_j} \right|_{n+1}^{(k+1)}$ \\
\bigskip

$\widehat{\alpha_j} =  \mathcal{F} \left(  \left. {\alpha}_j \right|_{n+1}^{(k+1)} \right) $ \\
\medskip
$\widehat{\overline{\alpha_j}}$ = {{\tt{ConjGrad}}} ( {\tt{GC}},\,  {\tt{M}}, \, $\widehat{{\alpha}_j}$, \, $TOL$) \\
\medskip
$\left. \overline{\alpha_j} \right|_{n+1}^{(k+1)} = \mathcal{F}^{-1} \left( \widehat{\overline{\alpha_j}} \right) $ \\

\bigskip

\SetKwFunction{FMain}{GC} 

\SetKwProg{Fn}{Function}{:}{}
    \Fn{\FMain{$\widehat{\overline{\alpha_j}}$}}{
Linear operator for the generalized Helmholtz-type equation in Fourier space (Eq. \ref{eq:LhelmFourier}) \\
    \smallskip
    $\nabla \overline{\alpha_j} = \mathcal{F}^{-1} \left( \vec{\xi} \, \,  \widehat{\overline{\alpha_j}} \right)$  \\
    $\vec{h} = \ell_j^2 \left( \vec{x} \,  \right) \nabla \overline{\alpha_j}$ \\
    $\widehat{\vec{h}} = \mathcal{F} \left( \vec{h} \right)$ \\
    \smallskip
    {\bf{return}} $ \widehat{\overline{\alpha_j}} + \vec{\xi} \cdot \widehat{\vec{h}} $
   }
   
   \bigskip
   
   \SetKwFunction{FMain}{M} 

\SetKwProg{Fn}{Function}{:}{}
    \Fn{\FMain{$\widehat{\overline{\alpha_j}}$}}{
Linear operator for the preconditioner of the Helmholtz-type equation in Fourier space (Eq. \ref{eq:preconditioner}) \\
    \smallskip

    {\bf{return}} $\left[ \frac{1  }{1 + < \ell^2 > \, \vec{\xi} \cdot  \vec{\xi} }\right]  \widehat{\overline{\alpha_j}} $
   }
   
   \bigskip
 \SetKwFunction{FMain}{ConjGrad}
\SetKwProg{Fn}{Function}{:}{}
    \Fn{\FMain{$\mathcal{A}, \mathcal{M},\, b , \, TOL $}}{
Conjugate gradient solver for linear operator $\mathcal{A}$, preconditioner $\mathcal{M}$ and the independent term $b$\\
    \smallskip
    {\bf{return}} $ \widehat{\overline{\alpha_j}} \, \,   | \, \,  \mathcal{A} ( \,  \widehat{\overline{\alpha_j}}  \, )  = b \, $  and  $\| \mathcal{A} ( \,  \widehat{ \overline{\alpha_j}}  \, ) - b  \| < TOL \cdot \|  b \| $ 
  }  
  \smallskip
\caption{Spectral solver for the Helmholtz-type equation.}\label{al:Helmholtz_algorithm}

\end{algorithm}

 \section{Numerical examples}  \label{sec:examples}

 \subsection{2D examples} \label{sec:2Dexamples}

\begin{figure}[htbp]
\centering
\includegraphics[width=7cm]{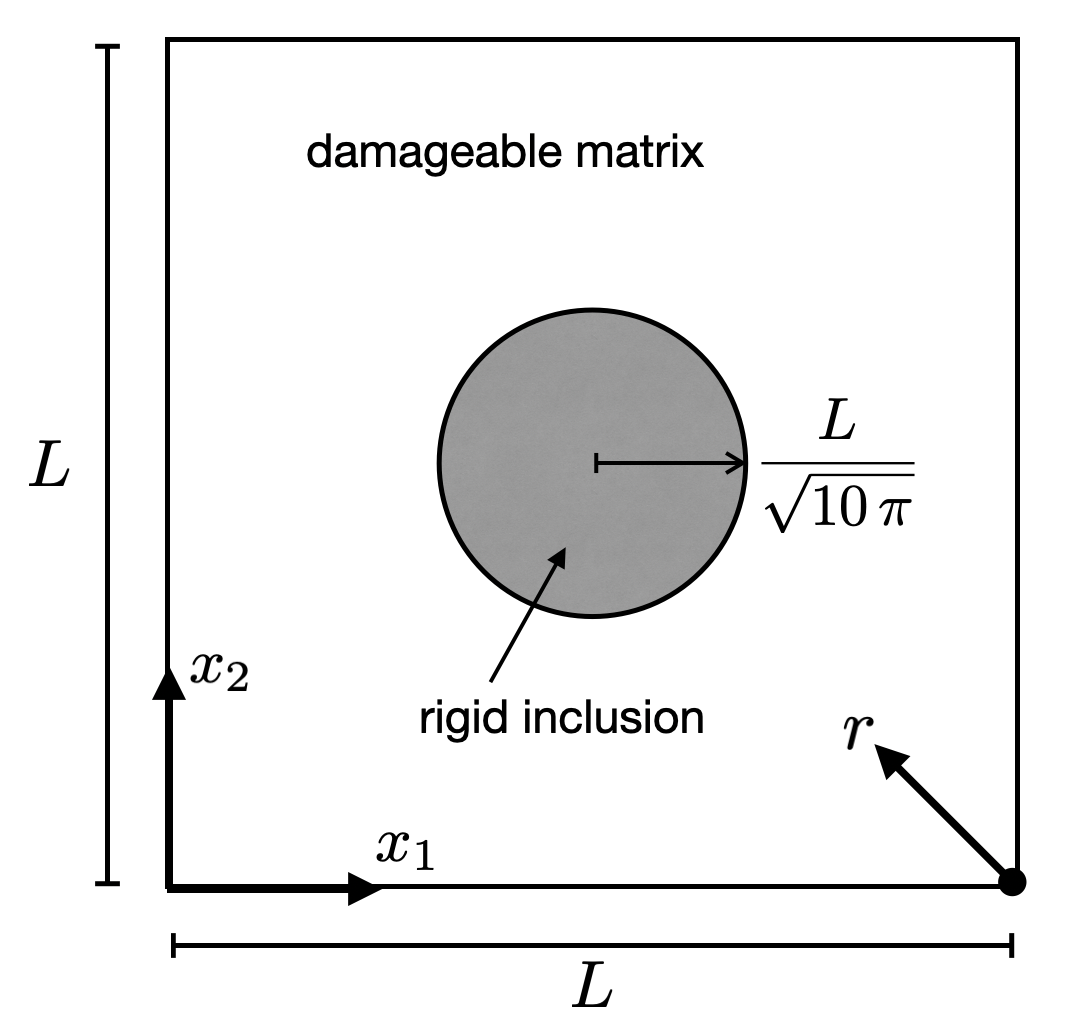}
\caption{\em{Geometry of the composite material considered in the 2D numerical simulations.}}
\label{fig:geom2D}
\end{figure}

The proposed algorithm is first tested on a simple two dimensional periodic microstructure under plane strain condition. The geometry of the problem consists of a squared RVE of size $L \times L$ containing a circular inclusion of radius $L / \sqrt{10 \pi}$, thus occupying 10\% of the total area. These particular conditions allow for a readable evaluation of the impact of the non-local regularization as well as a sensible reduction of the computational cost. A sketch of the geometry is reported in Fig \ref{fig:geom2D}. The sample is subjected to uniaxial tensile loading in which a macroscopic strain is prescribed in the direction $x_1$, while a stress free condition is enforced in the remaining components, namely

\begin{equation*}
{\boldsymbol{E}} = 
\begin{pmatrix}  E_{11} 
 & * \\ * & * \end{pmatrix} 
 \qquad \text{and} \qquad
{\boldsymbol{\Sigma}} =  
\begin{pmatrix} * & 0 \\ 0 & 0 \end{pmatrix} \, .
\end{equation*}

\noindent 
The macroscopic strain is prescribed incrementally until final failure of the matrix. Specification of the adopted rate of applied strain is provided while analyzing the numerical outcomes. 

The mechanical response of the material matrix is dictated by the theories described in Section \ref{sec:non-local-theories}. In both models, the elastic constants, given in terms of Young's  modulus $E$ and Poisson's coefficient $\nu$, are taken as $E_M=300$ GPa and $\nu_M=0.3$, while the yield stress is $\sigma_Y = 1$ GPa. For the Gurson model, the damage related parameters are specified by $q_1=1.5$, $q_2=1$, $q_3=2.25$, $f_C=0.15$, $f_F=0.25$, $f_N=0.04$, $\varepsilon_N=0.3$, and $s_N=0.1$. The initial void volume fraction is zero. The hardening law  $\sigma_0 = \sigma_0 \left( \varepsilon_0^p \right)$ is specialized according to Aravas \cite{Aravas1987} in the form 

\begin{equation} \label{eq:aravas}
\frac{\sigma_0}{\sigma_Y} = \left(\frac{\sigma_0}{\sigma_Y} + \frac{3 \, \mu_M }{\sigma_Y} \, \varepsilon_0^p \right)^N \, , 
\end{equation}

\noindent 
where $\mu_M$ is the matrix shear modulus and $N=0.1$.
On the other hand, for the Lemaitre model, the material parameters controlling damage are $\epsilon_C=0.03$ and $\epsilon_R=0.2$, while the flow stress holds $\sigma_0 \left( \epsilon_p \right)= \sigma_Y + k \, \epsilon_p$, with $k=10$ GPa. 

The circular inclusion is idealized as an elastic reinforcement not subjected to damage. Its mechanical behavior corresponds to an elastic isotropic solid with parameters  $E_I=900$ GPa and $\nu_I=0.3$. Although the elastic inclusion remains undamaged, a value for the characteristic length $\ell$ must be assigned to both phases of the composite since equations \eqref{eq:helmholtz_abstract} are solved in the whole domain of the RVE. As discussed in Section \ref{sec:Implicitgradientregularizationforheterogeneous materials}, to prevent undesired diffusion of plastic internal variables in the elastic phase, the length parameter $\ell \left( \vec{x} \right)$, specialized as $\ell_M$ for the matrix material and $\ell_I$ for the inclusion, must obey the inequality $\ell_I \ll \ell_M$. To avoid excessively penalizing the convergence of the Helmholtz-type problem, a large but finite ratio $\ell_M / \ell_I$ will be considered as discussed below. 

In the considered damage models, the local bearing capacity of the matrix material is lost as soon as the damage indicator reaches its upper limit, i.e. $f_*=f_V^*$ in Gurson model and $D=1$ in the Lemaitre model. If such a condition is attained, the convergence of the FFT-Galerkin scheme used to solve $\eqref{eq:Lmimentum_abstract}$ is compromised since the contrast between the stiffness of matrix and inclusion becomes too high. Therefore, for numerical convenience, the upper limit of the damage indicators are limited to $f_*=0.9 \, f_V^*= 0.6 $ and $D=0.99$. Accordingly, a low residual stress capacity is left to the material matrix allowing the simulation of the entire  macroscopic failure process.

 \subsubsection{Grid sensitivity}

In this section, we analyze the impact of the considered non-local regularization via the proposed FFT numerical implementation. To this end, different spatial discretizations have been adopted, i.e. $32\times32$, $64\times64$, and $128\times128$ grid points. To provide a fair analyses of the grid size dependence of the model, a constant macroscopic strain rate $\dot{E}_{11} = 1 \times 10^{-4}  \, 1/\text{s}$ is applied to all the analyses. 

\begin{figure}[htbp]
\centering
		\begin{tabular}{rc}
		\subfigure[]{ \includegraphics[height=5.5cm]{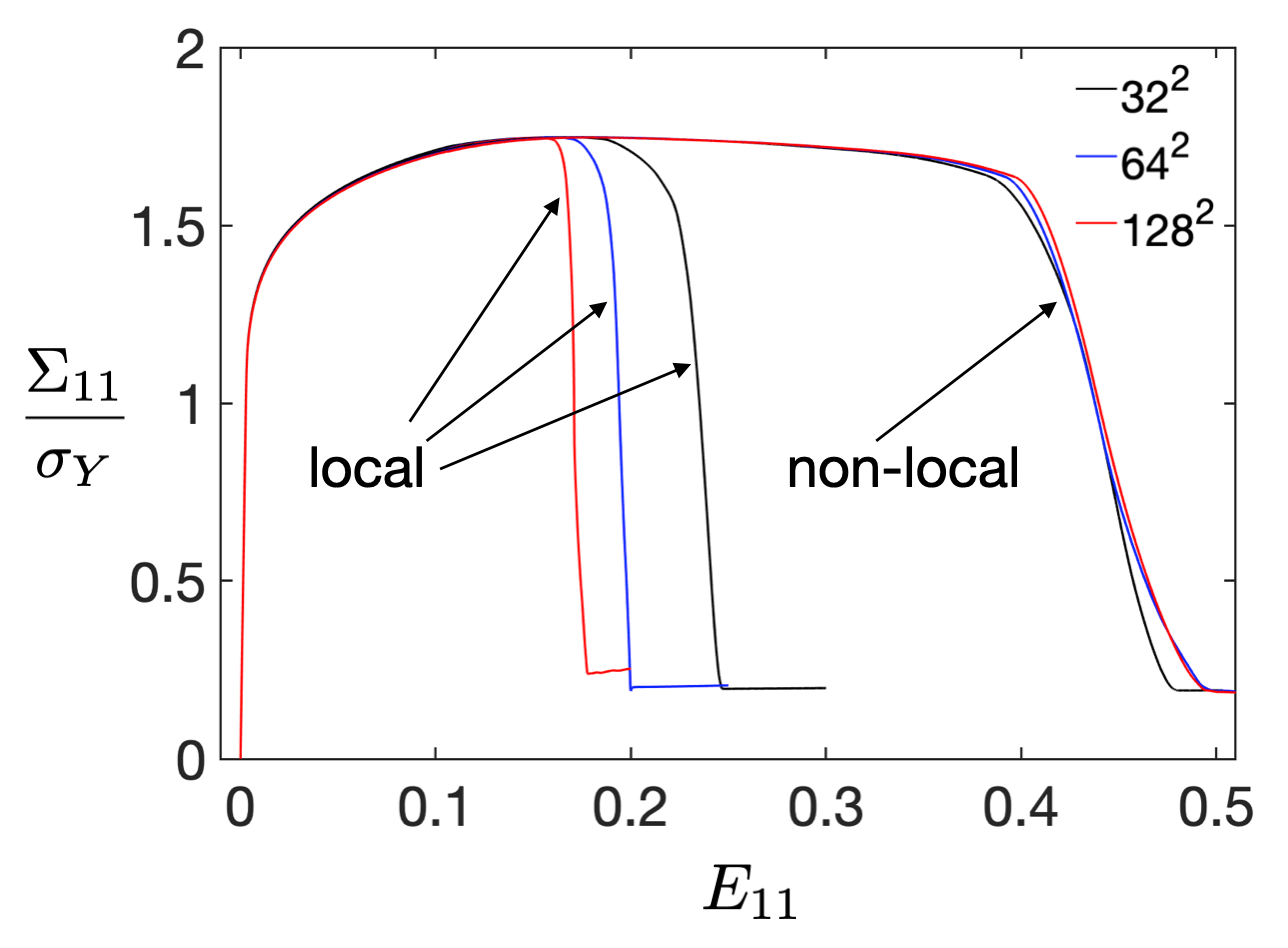} }
		&
		\subfigure[]{ \includegraphics[height=5.5cm]{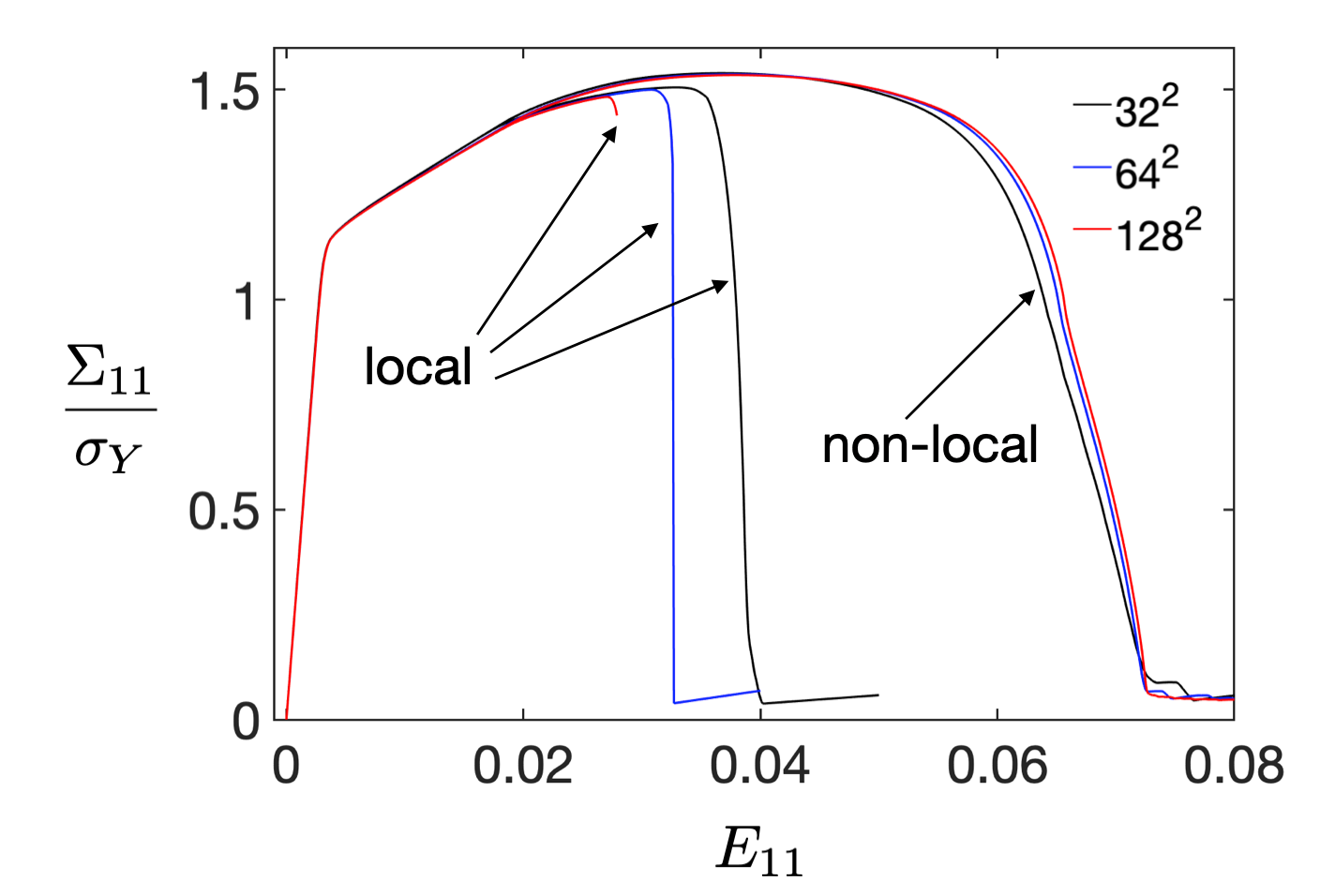}}
		\end{tabular}
\caption{\em{Plot of the normalized average stress component $\Sigma_{11}$  against the average strain component $E_{11}$ resulting from the considered Gurson (a) and Lemaitre models (b)  for different grids. In each graph, the curves obtained from the non-local models are plotted along with the results of the local counterparts for an immediate understanding of the impact of the non-local regularization on the grid dependence of the outcomes.}}
\label{fig:stressstraincurves}
\end{figure}

Figure \ref{fig:stressstraincurves} gathers the simulated stress-strain curves in the cases of Gurson (a) and Lemaitre (b) models. In order to highlight the effect of the non-local regularization, the results of the original local version of each model have been plotted as well. The latter has been simply obtained by taking both $\ell_M$ and $\ell_I$ smaller than the considered grid size for each spatial discretization, i.e. $\ell_M = \ell_I = 1 \times 10^{-4} \, L $. On the other end, $\ell_M=0.05 \, L $  and $\ell_I=0.001 \, L $ is considered for the non-local models since a regularizing effect in the matrix can be obtained only for characteristic lengths $\ell$ greater than the grid size. Accordingly, $\ell_M/\ell_I=50$. 

The numerical outcomes demonstrate the benefit of the implicit gradient regularization for both non-local models, as the stress-strain curve converges towards a unique solution upon grid refinement. Conversely, a marked grid dependence results from the local models. In all the cases, the simulated stress-strain curve is characterized by an initial elastic loading followed by a broad inelastic stage that results from the competing processes of strain-hardening and damage evolution during the plastic degradation of the material matrix. Once the damage mechanism prevails, a sudden stress drop results in correspondence to final rupture. It is worth nothing that a low residual stress remains at the end of the simulation due to the assumptions made on the damage parameters. 

\begin{figure}[htbp]
\centering
		\begin{tabular}{lcrc}
		\subfigure[]{ \includegraphics[height=5.5cm]{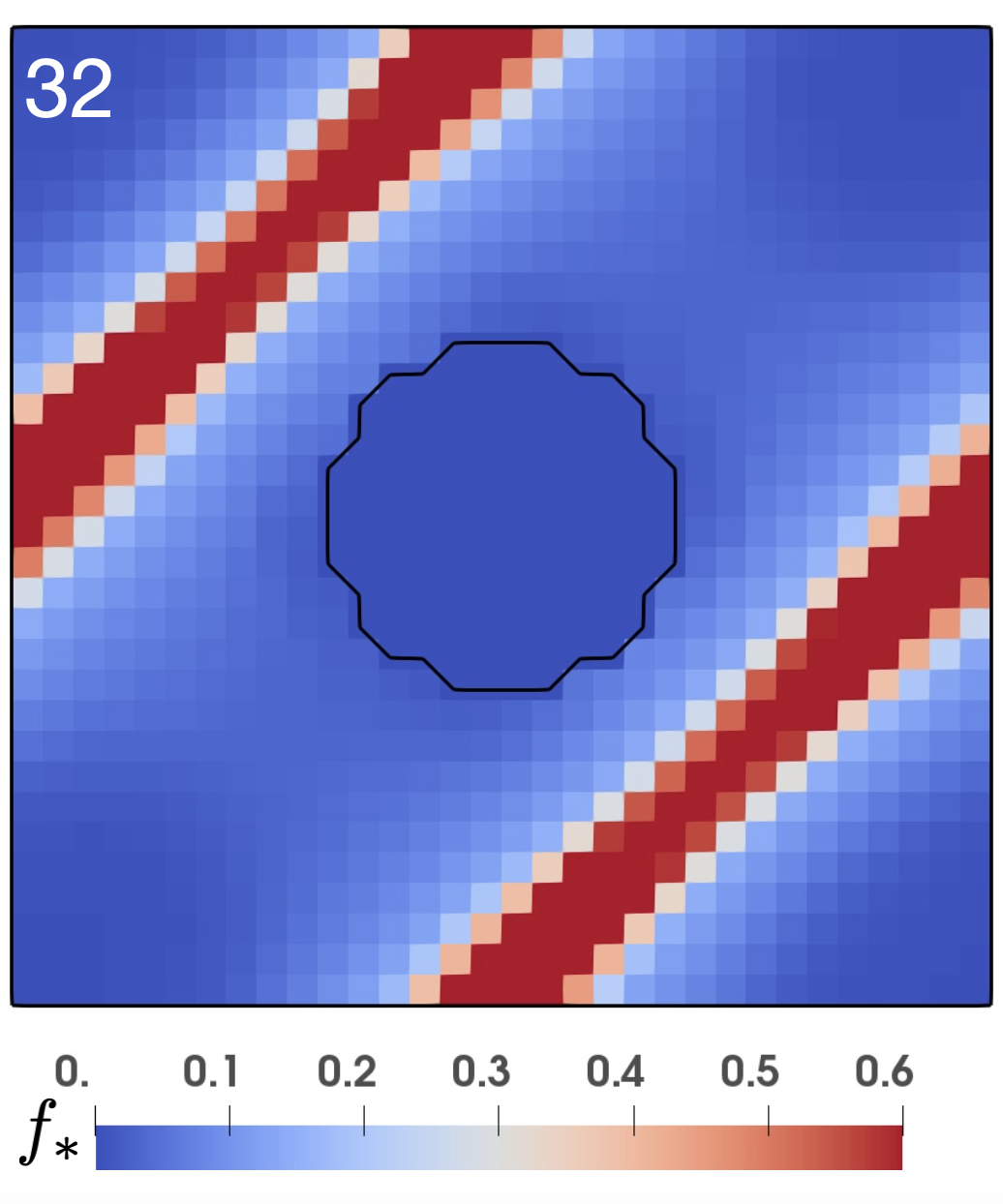}} 
		&
		\subfigure[]{ \includegraphics[height=5.5cm]{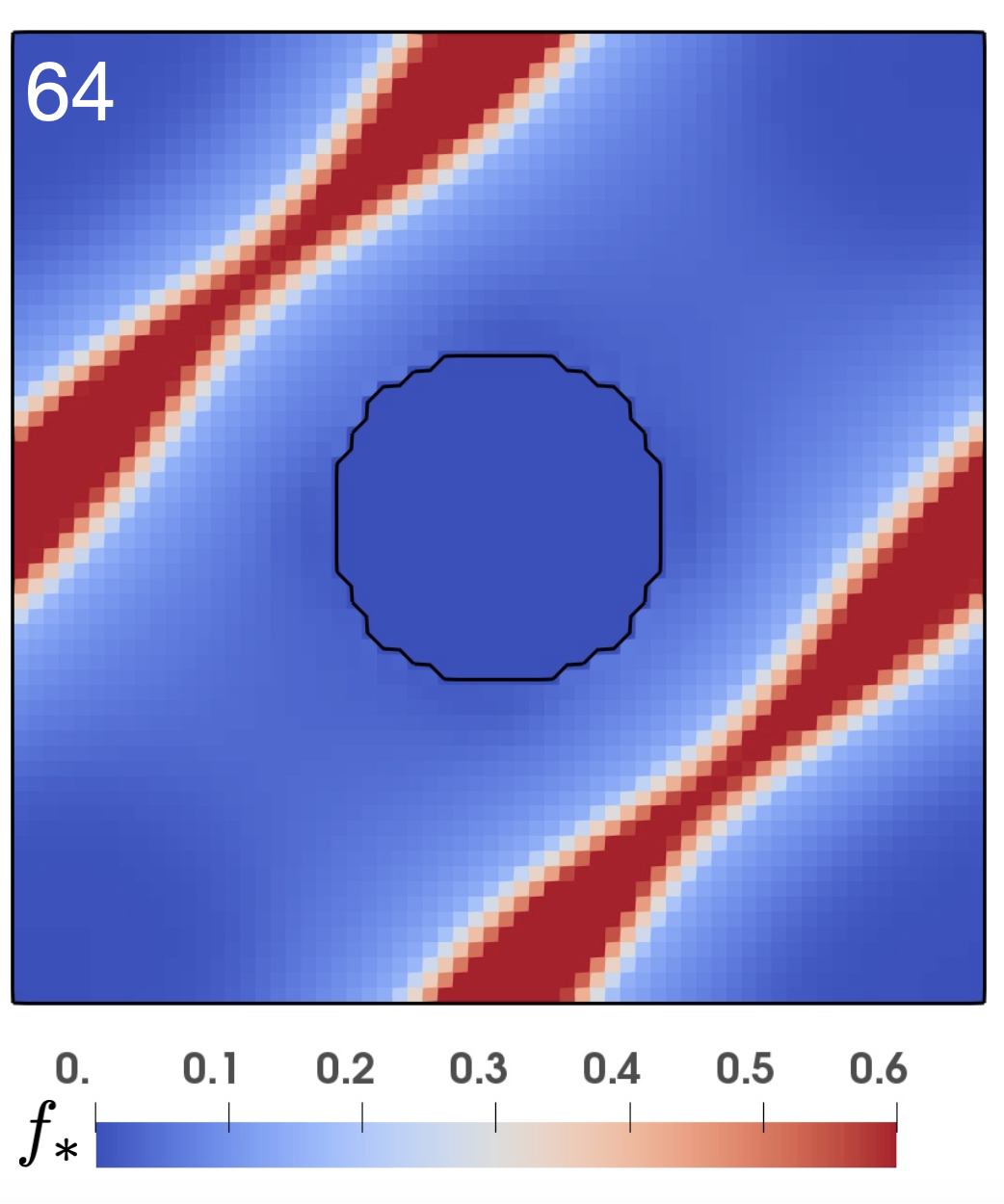}} 
		&
		\subfigure[]{ \includegraphics[height=5.5cm]{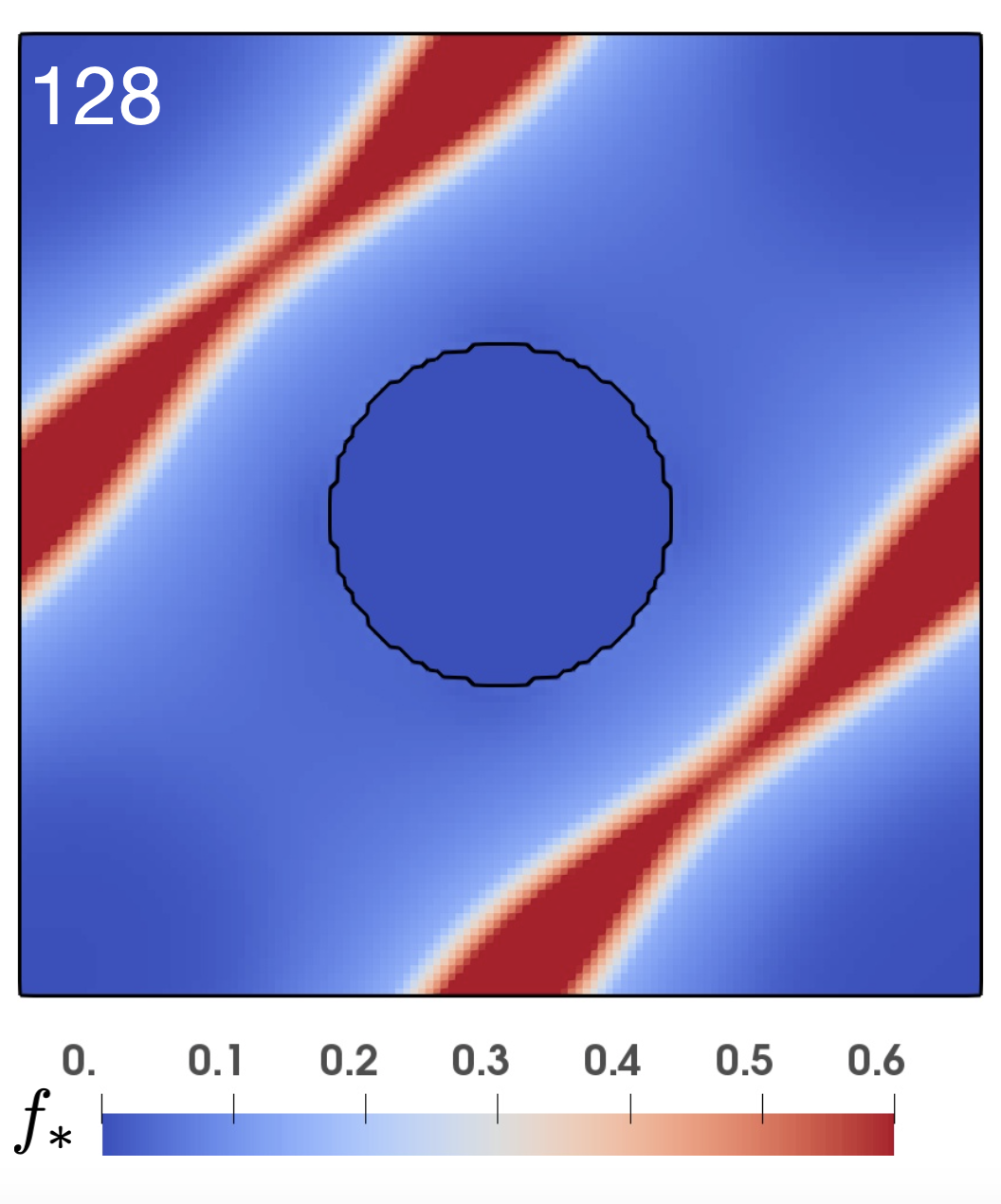}} 
		\end{tabular}
\caption{\em{Distribution of effective damage variable $f_*$ at final fracture ($E_{11}=0.5$) simulated in case of the non-local Gurson model with $\ell_m= 0.05 \, L$ and $\ell_I=0.001 \, L$ for (a) $32^2$, (b) $64^2$, and (c) $128^2$ grid points.}}
\label{fig:contoursNG}
\end{figure}

\begin{figure}[htbp]
\centering
		\begin{tabular}{lcr}
		\subfigure[]{ \includegraphics[height=5.5cm]{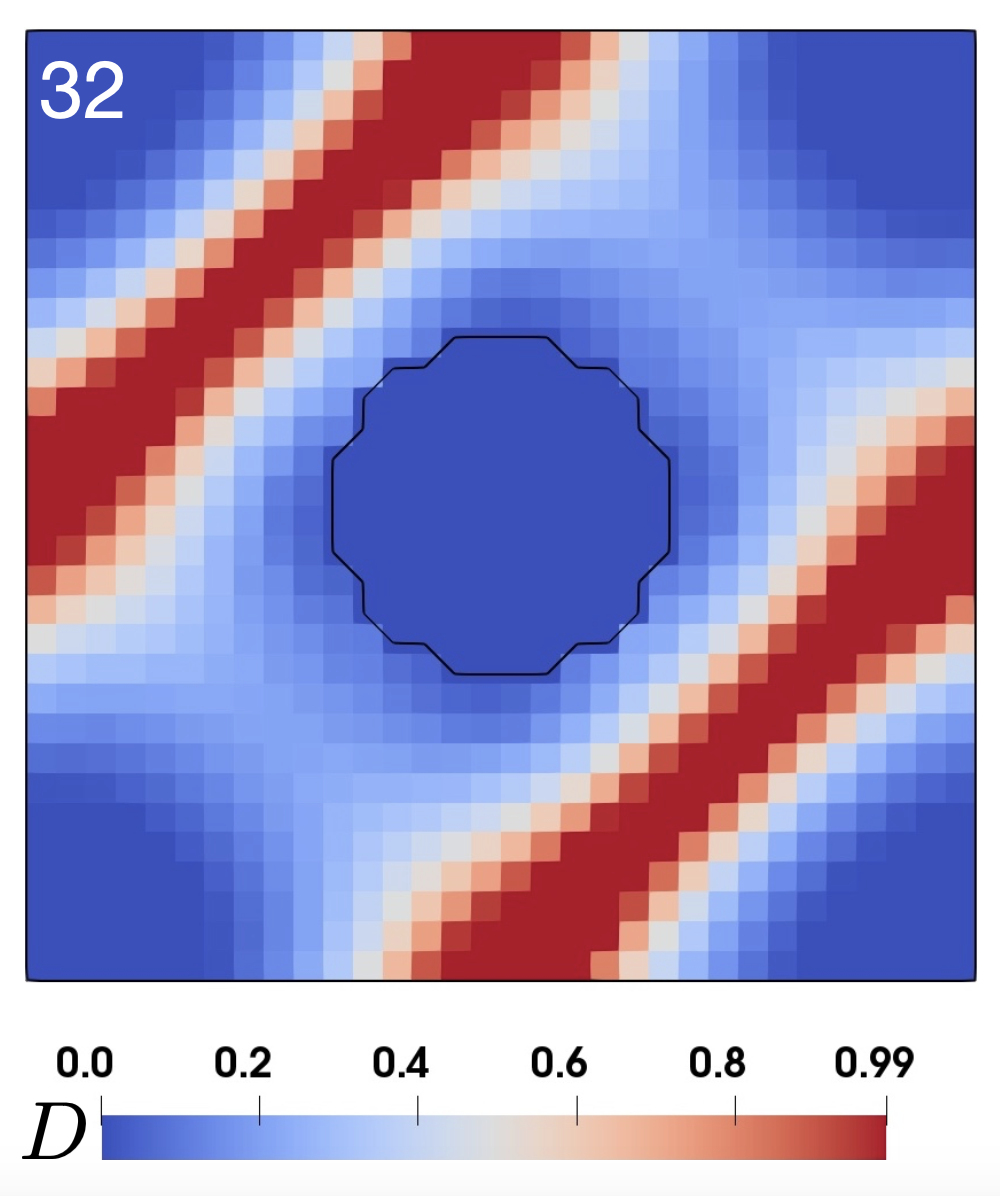}} 
		&
		\subfigure[]{ \includegraphics[height=5.5cm]{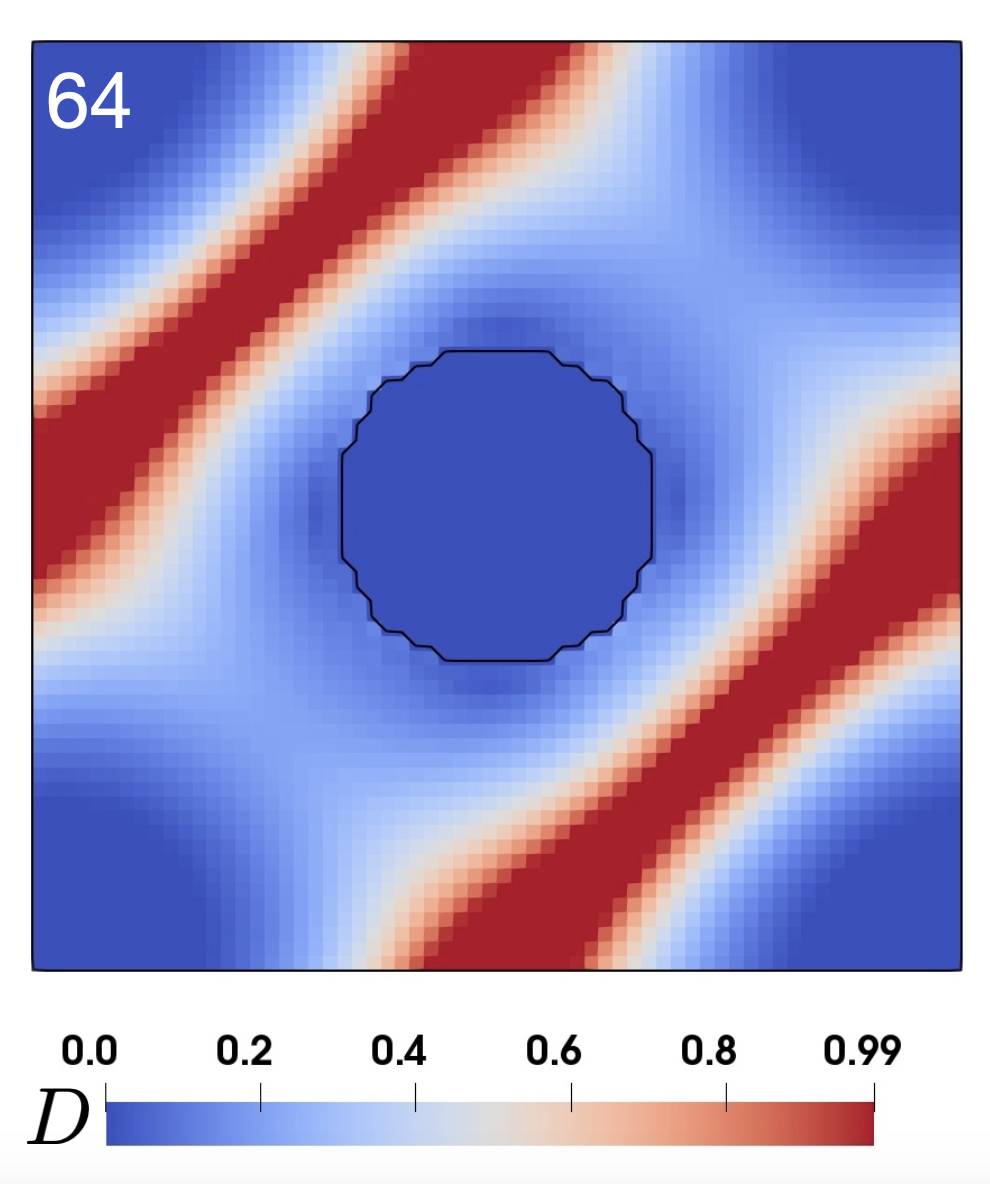}} 
		&
		\subfigure[]{ \includegraphics[height=5.5cm]{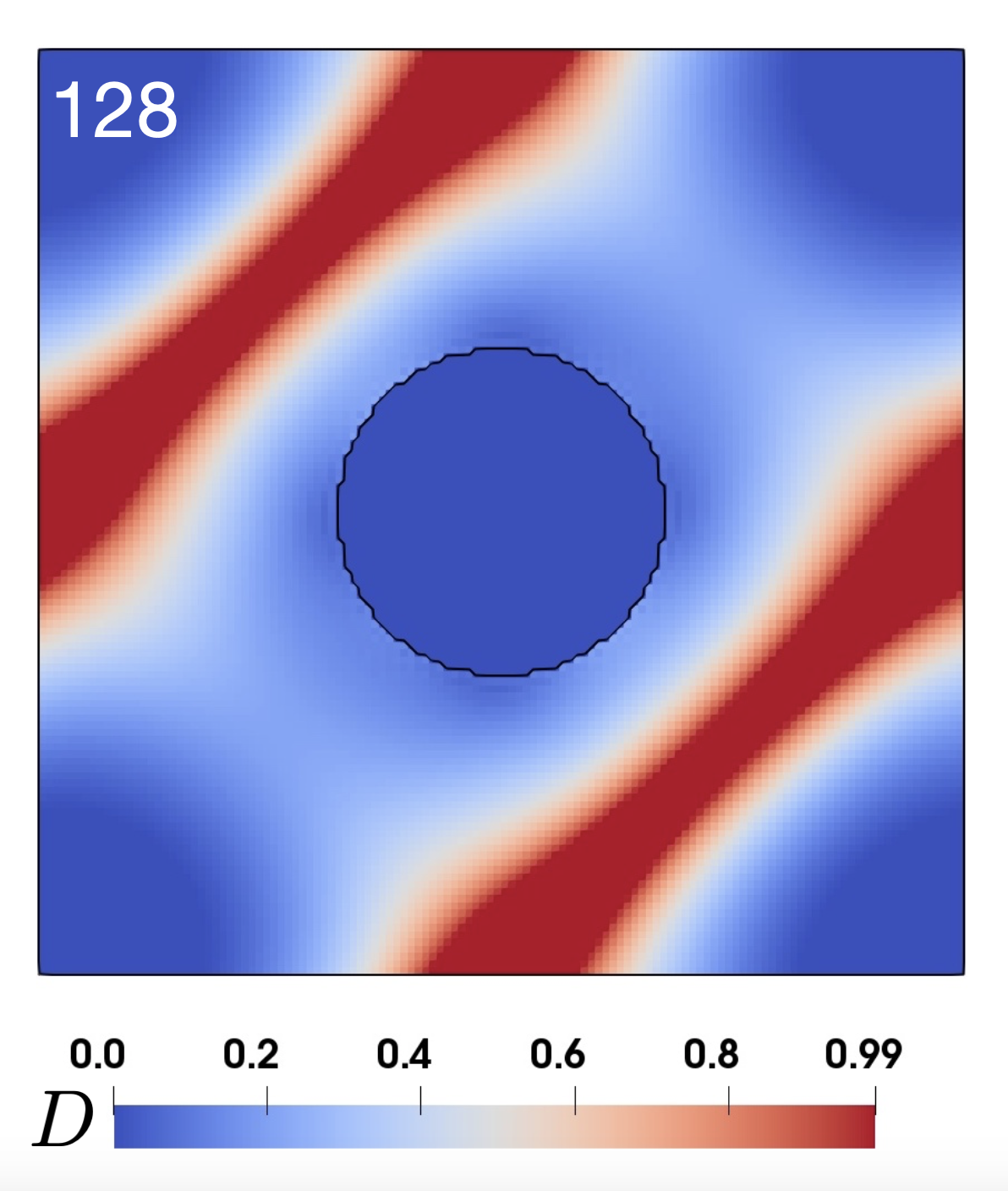}} 
		\end{tabular}
\caption{\em{Distribution of intrinsic damage variable $D$ at final fracture ($E_{11}=0.08$) simulated in case of the non-local Lemaitre model with $\ell_M=0.05 \, L$ and $\ell_I=0.001 \, L$ for (a) $32^2$, (b) $64^2$, and (c) $128^2$ grid points.}}
\label{fig:contoursNL}
\end{figure}

\begin{figure}[htbp]
\centering
		\begin{tabular}{lcr}
		\subfigure[]{ \includegraphics[height=5.5cm]{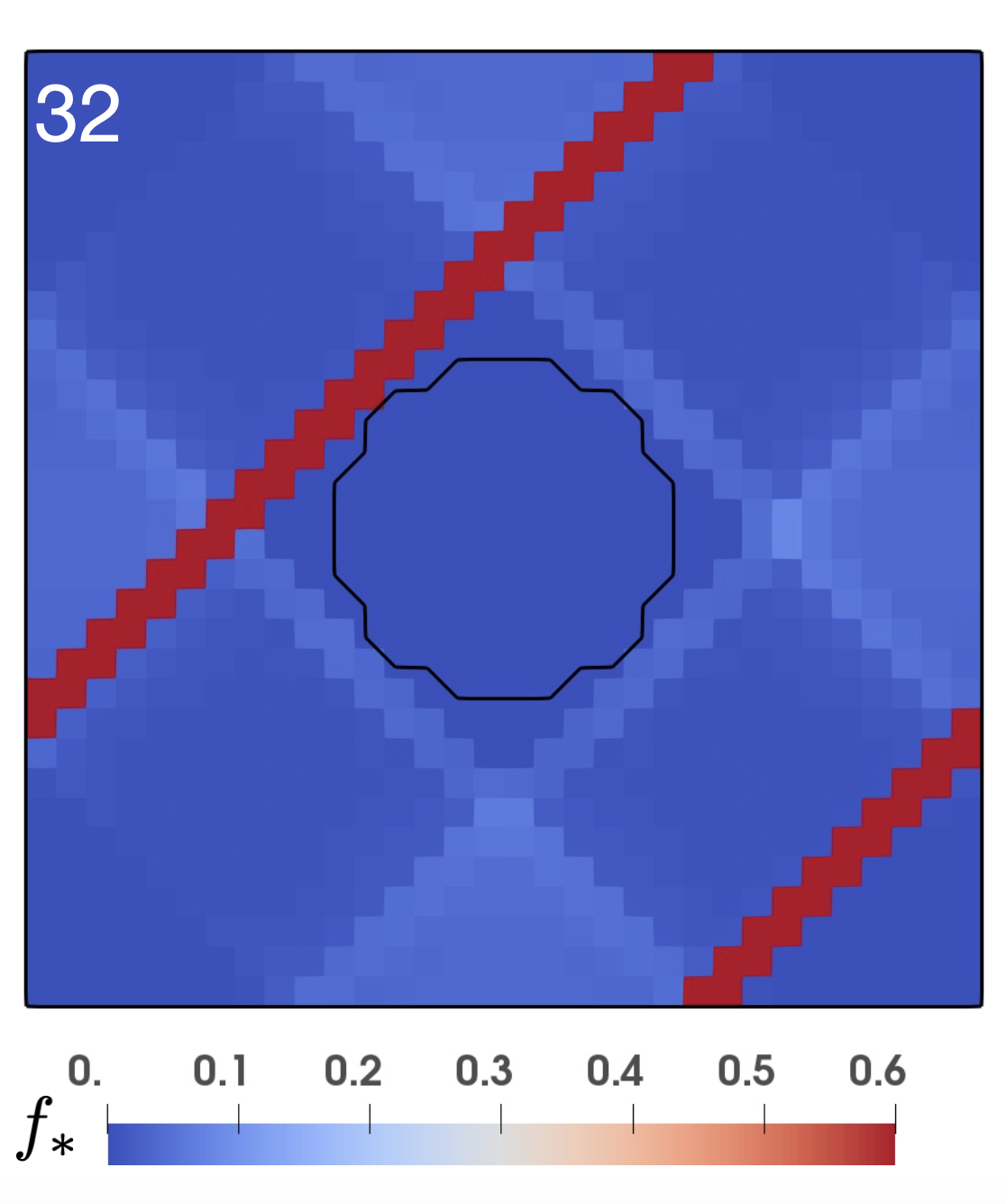}} 
		&
		\subfigure[]{ \includegraphics[height=5.5cm]{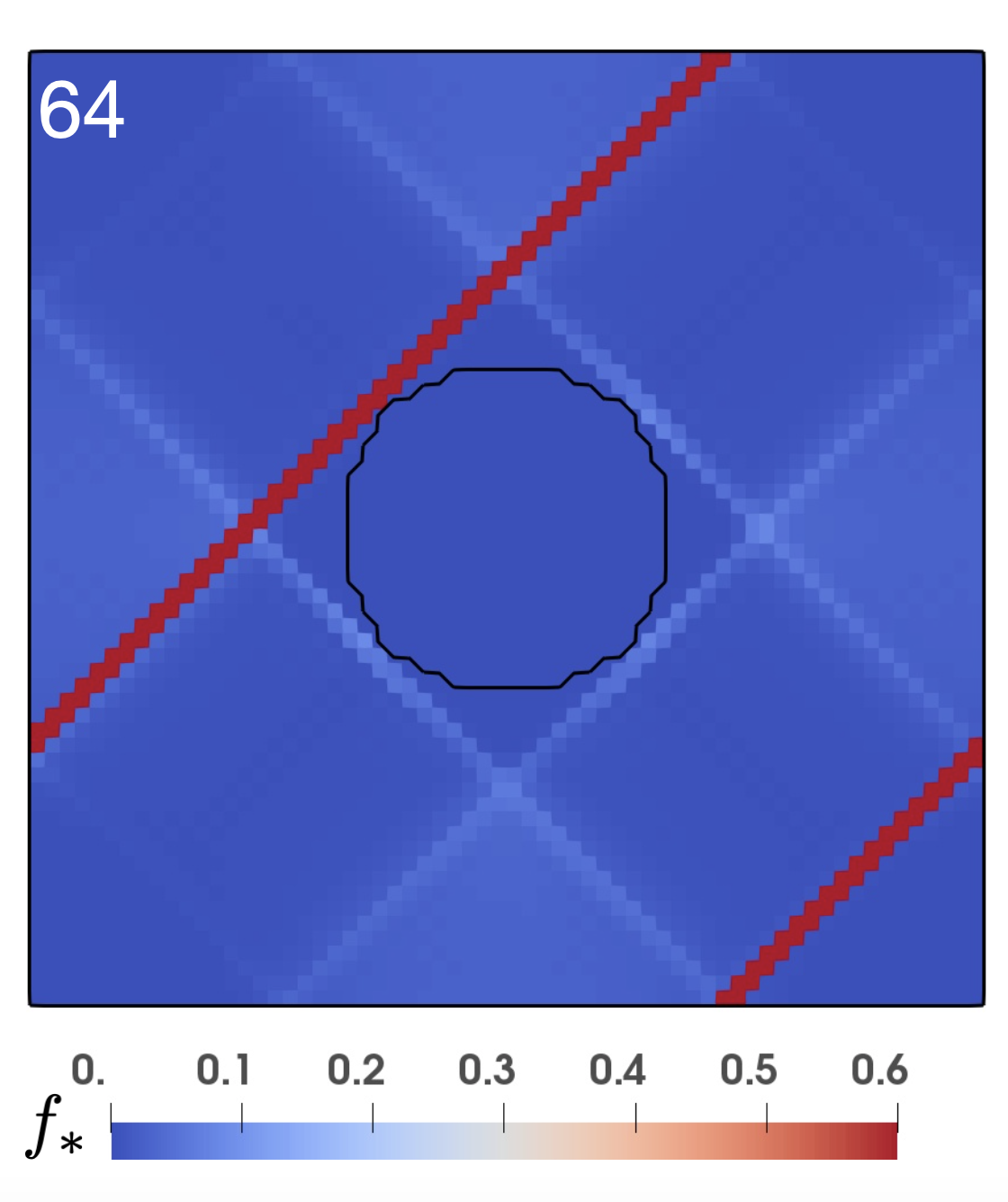}} 
		&
		\subfigure[]{ \includegraphics[height=5.5cm]{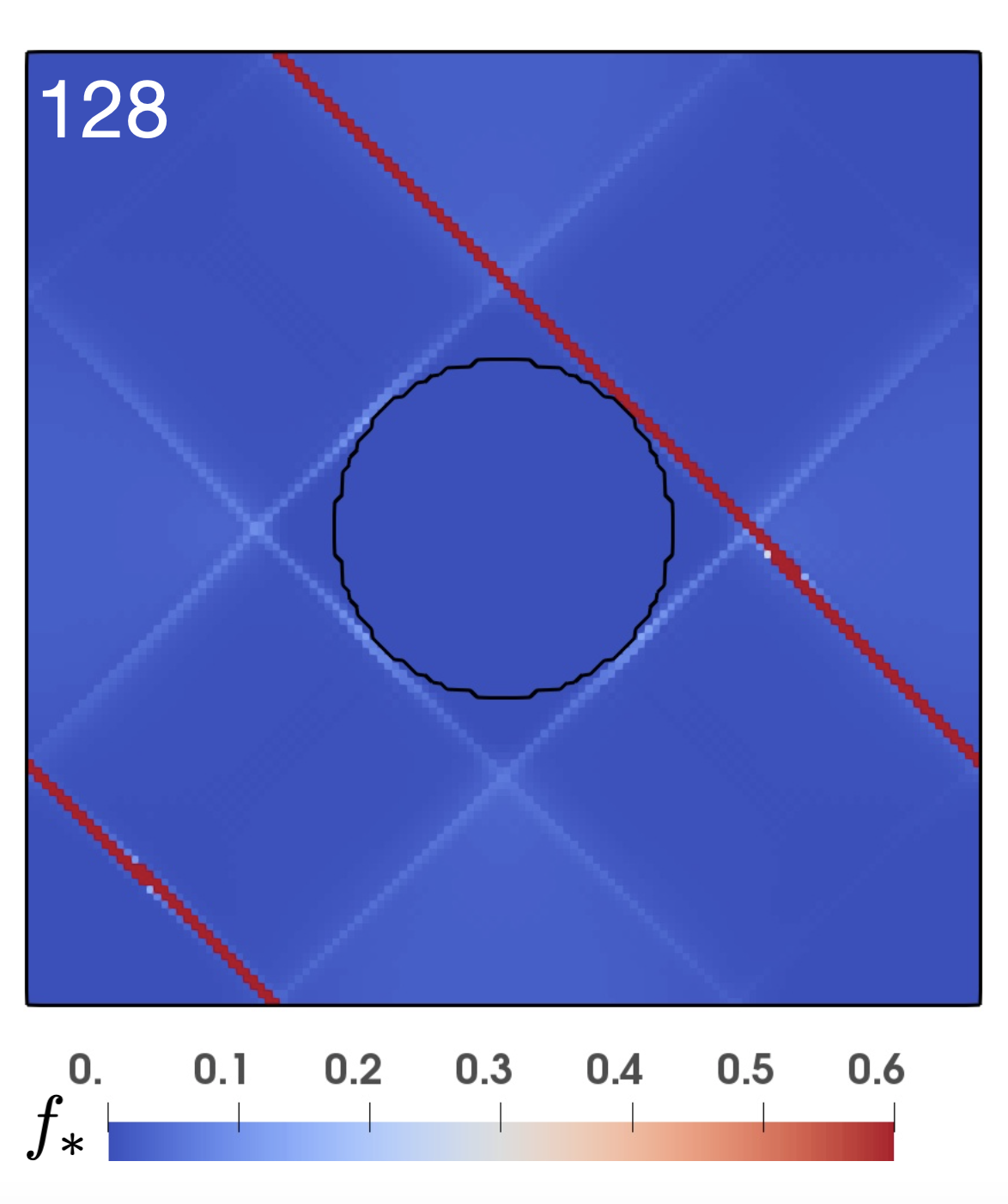}} 
		\end{tabular}
\caption{\em{Distribution of effective porosity $f_*$ at final fracture  simulated in case of the local Gurson model for grids of (a) $32^2$, (b) $64^2$, and (c) $128^2$. For all the grids it has been prescribed $\ell_M=\ell_I=1 \times 10^{-4} \, L$. }}
\label{fig:contoursLG}
\end{figure}

The characteristic grid independence of the solution of the non-local damage models can also be appreciated by looking at the distribution of the damage variable at final fracture for different grids, as reported in Fig. \ref{fig:contoursNG} and Fig. \ref{fig:contoursNL}. For all the simulated grids, the damage eventually condensate in two slip bands oriented at 45 degrees with similar width and location. Additionally, it can be observed how the localization band is distorted in the neighborhood of the elastic inclusion as a result of the generalized Helmholtz-type equation for a heterogeneous medium. This distortion is equivalent to the effect of having an internal Neumann-free boundary for the non-local variable in the Helmholtz-type equation, and shows the ability of the approach proposed to avoid artificial damage diffusion in non-damaging phases.
Conversely, as shown in Fig. \ref{fig:contoursLG}, the final distribution of damage resulting from the original local models reflects the pathological grid dependence due to loss of ellipticity of the problem. In this case the damage localizes in a slip band whose width is equal to the selected grid size. 

 \subsubsection{Impact of the characteristic length of the regularization}

\begin{figure}[htbp]
\centering
		\begin{tabular}{rc}
		\subfigure[]{ \includegraphics[height=5.5cm]{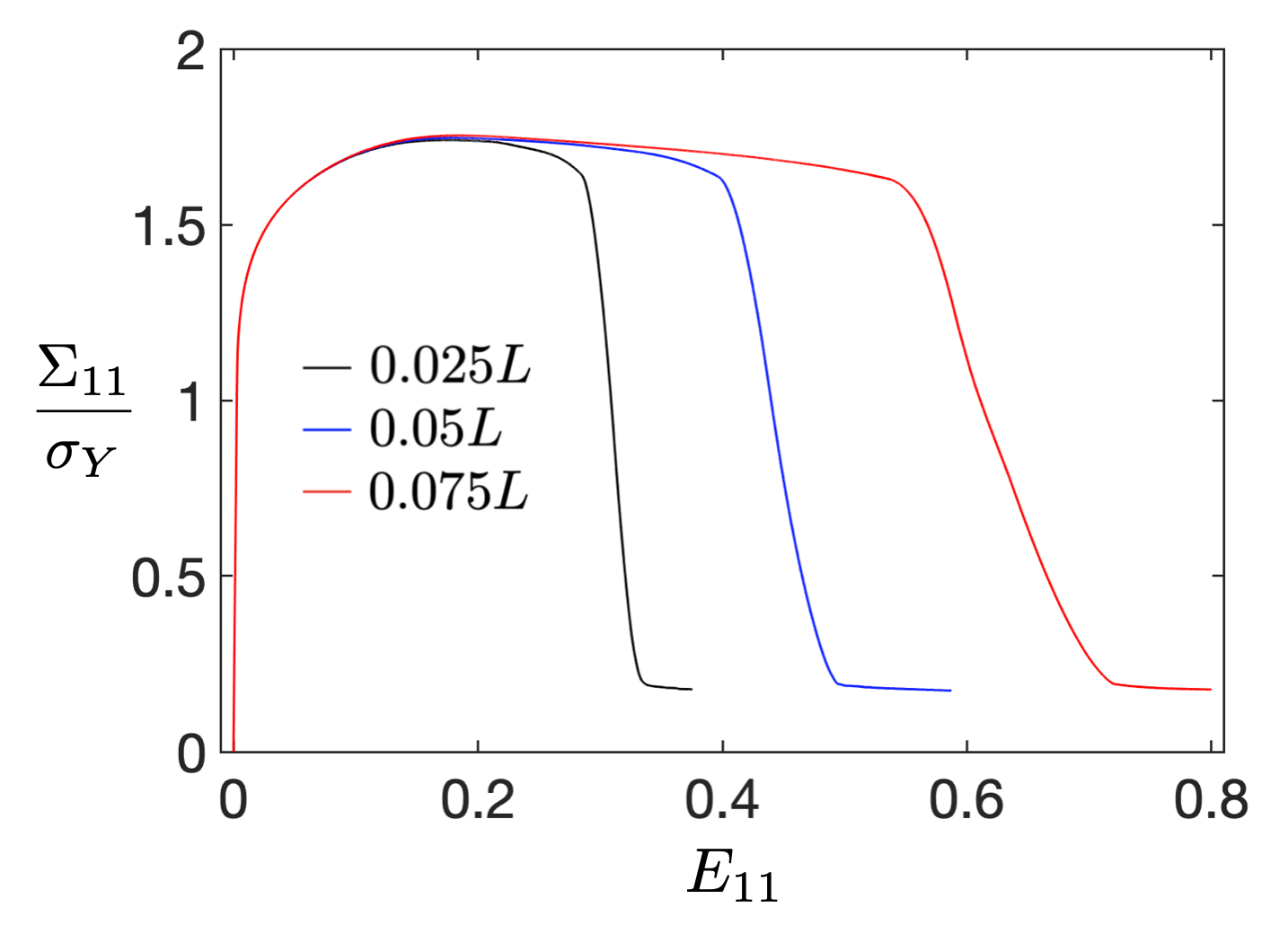} }
		&
		\subfigure[]{ \includegraphics[height=5.5cm]{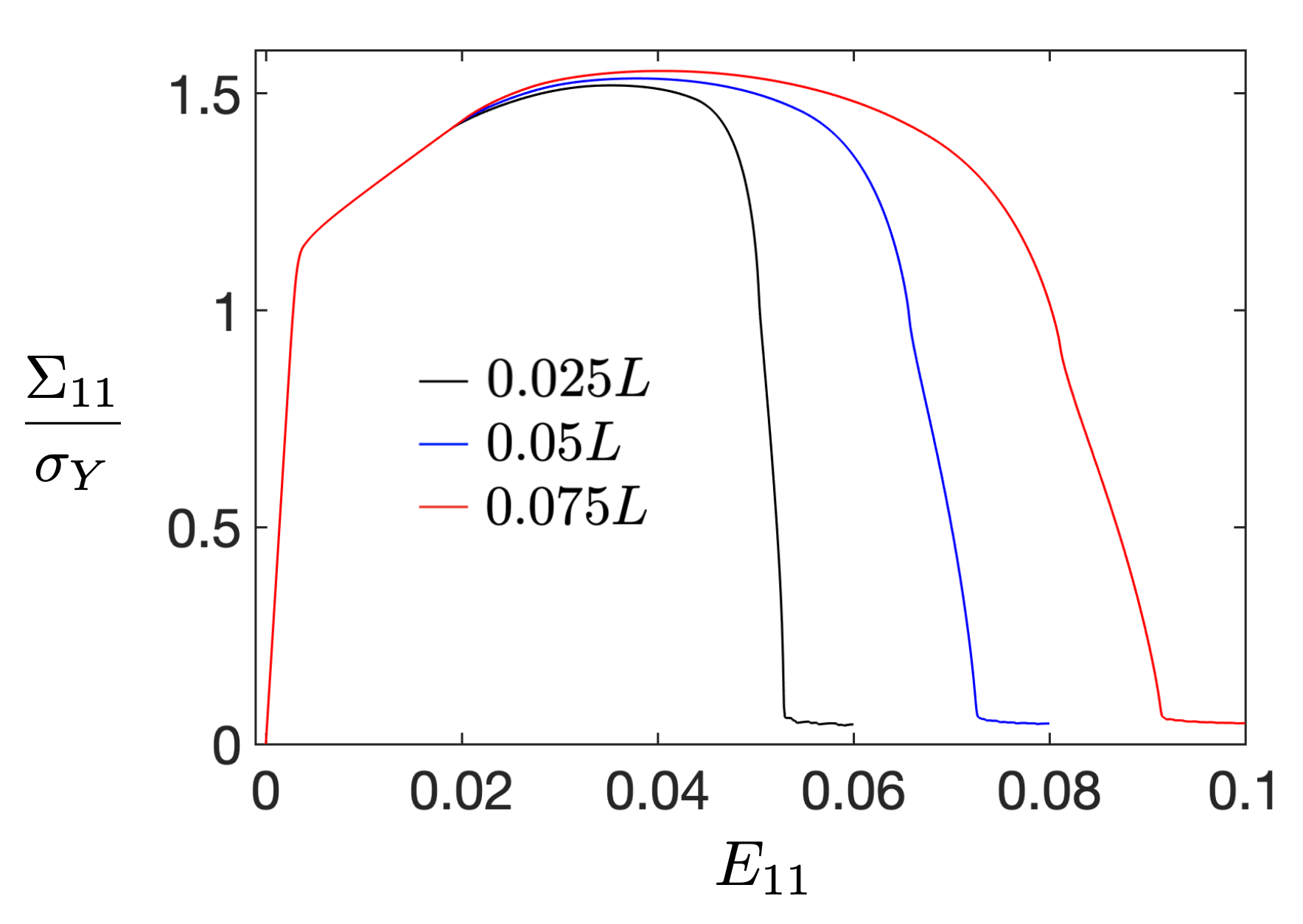}}
		\end{tabular}
\caption{\em{Impact of the matrix characteristic length $\ell_M$ on the simulated stress strain curves for non-local Gurson (a) and Lemaitre (b) models. For all the reported results, $\ell_I=0.001L$ has been assigned in the region of the elastic reinforcement.}}
\label{fig:ellstressstrain}
\end{figure}

The considered implicit gradient regularization introduces an internal length scale in the non-local damage models through the parameter $\ell$ in the Helmholtz-type equations  \eqref{eq:helmholtz_abstract}. Such a parameter basically affects the magnitude of the spreading of a local variable towards neighbouring grid points. For the considered 2D composite material, the characteristic length $\ell$ is not constant in space, but different values are assigned in the matrix material, i.e. $\ell_M$, and in the elastic reinforcement, i.e.$\ell_I$. Figure \ref{fig:ellstressstrain} plots the average mechanical response of the RVE for different assigned internal lengths in the matrix material at constant value $\ell_I=0.001L$. For both Gurson (Fig. \ref{fig:ellstressstrain}a) and Lemaitre (Fig. \ref{fig:ellstressstrain}b) models, the higher the parameter $\ell_M$ the higher the attained strain at final failure, e.g. ductility is increased. Indeed, a higher internal length scale implies a broader diffusion of the scalar plastic strain variables responsible of damage evolution. The latter thus makes the matrix material capable of dissipating more mechanical energy during the combined damage-plastic process. 

\begin{figure}[htbp]
\centering
		\begin{tabular}{lr}
		\subfigure[]{ \includegraphics[height=5.5cm]{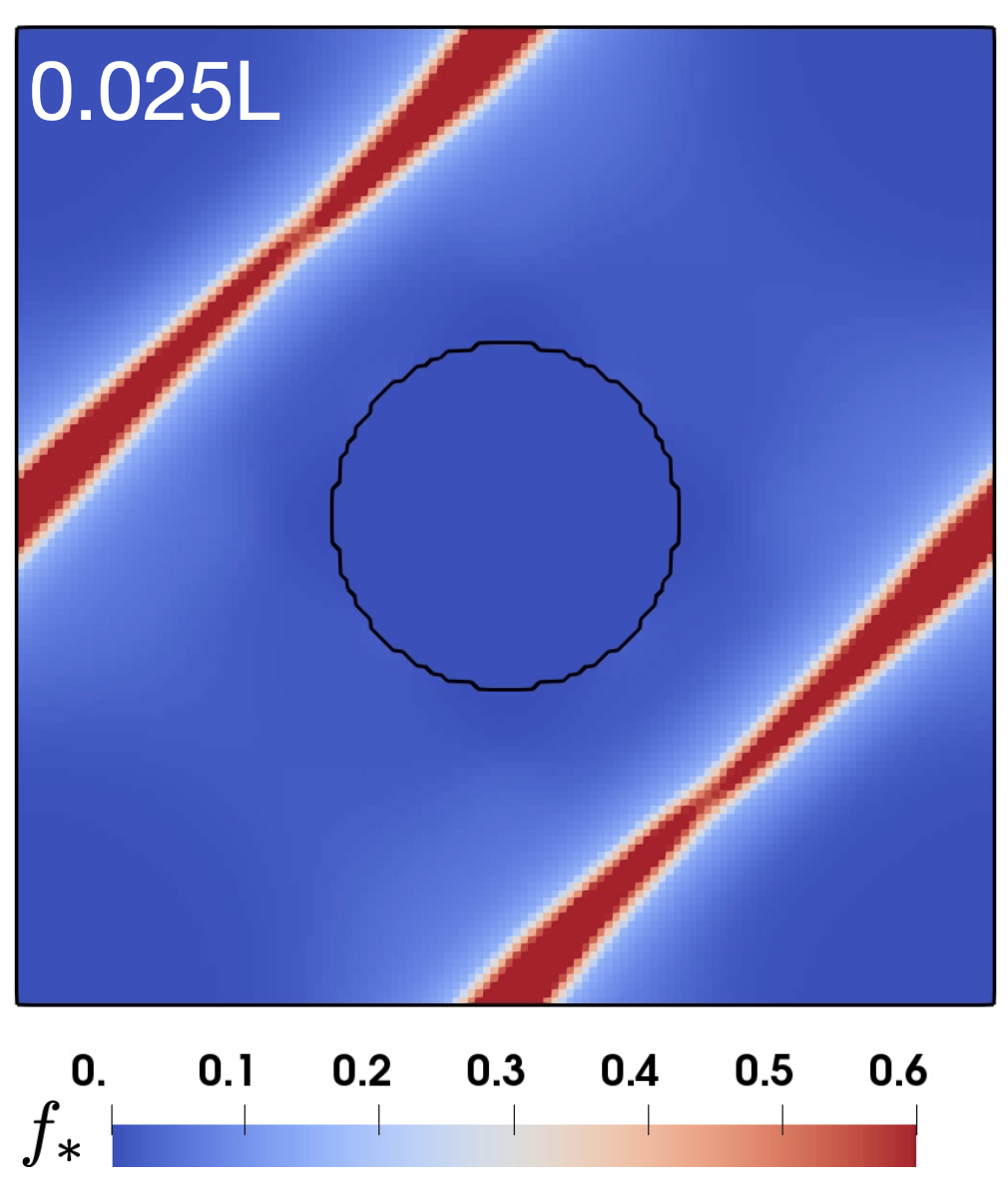}}  \hspace{2cm}
		&
		\subfigure[]{ \includegraphics[height=5.5cm]{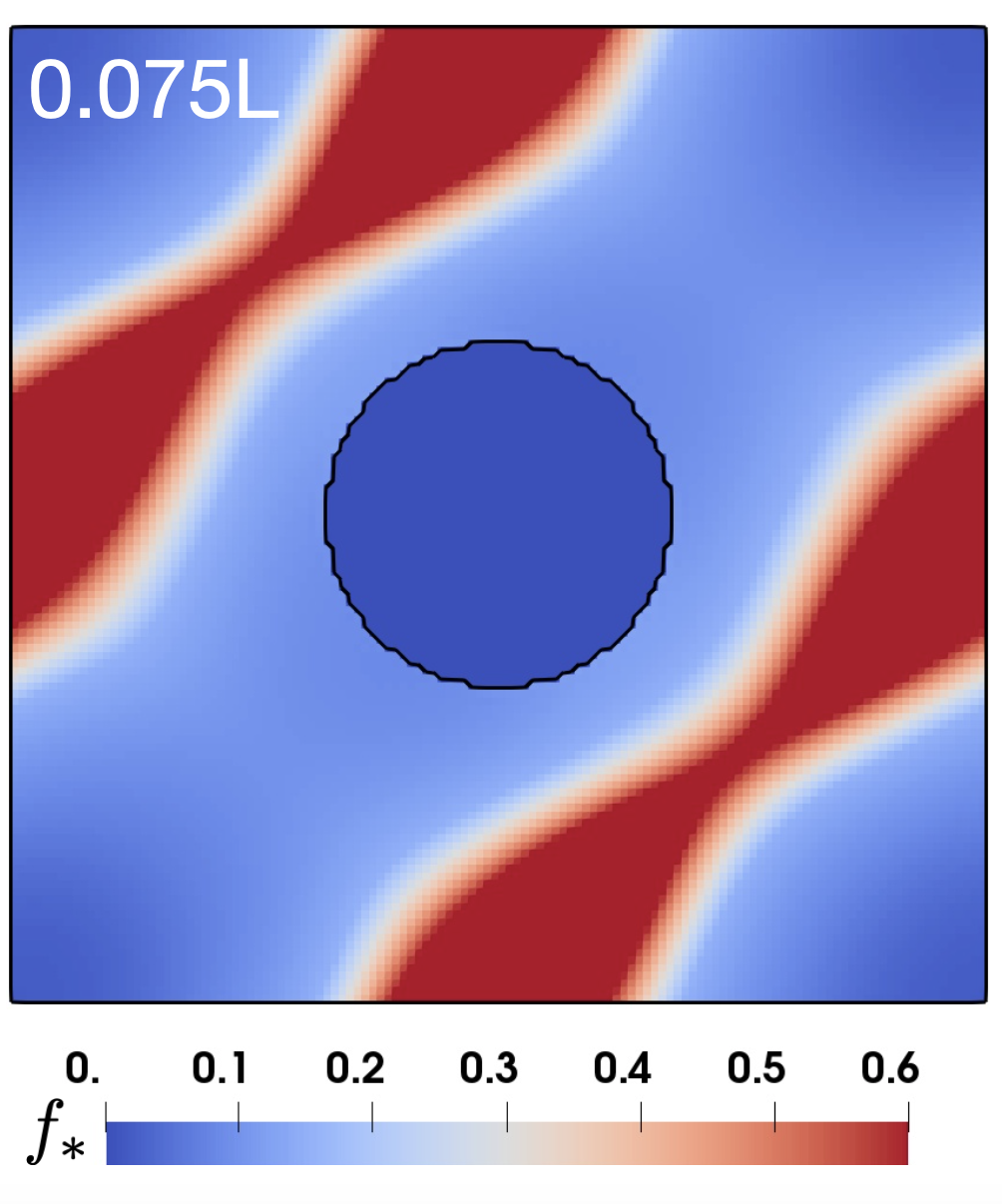}} 
		\end{tabular}
\caption{\em{Distribution of effective porosity $f_*$ at final fracture simulated in case of the non-local Gurson model for (a) $\ell_M=0.025 L$ and (b) $\ell_M=0.075$ with $128^2$ grid points. For all the reported results, $\ell_I=0.001$ has been assigned in the region of the elastic reinforcement.}}
\label{fig:contoursNGell}
\end{figure}

\begin{figure}[htbp]
\centering
		\begin{tabular}{lr}
		\subfigure[]{ \includegraphics[height=5.5cm]{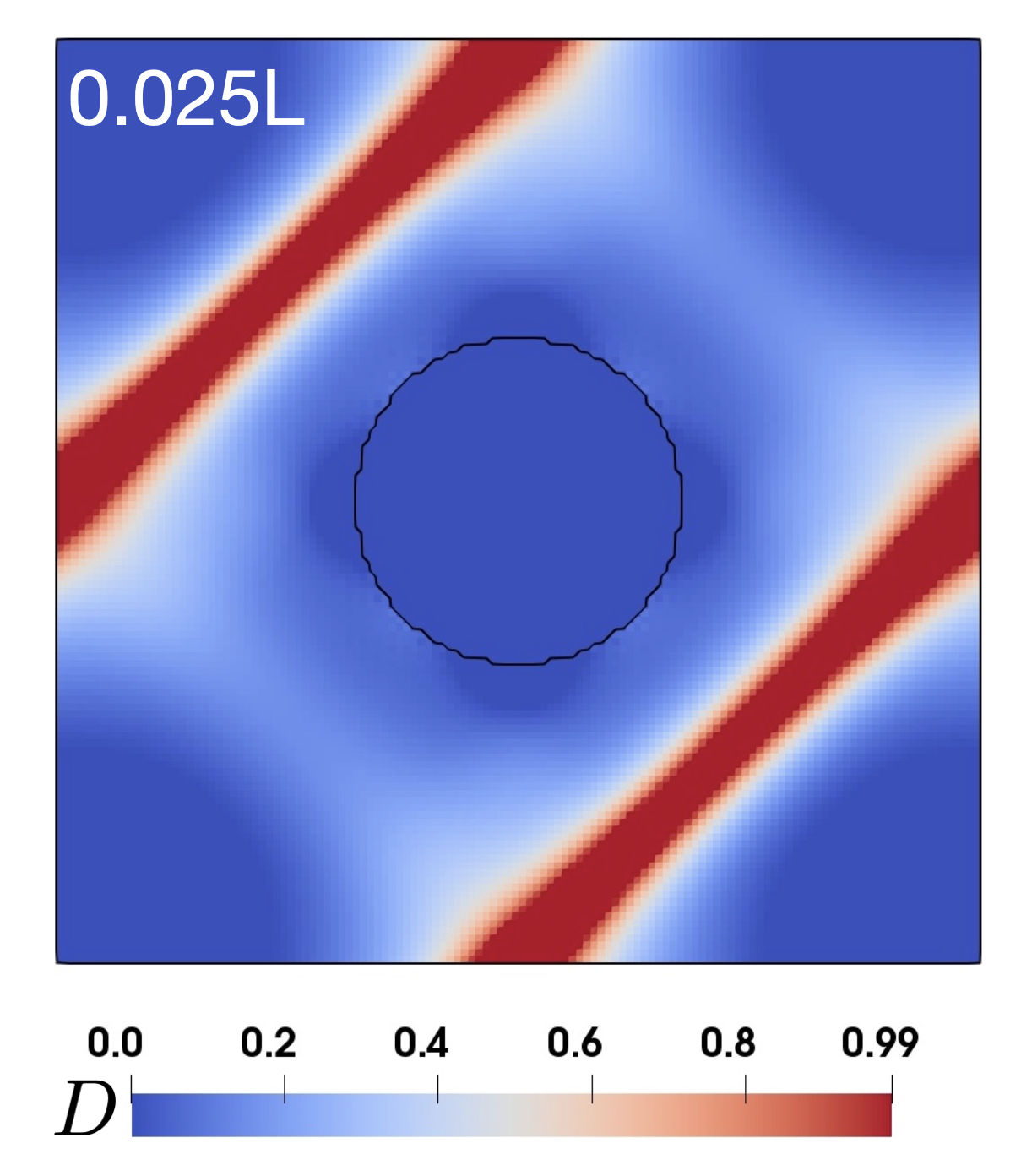}}  \hspace{2cm}
		&
		\subfigure[]{ \includegraphics[height=5.5cm]{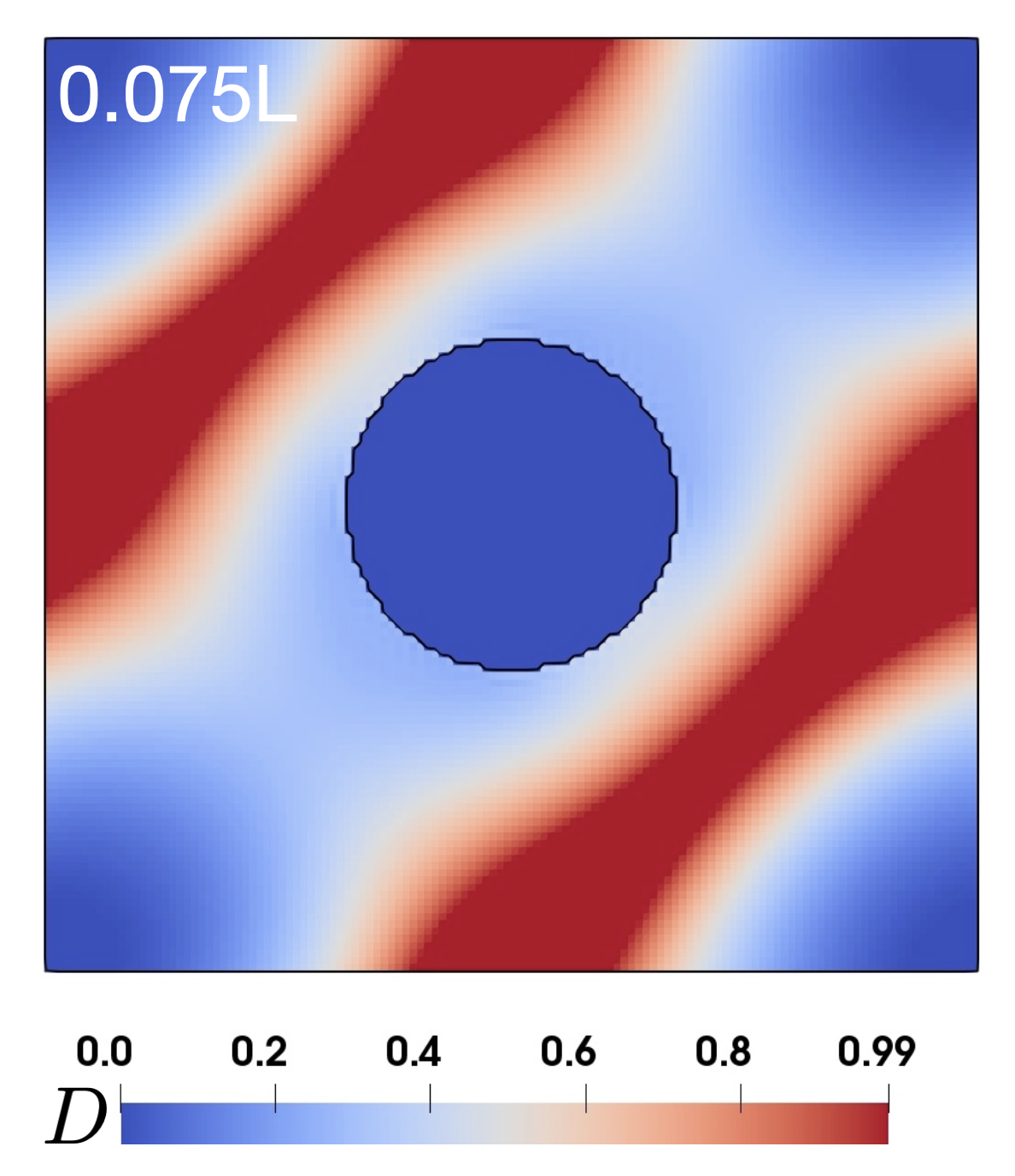}} 
		\end{tabular}
\caption{\em{Distribution of the intrinsic damage variable $D$ at final fracture simulated in case of the non-local Lemaitre model for (a) $\ell_M=0.025 L$ and (b) $\ell_M=0.075$ with $128^2$ grid points. For all the reported results, $\ell_I=0.001$ has been assigned in the region of the elastic reinforcement.}}
\label{fig:contoursNLell}
\end{figure}

The impact of $\ell_M$ on the spatial distribution of the damage variables at final fracture is reported in Fig. \ref{fig:contoursNGell} and Fig. \ref{fig:contoursNLell} for the non-local Gurson and Lemaitre models, respectively. As expected, the simulated slip band strongly depends on the matrix internal length resulting in thinner bands as $\ell_M$ decreases. The characteristic length of the regularization seems to affect the shape of the slip band as well. It results that higher $\ell_M$ promote the formation of bands with a more pronounced change in thickness along its longitudinal axis. On the other hand, for $\ell_M=0.025$ the damage variable localizes in a band with approximately constant thickness. Moreover, for large $\ell_M$ the damage distribution highlights a higher level of damage in the matrix material in the surrounding of the inclusion with respect to the case of a small $\ell_M$. Finally, the simulated orientation and location of the longitudinal axes of the slip bands are independent on the characteristic length $\ell_M$.

\begin{figure}[htbp]
\centering
		\begin{tabular}{rc}
		\subfigure[]{ \includegraphics[height=5.5cm]{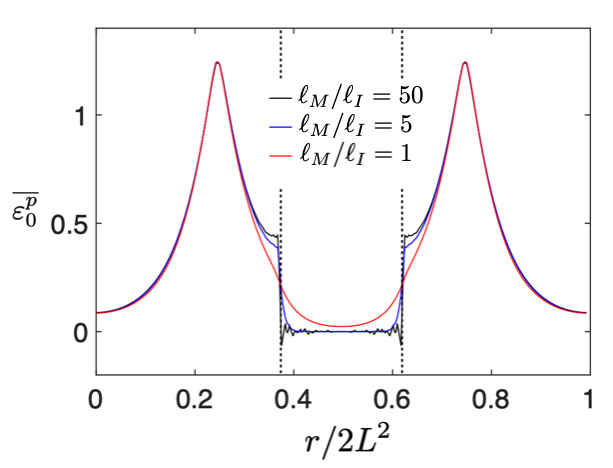} }
		&
		\subfigure[]{ \includegraphics[height=5.5cm]{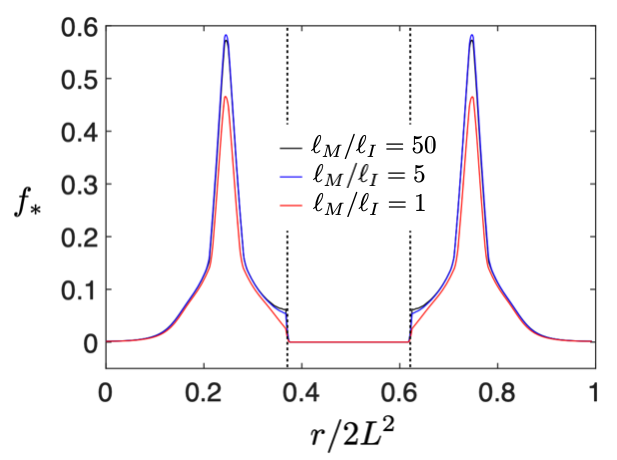}}
		\end{tabular}
\caption{\em{Plot of the profiles of the non-local equivalent plastic strain $\overline{\varepsilon_0^p}$ and effective porosity $f_*$ along direction $r$ (see Fig. \ref{fig:geom2D}) for different assigned ratios $\ell_M/\ell_I$. The reported curves refers to the non-local Gurson model in correspondence of $E_{11}=0.5$ for fixed $\ell_m=0.05$ and $128^2$ grid points. The dashed lines mark the boundary of the elastic inclusion.}} 
\label{fig:impactelli}
\end{figure}

The impact of the internal length $\ell_I$ in the elastic inclusion is evaluated in Fig. \ref{fig:impactelli} where the profiles of the simulated non-local equivalent plastic strain $\overline{\varepsilon_0^p}$ and effective porosity $f_*$ are plotted for different ratios $\ell_M/\ell_I$. Fig. \ref{fig:impactelli}a shows that $\ell_M/\ell_I>5$ effectively prevents the non-local equivalent plastic strain from diffusing inside the elastic matrix as desirable. On the contrary, a significant plastic deformation accumulates in the elastic reinforcement for $\ell_M/\ell_I=1$ providing a more gentle profile of $\overline{\varepsilon_0^p}$ at the matrix/inclusion interface. As plotted in Fig. \ref{fig:impactelli}b, the simulated profile of the effective porosity $f_*$ obviously reflects the distribution of the non-local equivalent plastic strain through the process of void nucleation. For higher values of the ratio $\ell_M/\ell_I=1$, the damage is more developed along the considered direction for the same applied macroscopic strain. 
It is worth nothing that very large ratios of $\ell_M/\ell_I$ lead to the appearance of Gibbs oscillations in the profile of the averaged field $\overline{\varepsilon_0^p}$, although the magnitude of these oscillations is quite limited thanks to the use of discrete derivatives in Fourier space \cite{WILLOT2015232}. Moreover, a closer look reveals that this oscillating behavior mostly affects the profile of $\overline{\varepsilon_0^p}$ inside the elastic phase, where damage is not developing. Therefore, the influence of these small oscillations in the damage nucleation/evolution in the matrix will be negligible. To further improve the accuracy of the Helmholtz-type equation for very high phase contrasts, the use of special enhanced FFT-formulations \cite{TO2020113160} might be used, but such aspects go beyond the scope of this paper and they will be studied in a devoted publication.

 \subsection{3D numerical examples} \label{sec:3Dexamples}

\begin{figure}[htbp]
\centering
		\begin{tabular}{lcr}
		\subfigure[]{ \includegraphics[height=5.5cm]{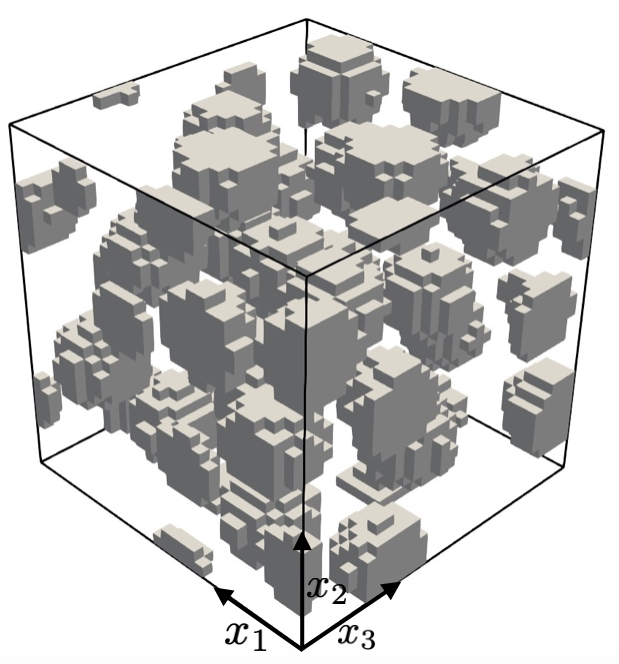}} 
		&
		\subfigure[]{ \includegraphics[height=5.5cm]{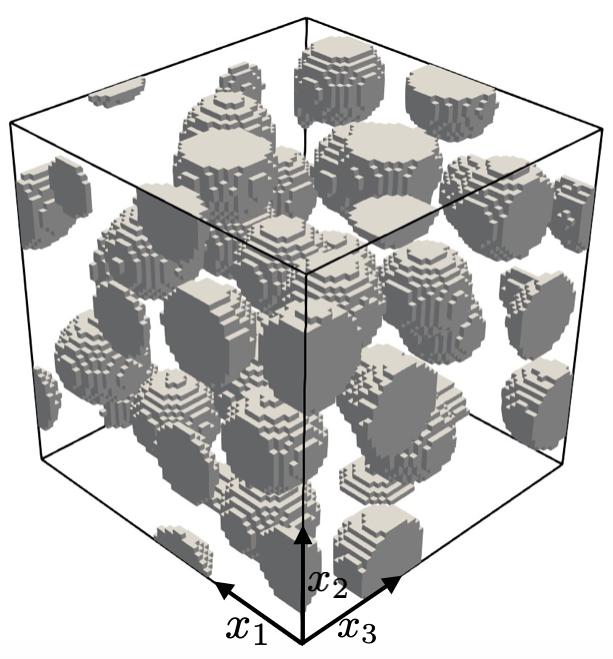}} 
		&
		\subfigure[]{ \includegraphics[height=5.5cm]{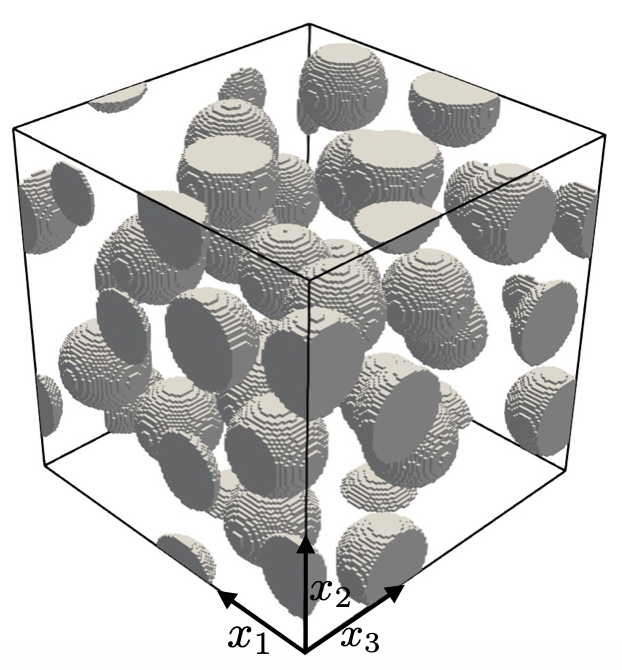}} 
		\end{tabular}
\caption{\em{Periodic multi-particle RVEs adopted in the 3D numerical simulations for (a) $32^3$, (b) $64^3$, and (c) $128^3$ grid points.}}
\label{fig:Geom3D}
\end{figure}

In this set of simulations the mechanical response of a three dimensional particle-reinforced composite is analyzed. The numerical study presented in \cite{LLORCA2004267}, which uses a local version of the Gurson model and finite elements, is reproduced here using the non-local FFT framework proposed. The RVE consists of thirty non-overlapping identical elastic spheres immersed in an elasto-plastic matrix undergoing damage. A cubic unit cell of size $L\times L \times L$ is considered with the elastic reinforcement occupying 20\% of the overall volume. The arrangement of the elastic particles is generated randomly and discretized with different raster resolution, namely  $32\times32\times32$, $64\times64\times64$, and  $128\times128\times128$ grid points. A representation of adopted geometry of the unit cell, for all the considered resolutions, is reported in Figure \ref{fig:Geom3D}. 

The sample is subjected to uniaxial tensile loading in which a macroscopic strain is prescribed in the direction $x_1$, while a stress free condition is enforced in the remaining components

\begin{equation*}
{\boldsymbol{E}} = 
\begin{pmatrix} E_{11}  & * & * \\ * & * &* \\  * & * &*  \end{pmatrix} 
 \qquad \text{and} \qquad
{\boldsymbol{\Sigma}} =  
\begin{pmatrix} * & 0 & 0 \\ 0 & 0  & 0 \\  0 & 0  & 0 \end{pmatrix} \, .
\end{equation*}

\noindent
In the matrix material, $E_M=70$, $\nu_M=0.33$, and $\sigma_Y=200$ MPa are assigned as typical values for Al alloys. As carried out in Section \ref{sec:2Dexamples}, the proposed algorithm is tested for both the considered Gurson and Lemaitre models. In the former, the properties of the combined plastic-damage process specify as $q_1=1.5$, $q_2=1$, $q_3=2.25$, $f_C=0.15$, $f_F=0.25$, $f_N=0.04$, $\varepsilon_N=0.1$, and $s_N=0.05$. The initial void volume fraction is zero. Moreover, the same hardening law Eq. \eqref{eq:aravas} adopted in the 2D simulations is considered. On the other hand, the set of parameters characterizing the Lemaitre model are $\epsilon_C=0.03$ and $\epsilon_R=0.2$ along with isotropic linear hardening $\sigma_0 \left( \epsilon_p \right)= \sigma_Y + k \, \epsilon_p$ with hardening modulus $k=10$ GPa. Following the same path of reasoning of Section \ref{sec:2Dexamples}, the upper limit value of the damage indicators is set to $f_V^*=0.6$ and $D=0.99$ for Gurson and Lemaitre models, respectively. 

The elastic inclusions do not undergo damage and their elastic properties correspond to $E_I=400$ GPa and $\nu_I=0.2$. In the non-local numerical simulations, the parameter appearing in the Helmholtz-type equations is specified as $\ell_M=0.05 \,  L$ for the matrix material and $\ell_I=0.001 \, L $ for the elastic inclusions. The latter choice prevent spurious diffusion of inelastic fields through the interface between matrix and elastic particles as demonstrated in the two dimensional examples.

\subsubsection{Grid sensitivity}

\begin{figure}[htbp]
\centering
		\begin{tabular}{rc}
		\subfigure[]{ \includegraphics[height=5.5cm]{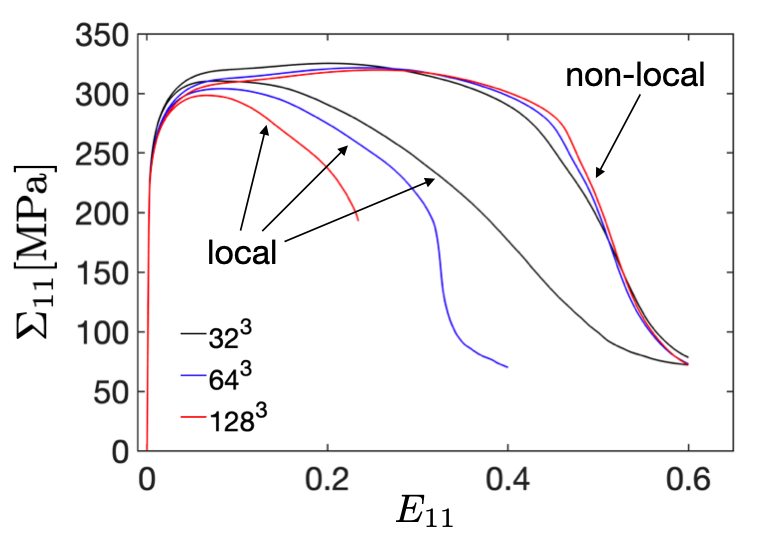} }
		&
		\subfigure[]{ \includegraphics[height=5.5cm]{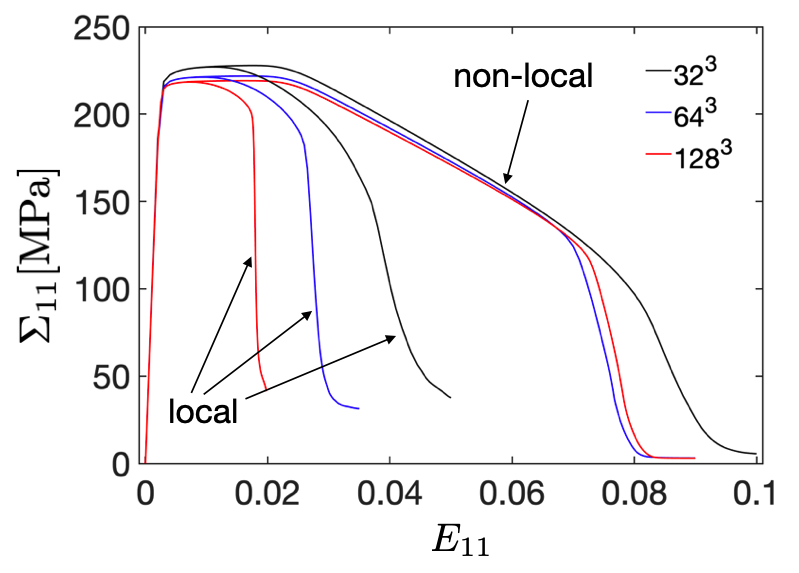}}
		\end{tabular}
\caption{\em{Plot of the average stress component $\Sigma_{11}$  against the average strain component $E_{11}$ resulting from the 3D multi-particle RVEs reported in Fig. \ref{fig:Geom3D} for the considered Gurson (a) and Lemaitre models (b). In each graph, the curves obtained from the non-local models are plotted along with the results of the local counterparts. }}
\label{fig:3D-stress-strain}
\end{figure}

The simulated macroscopic response of the considered three-dimensional microstructure is plotted in Figure \ref{fig:3D-stress-strain} for both Gurson (a) and Lemaitre (b) models. To emphasize the impact of the non-local formulation, the results of the non-local damage models are plotted together with the classical local counterpart. The latter are simply recovered by assigning a sufficiently small value of the characteristic length $\ell$ to avoid spreading of plastic local variables to neighbouring grid points, i.e. $\ell_M=\ell_I=1 \times 10^{-4} L $. A variable strain increment is applied due to requirements of smaller strain increments after a certain macroscopic strain related to the softening part of the curve. It turned out that the model resolution affected the total number of increments needed to reach the final deformation. Considering the non-local Gurson model, the number of increments necessary to achieve the final microscopic strain ($E_{11}=0.6$) were 600, 605, and 730 for $32^3$, $64^3$, and $128^3$ grids, respectively. For the non-local Lemaitre model the strain increments grow from 109,  for the $32^3$ grid, to 179 and 279 for $64^3$ and $128^3$ discretizations. A similar trend in the convergence was observed for the local models as well, and the average strain increment decreased with the number of voxels of the model.  Moreover, for the highest grid resolution the analyses of the local models could not reach the target macroscopic strain since the strain increment was reduced below the minimum admissible value, i.e. $\Delta E_{11} = 1 \times 10^{-5}$.


\begin{figure}[htbp]
\centering
		\begin{tabular}{lcr}
		\subfigure[]{ \includegraphics[height=6cm]{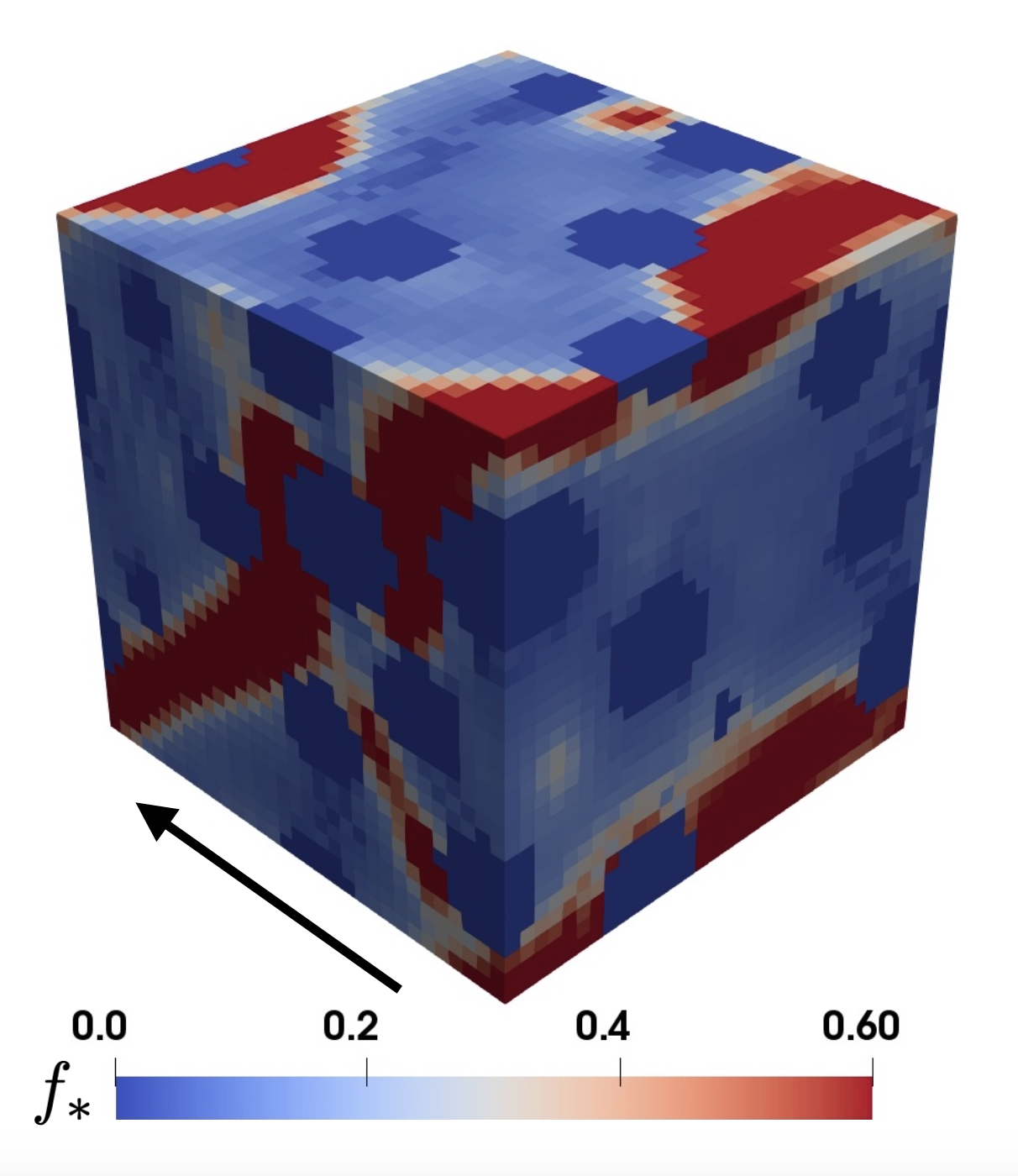}} 
		&
		\subfigure[]{ \includegraphics[height=6cm]{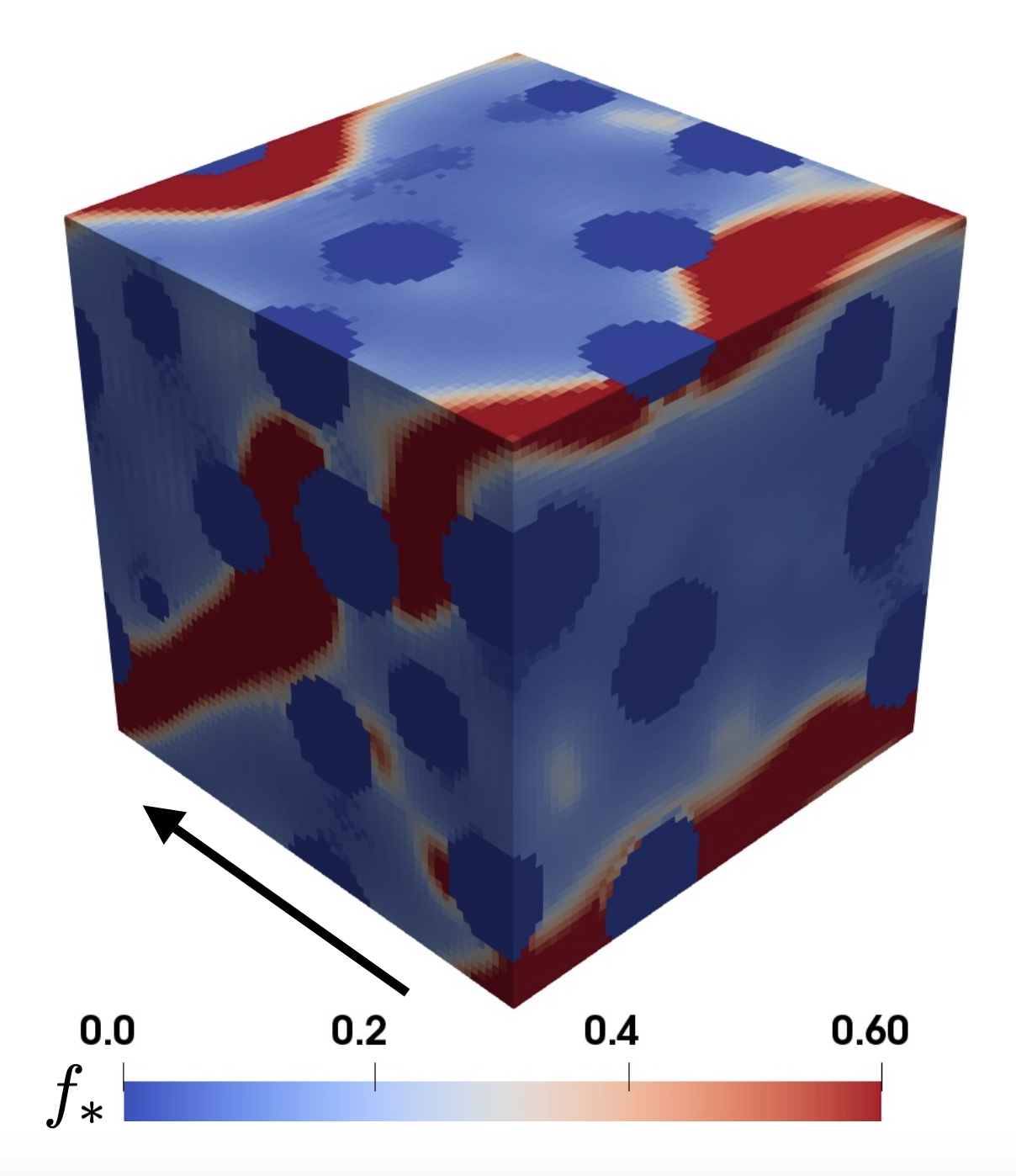}} 
		&
		\subfigure[]{ \includegraphics[height=6cm]{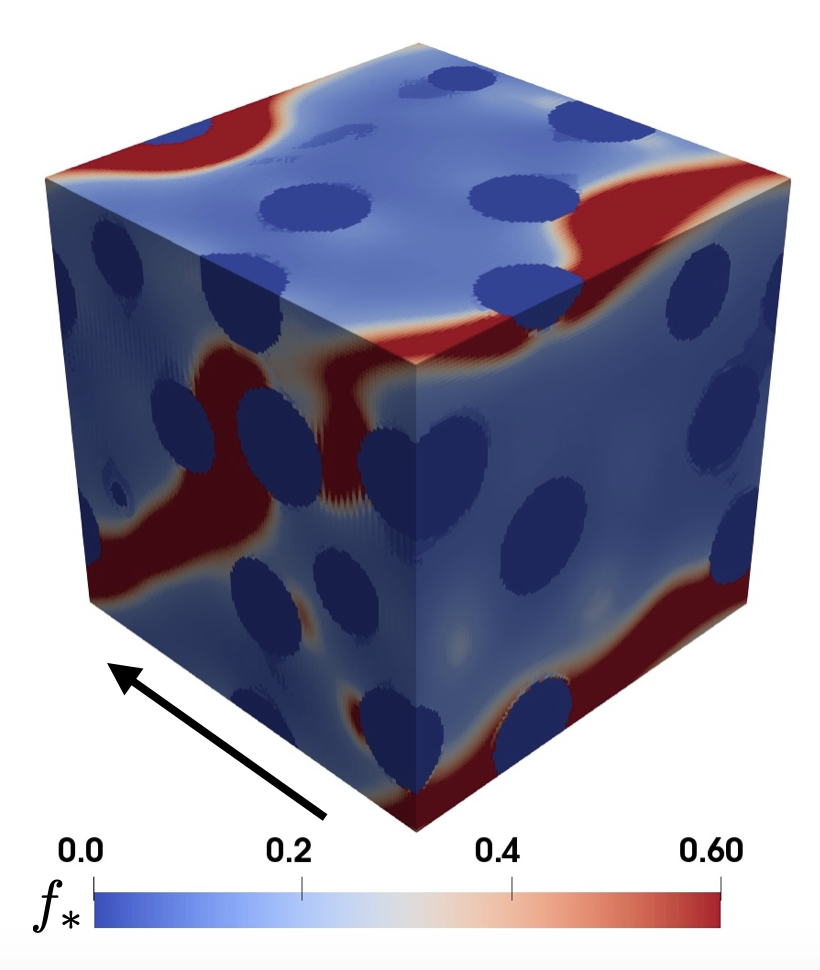}} 
		\end{tabular}
\caption{\em{Distribution of the effective porosity $f_*$ at final fracture ($E_{11}=0.6$) simulated in case of the non-local Gurson model with $\ell_M=0.05 \, L$ and $\ell_I=0.001 \, L$ for (a) $32^3$, (b) $64^3$, and (c) $128^3$ grid points. For all the considered grid resolutions, the black arrow indicates the loading direction. }}
\label{fig:contoursNG_3D}
\end{figure}

\begin{figure}[htbp]
\centering
		\begin{tabular}{lcr}
		\subfigure[]{ \includegraphics[height=6cm]{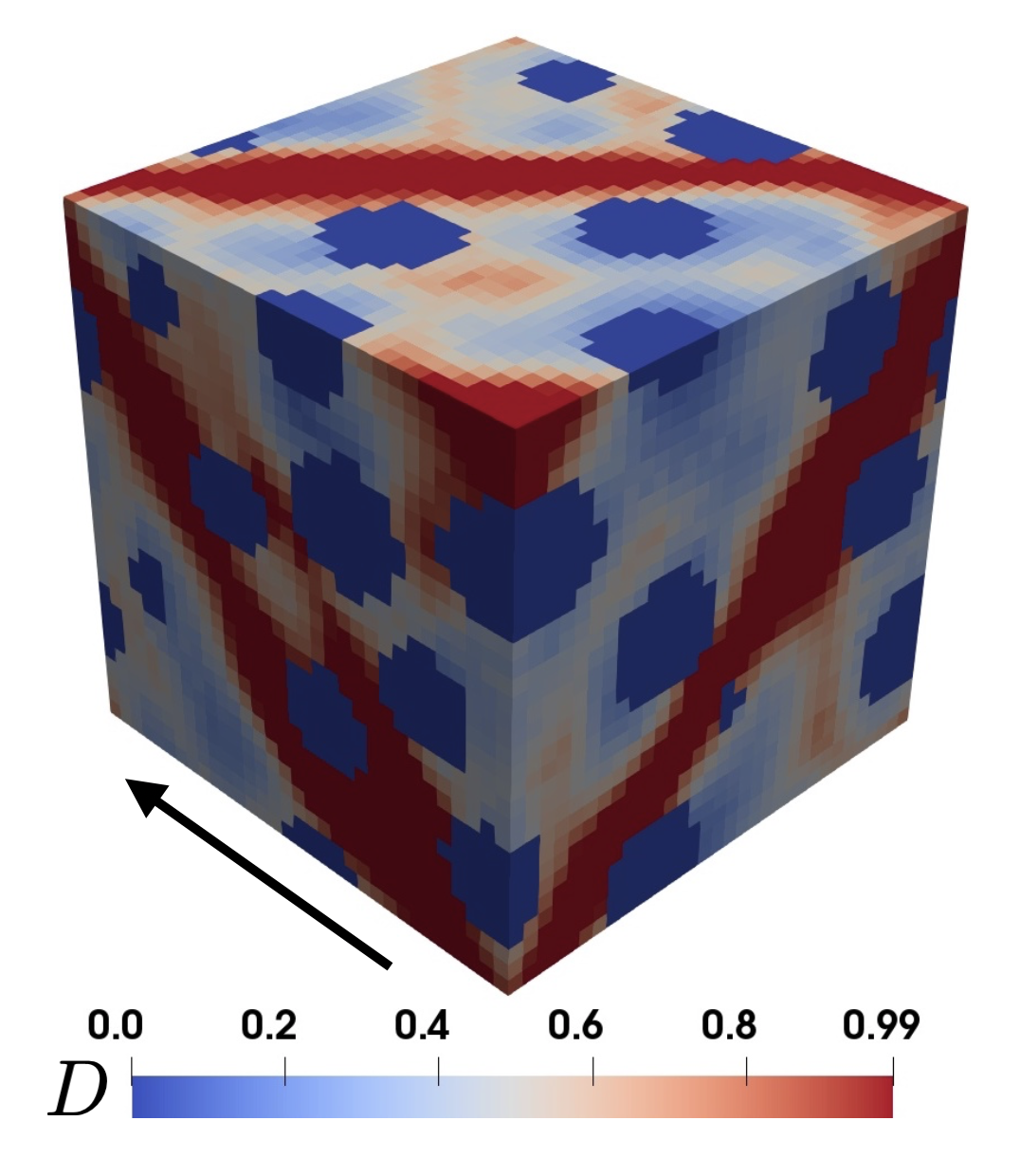}} 
		&
		\subfigure[]{ \includegraphics[height=6cm]{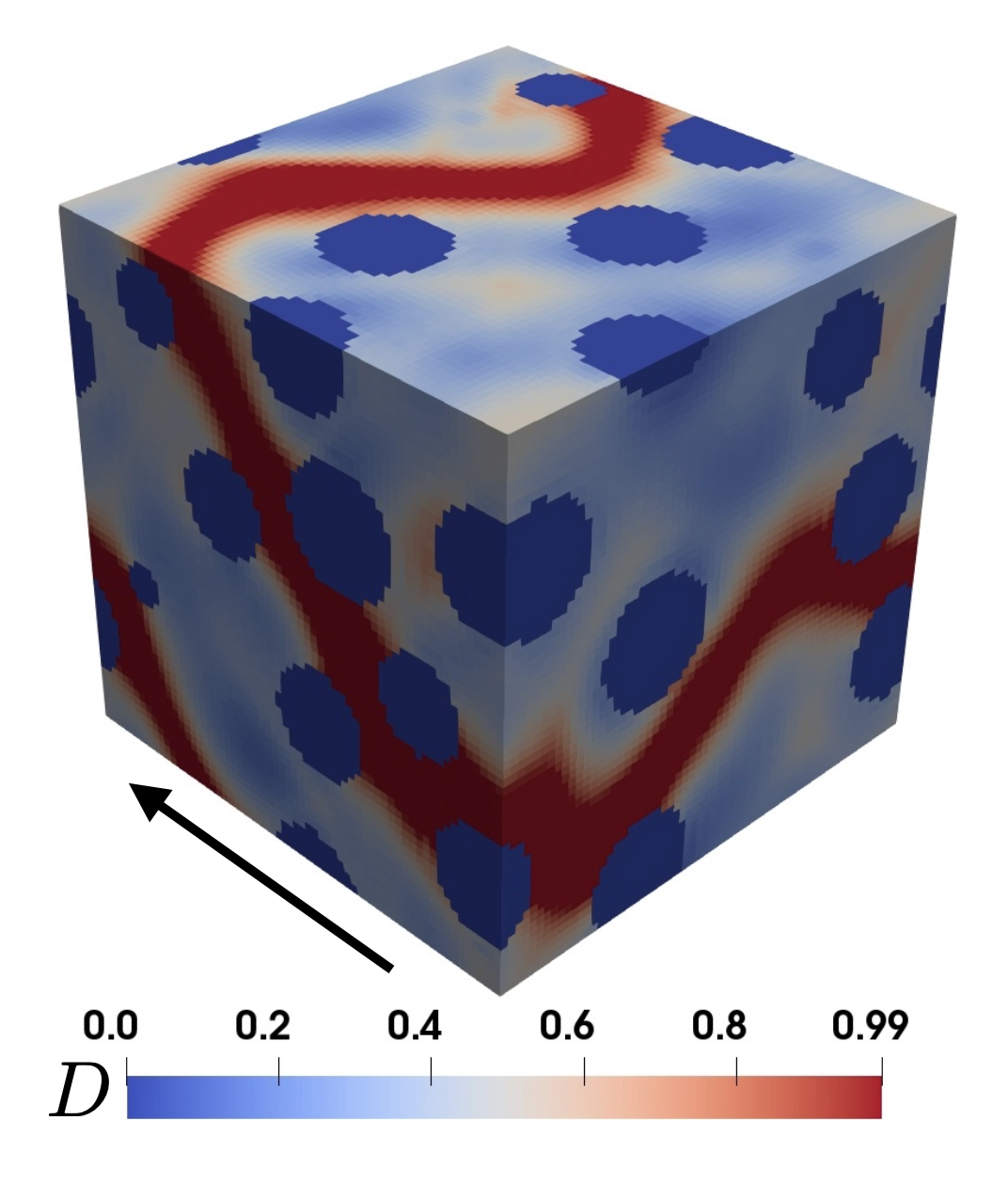}} 
		&
		\subfigure[]{ \includegraphics[height=6cm]{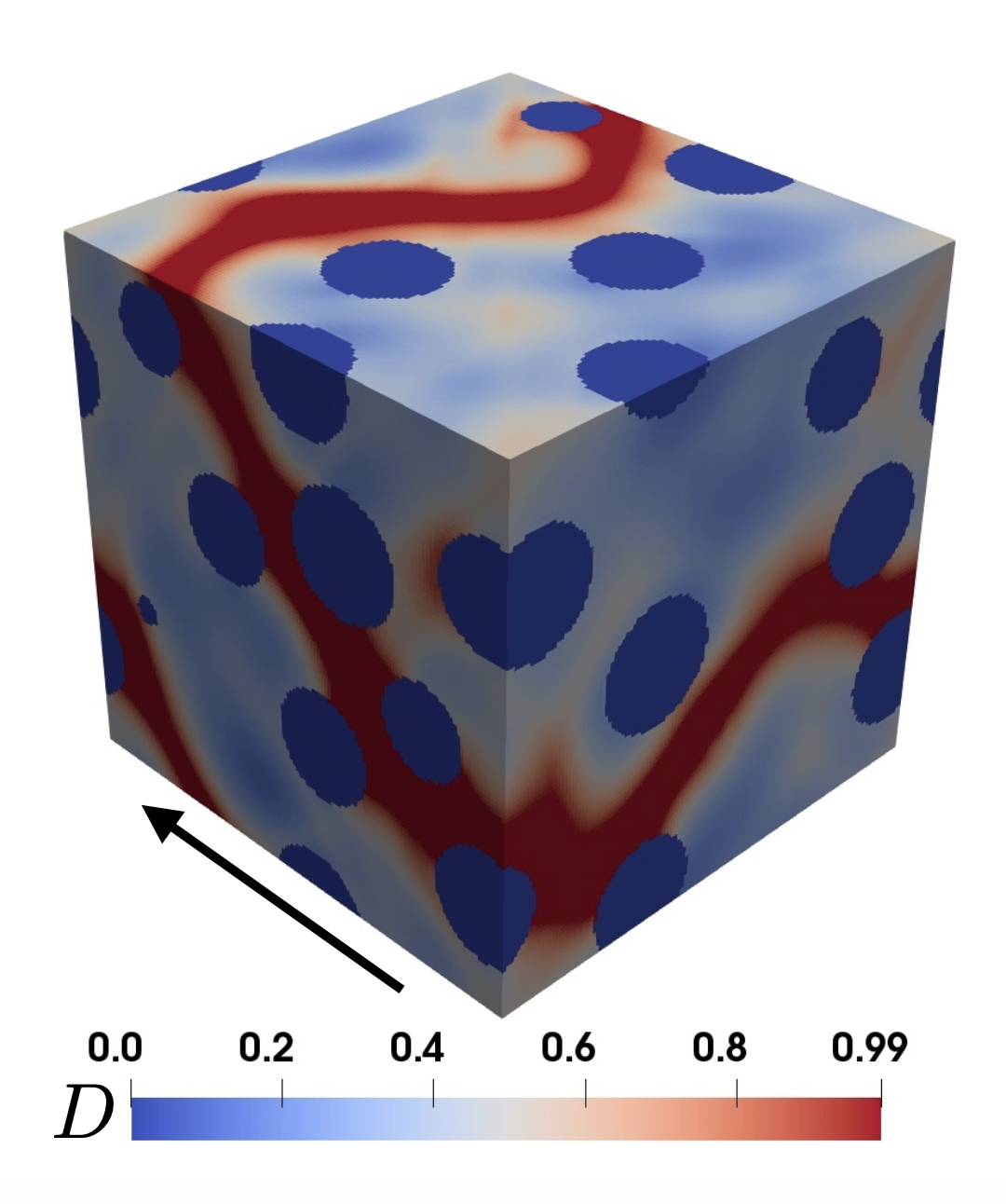}} 
		\end{tabular}
\caption{\em{Distribution of the intrinsic damage variable $D$ at final fracture ((a) $E_{11}=0.1$ - (b) and (c) $E_{11}=0.09$) simulated in case of the non-local Lemaitre model with $\ell_M=0.05 \, L$ and $\ell_I=0.001 \, L$ for (a) $32^3$, (b) $64^3$, and (c) $128^3$ grid points. For all the considered grid resolutions, the black arrow indicates the loading direction. }}
\label{fig:contoursNL_3D}
\end{figure}

The simulated macroscopic stress-strain curves (Figure \ref{fig:3D-stress-strain}) confirm the benefit of the non-local formulation in the regularization of the mechanical response during softening for both damage models. In the case of non-local Gurson model, the resulting stress strain curves were very close until the final fracture for all the different discretization levels considered. In the Lemaitre model, the results using grids with $64^3$ and $128^3$ voxels also lay very close to each other, while a significant difference at final rupture is found with the case of the $32^3$ grid. This dissimilarity is the consequence of the different evolution of the damage variable in the $32^3$ resolution, as shown in Fig. \ref{fig:contoursNL_3D}, which might be attributed to a poor spatial discretization of the RVE for the considered damage model. Nevertheless, the comparison between the final distribution of the variable $D$ between $64^3$ and $128^3$ grids remarks a satisfactory grid independence of the non-local formulation. On the other hand, the images reported in Fig. \ref{fig:contoursNG_3D}, testify a similar final distribution of the effective porosity for all the considered grid sizes simulated in case of non-local Gurson model. 

\begin{figure}[htbp]
\centering
\includegraphics[width=15cm]{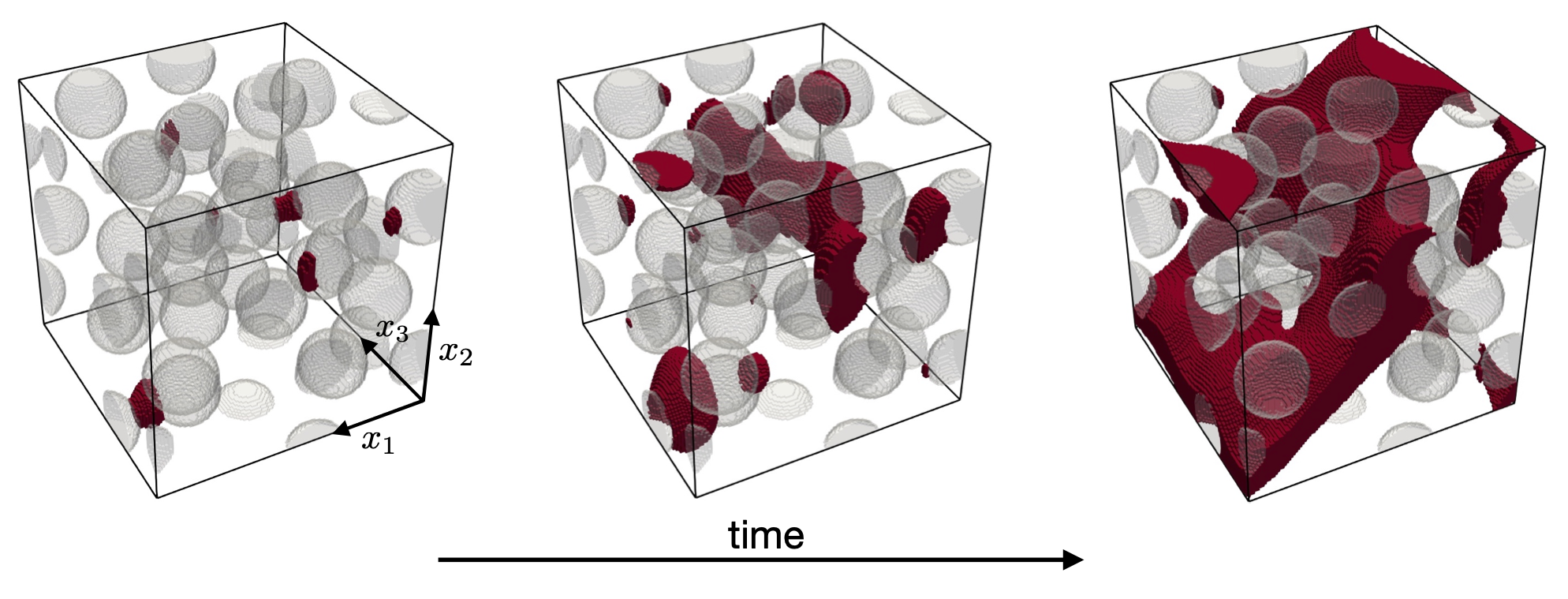}
\caption{\em{Three-dimensional propagation of ductile fracture predicted by the non-local Gurson model for the $128^3$ grid resolution. The distribution of the fracture has been identified by the collection of the grid points for which $f_*= 0.6$. }}
\label{fig:fractGN}
\end{figure}

\begin{figure}[htbp]
\centering
\includegraphics[width=15cm]{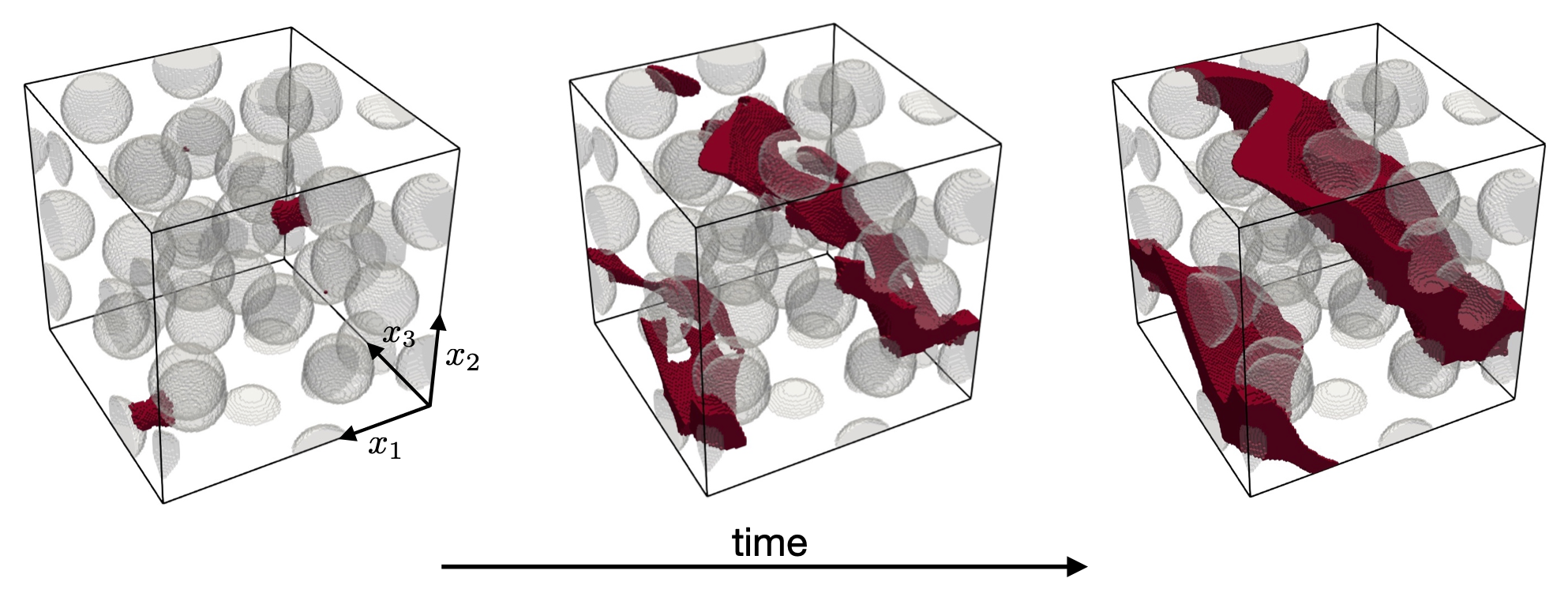}
\caption{\em{Three-dimensional propagation of ductile fracture predicted by the non-local Lemaitre model for the $128^3$ grid resolution. The distribution of the fracture has been identified by the collection of the grid points for which $D= 0.99$.}}
\label{fig:fractLN}
\end{figure}

Figure \ref{fig:fractGN} shows the three-dimensional evolution of the fracture simulated by the non-local Gurson model. The fracture initially nucleates in the regions of the material matrix where the distance among the reinforcements, along the loading direction, is small as observed in \cite{LLORCA2004267}. This fact reflects the plastic strain driven mechanisms underlying the evolution of the porosity in the Gurson model, as the matrix region between spheres closely packed along the deformation axis experiences higher plastic deformation. Subsequently, the fracture spreads to neighbouring regions, following the percolation path given by the presence of the elastic inclusions, to approximately resemble the shape of a plane at the end of the simulation. Such a plane represents a sort of shear plane that characterizes the collapse mechanism of the composite material for the considered loading condition. By the comparison  between Figure \ref{fig:fractGN} and \ref{fig:fractLN}, the nucleation of the fracture simulated by the non-local Lemaitre model takes place similarly to the case of Gurson model.  In addition, the collapse mechanism of the composite is still characterized by the presence of a shear plane that forms an angle of about 45 degrees with respect to the loading direction. However, the different fracture propagation in the two models leads to a distinct final distribution of the damage variable. It thus follows that the type of damage model, along with the choice of the relevant material parameters, can impact significantly on the prediction of the fracture propagation in complex heterogeneous media. 

\subsubsection{Convergence for a fixed topology}
\begin{figure}[htbp]
\centering
		\begin{tabular}{rc}
		\subfigure[]{ \includegraphics[height=5.5cm]{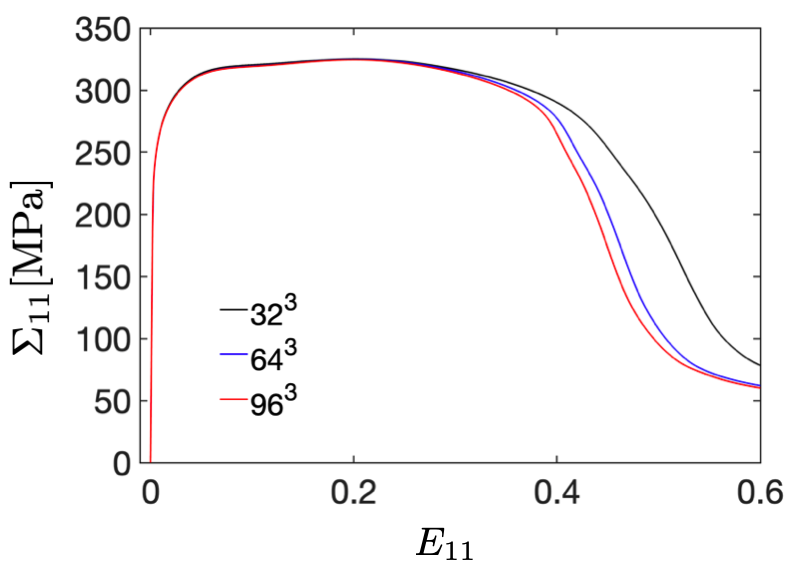} }
		&
		\subfigure[]{ \includegraphics[height=5.5cm]{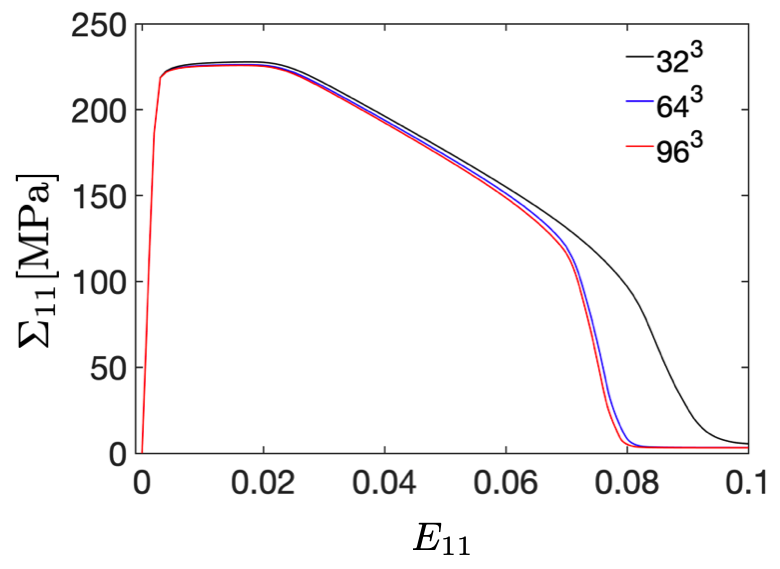}}
		\end{tabular}
\caption{\em{ Comparison between the macroscopic stress-strain curves simulated considering the same topology of the RVE (i.e. the one reported in Fig. \ref{fig:Geom3D}a) for different spatial discretizations. For both Gurson (a) and Lemaitre(b) non-local models, the results show a good convergence of the simulated mechanical behavior upon grid refinement. } }
\label{fig:3D-stress-straintopo}
\end{figure}

 In the RVEs considered in the aforementioned simulations, the spatial discretization strongly modifies the digitalized geometry of the elastic inclusions so that different grid resolutions imply topological changes as well (e.g. compare (a) and (c) in Fig. \ref{fig:Geom3D}). As a result of this fact, the simulated macroscopic stress strain curves reported in Fig. \ref{fig:3D-stress-strain} show a slight grid dependence in the elastoplastic response, i.e. when the applied deformation is low and the impact of damage has not yet come into play. In this section, for the sake of completeness, we report an additional converge study of the propose FFT algorithm in the case in which the level of the spatial discretization does not alter the topology of the RVE. This study is carried out considering the digitalized RVE with the lowest resolution, i.e. Fig. \ref{fig:Geom3D}a with $32^3$ voxels, as the actual geometry of the particle-reinforced composite to be simulated. The selected geometry was then discretized with different resolutions, namely $32^3$, $64^3$, and $96^3$ so that the RVE is perfectly discretized for all the considered grids. The relevant mechanical response has been simulated for both Gurson and Lemaitre models and the macroscopic stress strain curves are reported in Figure \ref{fig:3D-stress-straintopo}. As already observed in Fig. \ref{fig:3D-stress-strain}, the non-local formulation effectively alleviates the grid sensitivity typical of classical damage models. Nevertheless, when the topology is maintained and only the discretization is modified, the stress-strain response of the three different discretization level are almost indistinguishable until certain level of deformation ($E_{11} \simeq 0.3$ and $E_{11} \simeq 0.5$ for the Gurson and Lemaitre model, respectively). Subsequently, for higher applied strain, the simulated curves no longer overlap but clearly show a convergence of the predicted solution upon grid refinement.  
 
\subsubsection{Numerical efficiency}
The time required by each solver in the proposed iterative algorithm was analyzed in the considered 3D examples. From this analysis, the fraction of the total simulation time devoted to evaluate the constitutive equations is very limited and remained almost constant for all the discretization levels:  below $1.5$\% of the overall time for the Gurson model and below $0.2$\% for the Lemaitre model. 

Regarding the time spent to solve the set of coupled partial differential equations of the problem, the resolution of the mechanical equilibrium (a non-linear PDE solved using the FFT-Galerkin solver along with a Newton-Raphson algorithm) took the majority of time. In particular, the mechanical solver for the non-local Gurson model took from  $84$\% ($32^3$ grid) up to $92$\% ($128^3$ grid) of the total time while for the Lemaitre model it took from $95$\% to $97$\%. The remaining time was spent by the conjugate gradient solver for the heterogeneous Helmholtz-type equations. This difference in time is mainly due to the non-linearity of the equation of the mechanical equilibrium, opposite to the linear nature of the Helmholtz equation. This large difference in the time spent for the two solvers also suggests that the use of a brittle fracture model, which is linear for a fixed value of the damage parameter, could strongly reduce the computing time. 


 \section{Summary and conclusions} 

A general, robust and efficient FFT algorithm for the solution of non-local ductile damage in the field of computational micromechanics  has been proposed and particularized to two classical ductile damage models, namely Gurson \cite{GURSON1977} and Lemaitre \cite{Lemaitre1985} models.
 
To alleviate the typical grid dependence affecting the numerical results of classical local damage models, an implicit gradient regularization has been exploited. In the context of micromechanics, where different phases are found and some of them are not affected by damage, this approach needs to be modified. To this aim, the  Helmholtz-type equation of the implicit gradient approach has been properly generalized to the case of heterogeneous media by prescribing a non-uniform characteristic length $\ell(\vec{x})$ in accordance to the spatial arrangement of the microstructure. The choice of assigning different characteristic lengths to the phases constituting the microstructure reflects the physical mechanism underlying the non-local regularization in heterogeneous media. Indeed, it has been shown that the ratio of the characteristic length of the regularization between two neighbouring phases dictates the relevant interface conditions for the non-local variables. In this context, an interesting limit case is represented by the choice of an infinite contrast between two material phases since it theoretically leads to a free Neumann interface condition for the generalized Helmholtz-type equation.

The non-local extension of damage mechanics consists, in general, of an enriched continuum formulation where the classical balance of linear momentum is coupled with auxiliary equations of Helmholtz  type. Accordingly, additional degrees of freedom arise with respect to the conventional local models, thus requiring alternative solving schemes to be implemented. Therefore, a FFT based algorithm has been developed here due to its computational efficiency in the field of micromechanics. The proposed algorithm consists of an implicit iterative staggered scheme in which the governing equations are solved sequentially for any time increment. This approach facilitates the implementation of this class of coupled problems, since it allows for the usage of different spectral solvers for each different field equation. For the problem at hand, a FFT-Galerkin solver has been exploited for the solution of the purely mechanical problem, while the Helmholtz-type equation of the non-local regularization is solved using a conjugate gradient algorithm with a preconditioner.

The proposed non-local extension in the field of micromechanics and its numerical implementation has been analyzed on 2D and 3D numerical examples. In the 2D examples, the considered non-local damage models have been tested on a simple square RVE made up of an elastoplastic matrix with a circular reinforcement. The numerical solutions of the non-local damage models show a successful regularization and grid size independent results. It has also been shown how the characteristic length of the regularization impacts on the ductility of the composite as well as on the diffusion of non-local variables across the interface between material matrix and reinforcement. The efficiency of the proposed algorithm has been demonstrated in the simulation of the failure of a three dimentional multi-particle reinforced composite. To analyze the effect of the non-local regularization for such a complex problem, three different grid resolutions have been considered. Moreover, the development of the failure process, i.e. from nucleation to fracture propagation, has been analyzed for the different damage models considered in this paper.

In conclusion,  the FFT-based algorithm developed here constitutes a significant improvement in the prediction of ductile failure in the field of computational homegenization. Indeed, the efficiency of the proposed algorithm makes it possible the simulation of ductile damage evolution of complex microstructures with millions of degrees of freedom in a non-local context as never presented in previous publications, to the best of the author knowledge. Future possible developments of the present research may focus on the extension of the non-local formulation to anisotropic materials, e.g. for the regularization of ductile failure in polycrystals and fibres-reinforced composites. In addition, the present algorithm will be exploited for the numerical study of the size effect in the mechanical response of particle-reinforced composites.

 \section*{Acknowledgment}
The authors gratefully acknowledge the support provided by the Luxembourg National Research Fund (FNR), Reference No. 12737941. Javier Segurado acknowledges the European Union's Horizon 2020 research and innovation programme  for the project ``Multi-scale Optimisation for Additive Manufacturing of fatigue resistant shock-absorbing MetaMaterials (MOAMMM)'', grant agreement No. 862015, of the H2020-EU.1.2.1. - FET Open Programme.

\begin{appendices}

\section{Time discretization} \label{app:time_discretization}

\subsection{Gurson model}

For any time step $n=1, \, 2, \, ... \, N_t$, the time discretization of the non-local Gurson model presented in Section \ref{sec:non-local-gurson} yields

\begin{equation*}
\left\{
\begin{aligned}
& \text{div} \left[ \left. \boldsymbol{\sigma} \right|_{n+1} \right] = \vec{0}  \, \\
& \left. \overline{\varepsilon_0^p} \right|_{n+1} - \text{div} \left[ \ell^2 \left(  \vec{x}  \right)  \, \nabla \left.  \overline{\varepsilon_0^p} \right|_{n+1} \right] = \left. \varepsilon_0^p \right|_{n+1} \, , \\
& \left. \overline{\text{tr} \left[ \boldsymbol{\varepsilon}^p \right]}  \right|_{n+1} - \text{div} \left[ \ell^2 \left(  \vec{x}  \right) \, \nabla \left. \overline{\text{tr} \left[ \boldsymbol{\varepsilon}^p \right]} \right|_{n+1} \right] = \text{tr} \left[ \left. \boldsymbol{\varepsilon}^p \right|_{n+1} \right] \, ,
\end{aligned}
\right.
\end{equation*}

\noindent with 

\begin{subequations}
\begin{gather*}
\left. \boldsymbol{\sigma} \right|_{n+1} = \left. \boldsymbol{\sigma} \right|_{n} +  K \, \text{tr} \left[ \Delta {\boldsymbol{ \varepsilon }} - \Delta{\boldsymbol{ \varepsilon }}^p \right] \boldsymbol{I} \, + 2 \, \mu \, \text{dev} \left[\Delta{\boldsymbol{\varepsilon}}-  \Delta{\boldsymbol{ \varepsilon }}^p \right] \, , \\
\Delta {\boldsymbol{\varepsilon}}^p = \Delta \lambda \, \, \left[ \left. \frac{\partial \phi}{\partial p} \right|_{n+1}  \, \left.\frac{\partial p }{\partial \boldsymbol{\sigma}}\right|_{n+1} + \left. \frac{\partial \phi}{\partial s} \right|_{n+1} \,  \left. \frac{\partial s }{\partial \boldsymbol{\sigma}} \right|_{n+1} \right] \, , \\
\Delta {\varepsilon}_0^p = \frac{\left. \boldsymbol{\sigma} \right|_{n+1} : \Delta{  \boldsymbol{\varepsilon}}^p }{ \left( 1-  \left. f \right|_{n+1} \right) \, \sigma_0 \left( \left. \varepsilon_0^p \right|_{n+1} \right) } \, , \\
\Delta {f} =   \mathcal{A}_N  \,  \left( \left. \overline{\varepsilon_0^p} \right|_{n+1} \right) \,  \Delta{\overline{\varepsilon_0^p}} + \left( 1- \left. f  \right|_{n+1} \right) \,  \Delta {\overline{\text{tr} \left[ \boldsymbol{\varepsilon}^p \right]}} \, ,
\end{gather*}
\end{subequations}

\noindent and  $\Delta \lambda$ such that

\begin{equation*}
\phi \left( \left. \boldsymbol{\sigma} \right|_{n+1}, \left. \varepsilon_0^p \right|_{n+1} , \left. f_*  \right|_{n+1} \right) =   \left( \left. \frac{s}{\sigma_0} \right|_{n+1}\right) ^2 + 2 \, \left. f_*  \right|_{n+1} \, q_1 \, \text{cosh} \left( - \frac{3}{2} \left. \frac{q_2 \, p }{\sigma_0} \right|_{n+1} \right) - \left( 1 + q_3 \, \left. f_* \right|_{n+1} ^2  \right)  = 0 . 
\end{equation*}

\subsection{Lemaitre model}

For any time step $n=1, \, 2, \, ... \, N_t$, the time discretization of the non-local Lemaitre model presented in Section \ref{sec:non-local_lemaitre} yields

\begin{equation*}
\left\{
\begin{aligned}
& \text{div} \left[ \left. \boldsymbol{\sigma} \right|_{n+1} \right] = \vec{0}  \, \\
& \left. \overline{\epsilon_p} \right|_{n+1} - \text{div} \left[ \ell^2 \left(  \vec{x}  \right)  \, \nabla \left.  \overline{\epsilon_p} \right|_{n+1} \right] = \left. \epsilon_p \right|_{n+1} \, , \\
\end{aligned}
\right.
\end{equation*}

\noindent with

\begin{subequations}
\begin{gather*}
\left. \boldsymbol{\sigma} \right|_{n+1} = \left. \boldsymbol{\sigma} \right|_{n} +  \left( 1 - \left. D \right|_{n+1} \right)  \Delta  \left. \tilde{\boldsymbol{\sigma}} \right|_{n+1} - \Delta D \, \left. \tilde{\boldsymbol{\sigma}} \right|_{n+1}  , \\
\left. \tilde{\boldsymbol{\sigma}} \right|_{n+1}  = \left. \tilde{\boldsymbol{\sigma}} \right|_{n}  +  K \, \text{tr} \left[ \Delta {\boldsymbol{ \varepsilon }} - \Delta{\boldsymbol{ \varepsilon }}^p \right] \boldsymbol{I} \, + 2 \, \mu \, \text{dev} \left[\Delta{\boldsymbol{\varepsilon}}-  \Delta{\boldsymbol{ \varepsilon }}^p \right] \, , \\
\Delta\left. \boldsymbol{\varepsilon}^p \right|_{n+1} = \Delta\lambda \, \frac{ \left. \tilde{\boldsymbol{s}} \right|_{n+1} }{ \|  \left. \tilde{\boldsymbol{s}} \right|_{n+1} \|} \, ,  \\
\Delta{\epsilon_p} = \Delta \lambda \,  \sqrt{\frac{3}{2}} \, ,  \\
\end{gather*}
\end{subequations}

\noindent and  $\Delta \lambda$ such that

\begin{equation*}
\phi( \left. \tilde{\boldsymbol{\sigma}} \right|_{n+1}, \,\left.  \epsilon_p  \right|_{n+1} ) = \| \left.  \tilde{\boldsymbol{s}} \right|_{n+1} \| - \sqrt{\frac{3}{2}} \,\sigma_0 \left( \left. \epsilon_p \right|_{n+1} \right) = 0  \, .
\end{equation*}

 \end{appendices}
\end{document}